\documentclass[12pt]{article}
\usepackage{amsmath}
\usepackage{graphicx}
\usepackage{enumerate}
\usepackage[numbers]{natbib}
\usepackage{url} 
\usepackage{amsthm,amsmath,amsfonts,amssymb}

\newcommand{\blind}{1}

\usepackage{amsmath, mathrsfs, amsthm}
\usepackage{graphicx}
\usepackage{verbatim}
\usepackage{float}
\usepackage{subfigure}
\usepackage{booktabs}
\usepackage{multirow}
\usepackage{bbm}

\newtheorem{defin}{Definition}[section]
\newtheorem{condi}{Condition}[section]
\newtheorem{lem}{Lemma}[section]
\newtheorem{thm}{Theorem}[section]
\newtheorem{prop}{Proposition}[section]
\newtheorem{cor}{Corollary}[section]
\newcommand{\diag}{\mathrm{diag}}

\newcommand{\beq}{\begin{equation}}
\newcommand{\eeq}{\end{equation}}

\newcommand{\rom}[1]{%
  \textup{\uppercase\expandafter{\romannumeral#1}}%
}

\allowdisplaybreaks
\usepackage{hyperref}
\hypersetup{
    colorlinks=true,
    linkcolor=blue,
    filecolor=magenta,      
    urlcolor=cyan,
}

\begin{document}

\def\spacingset#1{\renewcommand{\baselinestretch}%
{#1}\small\normalsize} \spacingset{1}


\if1\blind
{
  \title{\bf Using SVD for Topic Modeling}
  \author{Zheng Tracy Ke\\
    Department of Statistics, Harvard University\vspace{.3cm}\\
    and \vspace{.3cm}\\
    Minzhe Wang\\
    Department of Statistics, University of Chicago}
  \maketitle
} \fi

\if0\blind
{
  \bigskip
  \bigskip
  \bigskip
  \begin{center}
    {\LARGE\bf Using SVD for Topic Modeling}
\end{center}
  \medskip
} \fi

\bigskip 

\begin{abstract}
The probabilistic topic model imposes a low-rank structure on the expectation of the corpus matrix. Therefore, singular value decomposition (SVD) is a natural tool of dimension reduction. We propose an SVD-based method for estimating a topic model. 
Our method constructs an estimate of the topic matrix from only a few leading singular vectors of the corpus matrix, and has a great advantage in memory use and computational cost for large-scale corpora. 
The core ideas behind our method include a pre-SVD normalization to tackle severe word frequency heterogeneity, a post-SVD normalization to create a low-dimensional word embedding that manifests a simplex geometry, and a post-SVD procedure to construct an estimate of the topic matrix directly from the embedded word cloud. We provide the explicit rate of convergence of our method. We show that our method attains the optimal rate in the case of long and moderately long documents, and it improves the rates of existing methods in the case of short documents. The key of our analysis is a sharp row-wise large-deviation bound for empirical singular vectors, which is technically demanding to derive and potentially useful for other problems. We apply our method to a corpus of Associated Press news articles and a corpus of abstracts of statistical papers.  
\end{abstract}

\noindent%

\vfill

\newpage
\spacingset{1.45}

\addtocontents{toc}{\protect\setcounter{tocdepth}{0}}

\section{Introduction} \label{sec:intro}
Topic modeling \citep{blei2003latent} is a useful tool for natural language processing, with many applications in digital humanities,  computational social science and  e-commerce \citep{blei2012probabilistic, fan2021much, wang2021bayesian}.  
Recently, it has also found applications in genetics \citep{dey2017visualizing}, bioinformatics, and personalized medicine \citep{li2021topic}. 
Suppose we observe $n$ documents written on a vocabulary of $p$ words. Let $N_i\in\mathbb{N}$ denote the length of document $i$, $1\leq i\leq n$. The {\it corpus} matrix $D \in \mathbb{R}^{p,n}$ is defined by 
\[
D(j,i) = \frac{\text{count of word $j$ in document $i$}}{N_i},  \qquad 1\leq i\leq n,1\leq j\leq p. 
\]
The {\it probabilistic Latent Semantic Indexing (pLSI)} model \citep{hofmann1999} is a popular topic model. Let $A_1,A_2,\ldots, A_K\in\mathbb{R}^p$ be $K$ topic vectors, where each $A_k$ is a probability mass function (PMF) on the vocabulary. Each document $i$ is associated with a nonnegative vector  $w_i\in\mathbb{R}^K$, where 
$w_i(k)$ is this document's weight on topic $k$, satisfying that $\sum_{k=1}^K w_i(k)=1$. Let $d_i\in\mathbb{R}^p$ denote the $i$th column of $D$. Then, $N_id_i$ is the vector of word counts in document $i$. The pLSI model assumes that $\{N_id_i\}_{1\leq i\leq n}$ are independently generated, with  
\beq \label{pLSI}
N_id_i \sim \mathrm{Multinomial}\Bigl(N_i,\;  \sum_{k=1}^K w_i(k)A_k \Bigr), \qquad 1\leq i\leq n. 
\spacingset{1} \footnote{$\mathrm{Multinomial}(N, v)$ denotes the multinomial distribution with $N$ being the number of trials and $v$ being the vector of event probabilities.} \spacingset{1.45} 
\eeq
Write $A = [A_1, A_2,   \ldots, A_K]\in\mathbb{R}^{p\times K}$ and $W=[w_1,w_2,\ldots,w_n]\in\mathbb{R}^{K\times n}$. We call $A$ and $W$ the {\it topic matrix} and {\it topic weight matrix}, respectively. Model~\eqref{pLSI} implies that 
\[
\mathbb{E}[D] \quad = \quad AW. 
\]
In real applications, $(n,p)$ are usually very large, but $K$ is small. The topic model imposes a low-rank structure on $\mathbb{E}[D]$. We are interested in estimating $A$ from $D$. 

In the literature, there are two major approaches to topic modeling. The first is Latent Dirichlet Allocation (LDA) \citep{blei2003latent}. It imposes a Dirichlet prior on $w_1,w_2,\ldots,w_n$ and estimates $A$ by variational EM algorithms. The second is the anchor word approach \citep{Ge}.  
It imposes the ``anchor word assumption" (see Definition~\ref{def:anchor}) on $A_1,A_2,\ldots,A_K$ and estimates model parameters by computing an approximate nonnegative factorization on $D$. 
In this paper, we focus on topic model estimation when the anchor word assumption \citep{donoho2003does,Ge} is satisfied.

\begin{defin} \label{def:anchor}
We call word $j$ an {\it anchor word} if row $j$ of $A$ has exactly one nonzero entry, and an {\it anchor word  for topic $k$} if the nonzero entry is located at column $k$, $1 \leq k \leq K$.  
\end{defin}

An anchor word is a unique `signature' of a topic. 
Evidence of anchor words was observed in real data. 
Ji et al. \citep{PaseII-paper2} analyzed abstracts of statistical papers and identified 11 topics. They discovered a list of anchor words for each topic. For example, for the topic ``Experimental Design", its anchor words are {\it aoptim}, {\it doptim}, {\it aberr}, ect.; for the topic ``Hypothesis testing", its anchor words are {\it stepdown}, {\it familywise}, {\it bonferroni}, etc.. 
Throughout this paper, we assume each topic has at least one anchor word. This is almost the necessary condition for identifiability of parameters in pLSI \citep{donoho2003does}.

The pLSI model is traditionally estimated by EM algorithm. Model \eqref{pLSI} is equivalent to a hierarchical data generating process where a latent topic label in $\{1,2,\ldots,K\}$ is drawn for every word in a document, and an EM algorithm   can be designed to maximize the likelihood \citep{mei2001note}. The pLSI model also induces a nonnegative matrix factorization (NMF) on $\mathbb{E}[D]$, so the NMF algorithms can be be used to estimate parameters of pLSI, with a proper normalization of the obtained factor matrices \citep{ding2008equivalence}. 
However,  these traditional methods do not explore the anchor word condition, so they face the identifiability issue and do not  guarantee to produce a consistent estimate of $A$.

The anchor word condition is equivalent to the separability condition in NMF literature, and ``separable NMF algorithms" have been widely used to fit the pLSI model. Arora et al. \citep{Ge} is one such method.
They started from the {\it word co-occurrence} matrix $DD'$ and applied a successive projection algorithm to rows of $DD'$ to find one anchor word per topic; they used these anchor words to re-arrange $DD'$ into four blocks, where the top left $K\times K$ block corresponds to the set of found anchor words; last, they estimated $A$ by taking advantage of the special structure in this block partition. Bing et al. \citep{bing2020fast} proposed another method for estimating $A$ based on the word co-occurrence matrix. They first identified a set of anchor words for each topic by alternatively checking the row maximum and column maximum of $DD'$, and then constructed an estimator of $A$ by pooling information in these rows/columns.

Despite of these interesting algorithms in the literature, one problem still remains open - how to use Singular Value Decomposition (SVD) to estimate a topic model. 
Since $\mathbb{E}[D]$ has a low rank, SVD is a natural and powerful tool for dimension reduction. 
It can be shown that information of $A$ is fully contained in the first $K$ left singular vectors of $D$. 
Working on this $p\times K$ matrix of singular vectors requires much less memory, compared with working on the corpus matrix $D$ or the word co-occurrence matrix $DD'$.  
Furthermore, SVD creates a projection of each row of $D$ into a {\it low-dimensional} space. If we can develop a method that estimates $A$ from these low-dimensional vectors, it will reduce the computational cost significantly. These advantages of SVD become prominent where $(n, p)$ get very large. In many real applications, $n$ is at the order of $10^4\sim 10^7$ and $p$ is at the order of $10^3\sim 10^4$, so  we expect to enjoy a great benefit from using an SVD-based method.   In fact, even before topic models were invented, SVD was already popular in ad-hoc semantic analysis \citep{deerwester1990indexing}.

Unfortunately, to our best knowledge, there has not yet been a rigorous method about using SVD for topic model estimation.   
There are two big hurdles. 
\begin{itemize} \itemsep -1pt
\item {\it The connection between singular vectors and the target quantity $A$ is opaque}. Even in the noiseless case, the population singular vectors are not explicit functions of $A$. It is unclear how to construct a valid estimate of $A$ from the singular vectors. 
\item {\it It lacks technical tools for analyzing the performance}. The analysis of an SVD-based method is technically challenging. It requires sharp large-deviation bounds for {\it each entry} of singular vectors, which is known to be sophisticated \citep{abbe2017entrywise,fan2020asymptotic}. 
Such results are rarely available in the literature. 
\end{itemize}
This paper tackles these challenges. Our main contributions are three-fold:
\begin{itemize} \itemsep -1pt
\item We propose a new SVD-based method for estimating a topic model. It constructs an estimate of $A$ using only the first $K$ left singular vectors of $D$. 
\item We give the error rate of our method. We show that our rate is minimax optimal for a wide parameter regime. We also show that our rate improves those in the literature, especially in the case of short documents and/or severe word frequency heterogeneity.  
\item We provide an entry-wise large-deviation bound for leading singular vectors. This is a technical tool that is potentially useful for other SVD-based analysis of text data. 
\end{itemize}

In the literature, there exist topic modeling methods that use SVD, but none of them meet our criteria of an ``SVD-based" method. They apply SVD to either construct a low-rank approximation of the data matrix \citep{bansal2014provable} or assist the anchor word selection in an existing method such as \cite{Ge}. These methods do not aim to estimate $A$ from singular vectors directly. They instead use SVD to ``de-noise" one step of the algorithm, but the core idea is still to estimate $A$ from either $D$ or $DD'$. Therefore, these methods do not enjoy the full advantage of dimension reduction by SVD. 

We propose a new method, Topic-SCORE, to estimate $A$ from leading singular vectors directly. The method contains several innovative ideas, including a {\it pre-SVD normalization} to deal with severe word frequency heterogeneity, a {\it post-SVD normalization} to create low-dimensional word embeddings that exhibit a simplex geometry, and a simple-to-implement {\it post-SVD procedure} to construct $\hat{A}$ from the word embeddings.  

The {\it pre-SVD normalization} aims to tackle severe word frequency heterogeneity in real corpora. As a consequence of frequency heterogeneity, the noise levels in different rows of $D$ are different. If we apply SVD directly, the signal-to-noise ratio is non-optimal. 
The pre-SVD normalization adjusts the noise levels in different rows to improve accuracy of SVD. 
The {\it post-SVD normalization} creates a low-dimensional word embedding that supports estimation of $A$. Although SVD natually creates a word embedding into $\mathbb{R}^K$, these embedded points are not ready to use. In the noiseless case, they are contained in a {\it simplicial cone} with $K$ supporting rays \citep{donoho2003does}. To facilitate estimation of $A$, we must normalize these embedded points properly, so that in the noiseless case the normalized points are contained in a {\it simplex} with $K$ vertices. 
We borrow the idea of SCORE normalization \citep{SCORE,Mixed-SCORE} in network data analysis to design a satisfactory post-SVD normalization. 
Given the word embedding, we then design a {\it post-SVD procedure} to construct $\hat{A}$. 
It has to coordinate with the pre-SVD and post-SVD normalizations and ``revert" these normalizations in a proper way. 
In our method, after SVD is done, we only operate on the low-dimensional word embeddings and never need to return to $D$ or $DD'$. Therefore, our method enjoys the full benefit of dimension reduction by SVD, especially on  memory use and computational cost (see Section~\ref{sec:Method} and Table~\ref{table:semi-synthetic}).

We provide the rate of convergence of our method under the $L^1$-loss: $\sum_{k=1}^K\|\hat{A}_k-A_k\|_1$. We let $n\to\infty$ and allow the vocabulary size $p$ and the average document length $N$ to grow with $n$. We show the optimality of our method by giving a matching lower bound. Our results cover both cases of long documents ($N>p$) and short documents ($N<p$), and we show that the error rate of our method is insensitive to severe word frequency heterogeneity. 
A key technical tool in our analysis is the {\it row-wise large-deviation bound} for empirical singular vectors. 
There have been some recent theoretical results about getting such row-wise bounds for eigenvectors of sub-Gaussian random matrices or network adjacency matrices \citep{abbe2017entrywise,fan2020asymptotic}, but their techniques do not apply to our setting, because the entries of $D$ have heavy tails 
and weak dependence.  
We prove the above bound using non-trivial new techniques.

While we primarily focus on estimating $A$ in this paper, our method also yields a simple approach to estimating $W$, where we run a weighted least-squares by regressing each column of $D$ on the columns of $\hat{A}$. We also give the error rate on estimating $W$ by this approach, as a by-product of our main results.

The remaining of this paper is organized as follows. In Section~\ref{sec:Method}, we describe our method and explain the rationale of each step. In Section~\ref{sec:Theory}, we present the theoretical results, including the error rate and the row-wise large deviation bounds for singular vectors. In Section~\ref{sec:Realdata}, we apply our method to two real corpora, one consisting of Associated Press news articles and the other consisting of paper abstracts from representative statistics journals. Section~\ref{sec:Simu} contains simulations. Section~\ref{sec:Discuss} contains discussions.

\section{An SVD-basd method for topic matrix estimation} \label{sec:Method}
We recall that $D\in\mathbb{R}^{p\times n}$ is the corpus matrix, $A=[A_1,A_2,\ldots,A_K]\in\mathbb{R}^{p\times K}$ contains the $K$ topic vectors, and $W=[w_1,w_2,\ldots,w_n]\in\mathbb{R}^{K\times n}$ contains the weight vectors of documents. By model~\eqref{pLSI}, $\mathbb{E}[D]=AW\equiv D_0$. Below, in Section~\ref{subsec:oracle-method}, we consider an oracle case, where $D_0$ is directly observed. We propose an oracle procedure for recovering $A$ from $D_0$. In Section~\ref{subsec:real-method}, 
we consider the real case where $D$, instead of $D_0$, is observed. We modify the oracle procedure to deal with stochastic noise, which gives our final method.

\subsection{The oracle case} \label{subsec:oracle-method}
In the oracle case, we observe the non-stochastic matrix $D_0$. Let $M_0\in\mathbb{R}^{p\times p}$ be an arbitrary diagonal matrix with strictly positive diagonals. We first normalize $D_0$ to $M_0^{-1/2}D_0$. 
This mimics the pre-SVD normalization to be used in the real case. 
Let $\sigma_1\geq\sigma_2\geq\cdots\geq \sigma_K>0$ be the nonzero singular values of $M_0^{-1/2}D_0$, and let $\xi_1,\xi_2,\ldots,\xi_K\in\mathbb{R}^p$ be the corresponding singular vectors. 
Write $\Xi=[\xi_1,\xi_2,\ldots,\xi_K]$.

\begin{defin} \label{def:cone+simplex}
A simplicial cone with $K$ supporting rays $u_1,u_2,\ldots,u_K$ is the set of points $x$ such that $x=\sum_{k=1}^K a_ku_k$, where $a_k\geq 0$ for $1\leq k\leq K$. A simplex with $K$ vertices $v_1, v_2, \ldots, v_K$ is the set of points $x$ such that $x=\sum_{k=1}^K b_ku_k$, where $b_k\geq 0$ and $\sum_{k=1}^K b_k=1$.
\end{defin}

The next lemma describes the geometry of the point cloud formed by rows of $\Xi$.

\begin{lem} \label{lem:simplicial-cone}
Suppose each topic has at least one anchor word. Denote by $x_j\in\mathbb{R}^K$ the $j$th row of $\Xi$, $1\leq j\leq p$. 
There exists a simplicial cone with $K$ supporting rays 
such that: (i) Each $x_j$ is contained in this simplicial cone.  (ii) If $j$ is an anchor word of topic $k$, then $x_j$ is located on the $k$th supporting ray of this simplicial cone. 
\end{lem}

\begin{figure}[tb!]
\centering
\includegraphics[height = .24\textwidth]{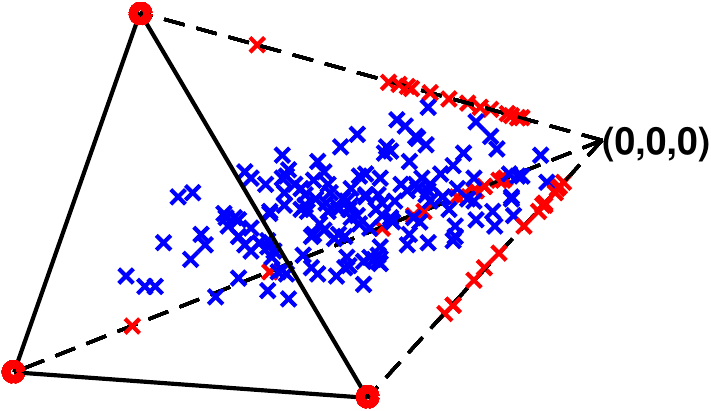} \hspace{.48 in}
\includegraphics[height = .26\textwidth]{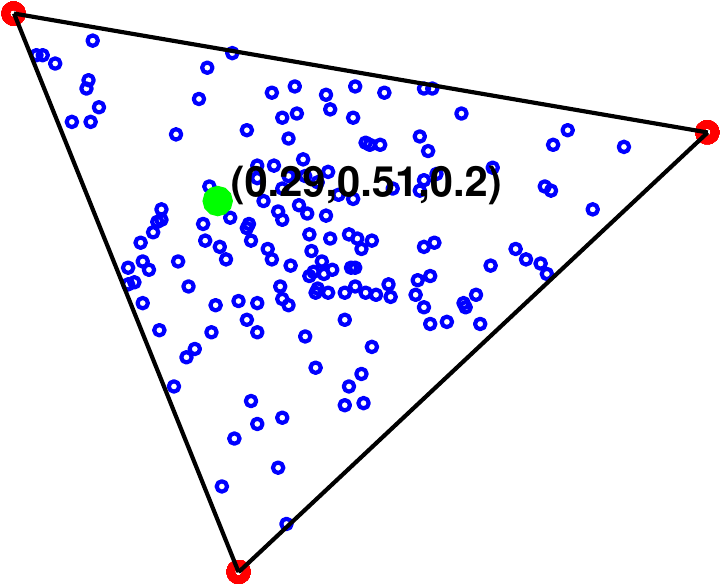} 
\caption{The geometry in the spectral domain ($K=3$). Left: rows of $\Xi$ and the simplicial cone. The red and blue crosses correspond to anchor rows and non-anchor rows, respectively. The dashed lines are the supporting rays of the cone. For visualization, we also plot a cross-section of the cone, which is the solid triangle (the red circles help visualize the shape of the cross-section but do not correspond to any row of $\Xi$). Right: rows of $R$ and the Ideal Simplex (solid triangle). The green dot shows one $r_j$, where its corresponding $\pi_j$ is $(0.29, 0.57, 0.2)'$.}\label{fig:simplex}
\end{figure}

An example with $K=3$ is given in Figure~\ref{fig:simplex} (left panel). We assume each topic has at least one anchor word. It means there is at least one $x_j$ located on each supporting ray of the simplicial cone. However, it is unclear how to use this geometry to assist the estimation of $A$. It is even unclear how to recovery this simplicial cone from the point cloud of $x_j$'s. We hope to conduct a normalization on each $x_j$, such that the simplicial cone is converted to a simplex, where each supporting ray is `compressed' into one vertex of the simplex. See Figure~\ref{fig:simplex} (right panel). 
Then, we can easily recover this simplex by computing the convex hull of $x_j$'s (there are many algorithms for computing the convex hull of a point cloud). 

What we desire here is a post-SVD normalization that produces a simplex. If all $x_j$'s are non-negative vectors, we can simply normalize each $x_j$ by its own $\ell^1$-norm. Unfortunately, the mutual orthogonality of singular vectors makes it impossible that all $x_j$'s are non-negative vectors, and so the naive normalization by $\ell^1$-norm does not work. We borrow the SCORE normalization \citep{SCORE} from network data analysis, where we normalize each $x_j$ by its first coordinate. In the normalized vector, the first coordinate is always equal to $1$ and is dropped. This gives rise to the following matrix
$R\in\mathbb{R}^{p\times (K-1)}$, where 
\begin{equation} \label{SCORE-oracle} 
R(j, k) = \xi_{k+1}(j) / \xi_1(j), \qquad 1 \leq j \leq p,   \; 1 \leq k \leq K-1. 
\end{equation} 
Write $R=[r_1,r_2,\ldots,r_p]'$. Then, $r_j$ is the low-dimensional embedding of word $j$ into $\mathbb{R}^{K-1}$, for each $1\leq j\leq p$. 
For \eqref{SCORE-oracle} to be well-defined, we need that each entry of $\xi_1$ is nonzero. 
Since $\xi_1$ is the first singular vector of a nonnegative matrix, this is guaranteed by Perron's theorem, under mild regularity conditions. The next lemma shows that the point cloud of $r_1,r_2,\ldots,r_p$ are indeed contained in a simplex:

\begin{lem}[Ideal Simplex] \label{lem:IdealSimplex}
Suppose each topic has at least one anchor word. Denote by $r_j\in\mathbb{R}^{K-1}$ the $j$th row of $R$, $1\leq j\leq p$. 
There exists a simplex ${\cal S}^*_K$ with $K$ vertices $v_1^*, v_2^*,\ldots,v_K^*$ such that: 
(i) Each $r_j$ is contained in ${\cal S}^*_K$ and can be written as a convex linear combination of the $K$ vertices: $r_j=\sum_{k=1}^K \pi_j(k)v_k^*$, where $\pi_j(k)\geq 0$ and $\sum_{k=1}^K \pi_j(k)=1$.  (ii) If $j$ is an anchor word of topic $k$, then $r_j$ is located at the vertex $v_k^*$, $1\leq k\leq K$.   
\end{lem}

We call ${\cal S}_K^*$ the Ideal Simplex (this simplex is uniquely determined by $\xi_1,\xi_2,\ldots,\xi_K$). See Figure~\ref{fig:simplex} (right panel).  
The original SCORE normalization \citep{SCORE} was applied to eigenvectors of a network adjacency matrix, in order to remove the effect of degree heterogeneity. We use a similar normalization here, but for a very different purpose: Our post-SVD normalization is applied to singular vectors of a text corpus matrix, in order to produce an Ideal Simplex.

\spacingset{1} 
\spacingset{1.45}

Given the embedded low-dimensional point cloud $r_1,r_2,\ldots,r_p$, we can simultaneously recover the $K$ vertices of the Ideal Simplex and the convex combination vectors $\pi_1,\pi_2,\ldots,\pi_p$ in Lemma~\ref{lem:IdealSimplex}. We first identify the vertices $v_1^*,v_2^*,\ldots,v_K^*$ by computing the convex hull of the point cloud. Next, for each $j$, we solve $\pi_j$ from the linear equation:
\beq \label{solving-Pi}
\begin{pmatrix}
1 & \cdots & 1\\
v_1^* & \cdots & v_K^* 
\end{pmatrix}\pi_j = \begin{pmatrix}1 \\ r_j\end{pmatrix}. 
\eeq
Write $\Pi=[\pi_1,\pi_2,\ldots,\pi_p]'\in\mathbb{R}^{p\times K}$. We now introduce an explicit procedure to recover $A$ from $\Pi$. It is based on the following lemma:

\begin{lem}[Recovery of $A$]  \label{lem:Pi-to-A} 
There exists a a positive vector $q\in\mathbb{R}^K$ such that 
$M_0^{-1/2} A \cdot \diag(q) =  \diag(\xi_1)\cdot \Pi$. 
\end{lem} 

By Lemma~\ref{lem:Pi-to-A}, we have 
\beq\label{recovering-A}
A\cdot \diag(q) = M_0^{1/2}\cdot \diag(\xi_1)\cdot \Pi. 
\eeq
On the right hand side of \eqref{recovering-A}, $M_0$ and $\mathrm{diag}(\xi_1)$ contain the normalizing factors in the pre-SVD normalization and post-SVD normalization, respectively, which are known. Therefore, we can obtain the right hand side of \eqref{recovering-A}, which gives an estimate of $A\cdot \diag(q)$. We then utilize the fact that each column of $A$ has a unit $\ell^1$-norm. We thus recover $A$ by dividing each column of $A\cdot \diag(q)$ by its own $\ell^1$-norm.  

Summarizing the above results gives an oracle procedure for recovering $A$ from $D_0$:
\begin{itemize} \itemsep -2pt
\item {\it (Pre-SVD normalization)}. Normalize $D_0$ to $M_0^{-1/2}D_0$, where $M_0$ can be any diagonal matrix with positive diagonal entries. 
\item {\it (SVD)}. Obtain $\xi_1,\xi_2,\ldots,\xi_K$, the left singular vectors of $M_0^{-1/2}D_0$. 
\item {\it (Post-SVD normalization)}. Obtain the matrix $R=[r_1,r_2,\ldots,r_p]'$ as in \eqref{SCORE-oracle}. 
\item {\it (Vertex hunting)}. Use the low-dimensional point cloud $r_1,r_2,\ldots, r_p$ to find the vertices $v_1^*, v_2^*,\ldots,v_K^*$ of the Ideal Simplex. 
\item {\it (Topic matrix estimation)}. For $1\leq j\leq p$, solve $\pi_j$ from \eqref{solving-Pi}. Write $\Pi=[\pi_1,\pi_2,\ldots,\pi_p]'$. 
Obtain the matrix $M_0^{1/2}[\diag(\xi_1)] \Pi$ and normalize each column to have a unit $\ell^1$-norm. The resulting matrix is exactly $A$. 
\end{itemize}
We call this method the {\it oracle Topic-SCORE}. 

{\bf Remark 1}. In this oracle procedure, the pre-SVD normalization, post-SVD normalization and post-SVD steps are designed carefully to coordinate with each other. 
For example, although the pre-SVD normalization affects the singular vectors, the post-SVD normalization guarantees to produce an Ideal Simplex, regardless of the choice of $M_0$. Furthermore, the normalizing factors we use in the pre-SVD and post-SVD normalizations are both incorporated in the last step of recovering $A$ from $\Pi$. The pre-SVD normalization uses $M_0^{-1/2}$ to normalize $D_0$, and the post-SVD normalization uses $[\diag(\xi_1)]^{-1}$ to normalize $\Xi$. In the last step, we ``revert" these normalizations by multiplying $\Pi$ by $M_0^{1/2}\cdot \diag(\xi_1)$.

\subsection{The real case} \label{subsec:real-method}
In the real case, we are given $D$, a noisy version of $D_0$. Most steps in the oracle procedure can be directly extended, except for {\it Pre-SVD normalization} and {\it Vertex hunting}. 

We first consider the pre-SVD normalization. In the oracle case, we are free to choose the diagonal matrix $M_0$. However, in the real case, we must choose $M_0$ carefully, in hopes of adjusting the noise level in different rows and boosting the signal-to-noise ratio in SVD. 
By model \eqref{pLSI}, $D(j,i)\sim  N_i^{-1}\mathrm{Binomial}(N_i, \, D_0(j,i))$, where $N_i$ is the length of document $i$. When $\|D_0\|_{\max}\leq 1-c$ and all $N_i$'s are the same order, for every $1\leq j\leq p$, 
\[
\sum_{i=1}^n \mathrm{Var}\bigl([M_0^{-1/2}D_0](j,i)\bigr) = \frac{\sum_{i=1}^n N_i^{-1}D_0(j,i)[1-D_0(j,i)]}{M_0(j,j)}\asymp \frac{\sum_{i=1}^n D_0(j,i)}{N\cdot M_0(j,j)}. 
\]
By choosing $M_0(j,j)\propto \sum_{i=1}^n D_0(j,i)$, we can make the sum of variances of each row to be at the same order. 
This motivates us to use 
\beq \label{M0}
M_0 = \diag\bigl(n^{-1}D_0{\bf 1}_n\bigr). 
\eeq
There is a deeper reason for choosing this $M_0$: It allows us to get the sharp row-wise large-deviation bounds for singular vectors of $M_0^{-1/2}D$ (to be presented in Section~\ref{sec:Theory}). 
In fact, we first derived these large-deviation bounds for an arbitrary $M_0$ and then picked the current $M_0$ to optimize these bounds; this motivation is buried in our theoretical analysis (see Section~\ref{sec:high-level-proof} of the supplementary material for a detailed explanation). 
In \eqref{M0}, $D_0$ is not observed. We replace $M_0$ by a stochastic proxy, $M$, where for $1\leq j\leq p$, 
\beq \label{M}
M(j,j) = \max\bigl\{\hat{\eta}_j, \;\mathrm{quantile}_{\tau}(\hat{\eta})\bigr\}, \qquad\mbox{with}\quad \hat{\eta} = n^{-1}D{\bf 1}_n. 
\eeq
Here, $\mathrm{quantile}_{\tau}(\hat{\eta})$ is the $\tau$-quantile of $\hat{\eta}_1,\hat{\eta}_2,\ldots,\hat{\eta}_p$. For theoretical results and simulations in this paper, we always set $\tau=0$, so that $M=\diag(n^{-1}D{\bf 1}_n)$. In real data analysis, it is sometimes beneficial to use a positive value of $\tau$, to avoid over-weighting those extremely-low-frequency words in the pre-SVD normalization.

\begin{figure}[tb!]
\centering
\includegraphics[width = .31\textwidth, trim=40 100 20 100, clip=true]{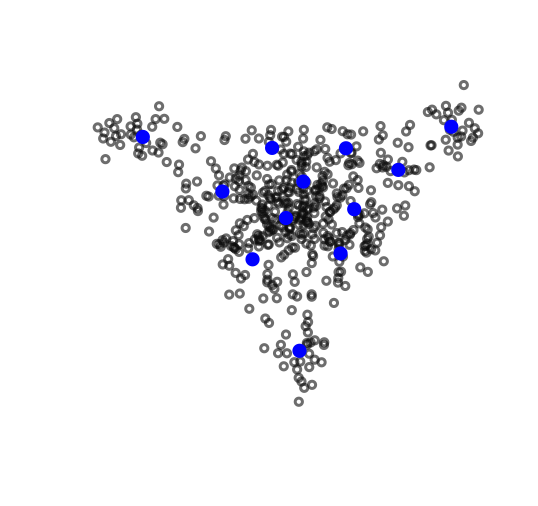}
\includegraphics[width = .31\textwidth, trim=40 100 20 100, clip=true]{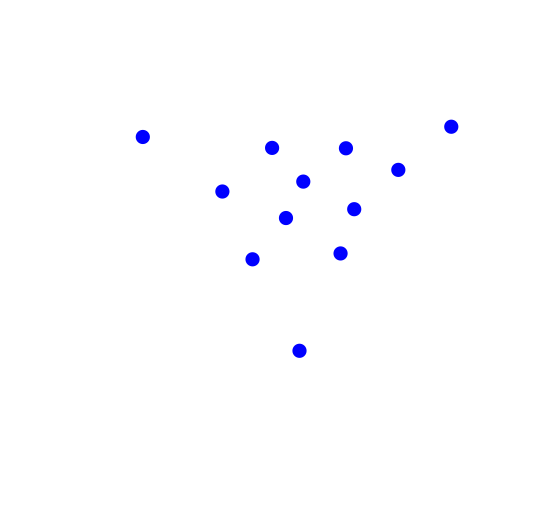}
\includegraphics[width = .315\textwidth, trim=40 100 20 100, clip=true]{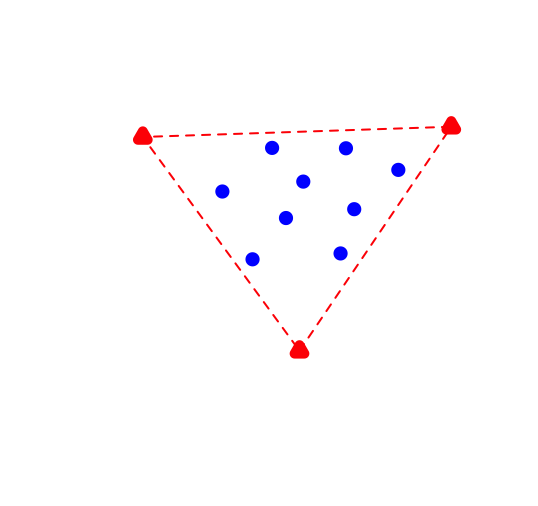}
\caption{The SVS algorithm for vertex hunting ($K=3$). Left: Apply the classical $k$-means to $\hat{r}_1,\ldots,\hat{r}_p$ and obtain the Euclidean centers of clusters (blue points). Middle: Remove $\hat{r}_1,\ldots,\hat{r}_p$ and only keep the cluster centers. Right: Fit a simplex using these cluster centers.}\label{fig:VH}
\end{figure}

Next, we consider the vertex hunting step. Let $\hat{\xi}_1,\hat{\xi}_2,\ldots,\hat{\xi}_K$ be the first $K$ left singular vectors of $M^{-1/2}D$. We define a stochastic proxy for the matrix $R$ in \eqref{SCORE-oracle}:
\beq \label{SCORE-real} 
\hat{R}(j,k) = 
\hat{\xi}_{k+1}(j) / \hat{\xi}_1(j),   \qquad 1 \leq k \leq K -1,  \; 1 \leq j \leq p.  
\eeq 
Let $\hat{r}_j'$ denote the $j$th row of $\hat{R}$, $1\leq j\leq p$. The point cloud $\hat{r}_1,\hat{r}_2,\ldots,\hat{r}_p$ gives a ``blurred" version of the Ideal Simplex (see Figure~\ref{fig:VH}, left panel). 
We can no longer find the vertices by computing the convex hull of the point cloud.

The problem here is how to learn a simplex from a noise-corrupted point cloud. 
Fortunately, this problem has been considered in the literature of linear unmixing analysis \citep{bioucas2012hyperspectral}, with many available algorithms. 
We thereby replace the {\it Vertex Hunting} step in the oracle Topic-SCORE by one of those existing algorithms.

We discuss two vertex hunting algorithms. The first is successive projection (SP) \citep{araujo2001successive}. It starts from finding $\hat{r}_j$ whose Euclidean norm is the largest and setting this $\hat{r}_j$ as the first estimated vertex $\hat{v}^*_1$. Then, for each $2\leq k\leq K$, it subsequently finds $\hat{v}_k$ from $\hat{v}_1,\ldots,\hat{v}_{k-1}$ as follows: Let $P_{k-1}$ be the projection matrix to the linear span of $\hat{v}^*_1,\ldots,\hat{v}^*_{k-1}$. The algorithm selects $j$ to maximize $\|(I - P_{k-1})\hat{r}_j\|$ and sets the corresponding $\hat{r}_j$ as $\hat{v}^*_k$. 
The SP algorithm is easy to implement and has a low computational cost. It works well when the noise level in $\hat{r}_j$'s is small. However, SP is not robust to strong noise or outliers. 

\spacingset{1} 
\spacingset{1.45}

The second is sketched vertex search (SVS) \citep{Mixed-SCORE}. 
This algorithm has a {\it de-noise} step and a {\it vertex search} step. The {\it de-noise} step processes $\hat{r}_1,\hat{r}_2,\ldots,\hat{r}_p$ in hopes of reducing the noise level significantly. This is done by applying a k-means algorithm pretending that there are $L$ clusters, where $L$ is a tuning integer that is much smaller than $p$ but considerably larger than $K$. Let $\hat{c}_1,\hat{c}_2,\ldots,\hat{c}_L$ be the cluster centers output by k-means. By nature of k-means, each $\hat{c}_\ell$ is an average of nearby $\hat{r}_j$'s. Therefore, these cluster centers are less noisy than the originally observed $\hat{r}_j$'s.  
Next, the {\it vertex search} step fits a simplex using $\hat{c}_1,\hat{c}_2,\ldots,\hat{c}_L$. This is done by searching all simplexes ${\cal S}$ whose $K$ vertices are located on these cluster centers and selecting the simplex such that the maximum distance from any $\hat{c}_\ell$ to ${\cal S}$ is minimized. 
An illustration for $K=3$ is in Figure~\ref{fig:VH}. 
SVS performs especially well under strong noise. 

Since these vertex hunting algorithms are not the main contributions of this paper, we relegate the pseudo-code of SP and SVS to the supplementary material. In principle, we can plug in any vertex hunting algorithm. 

We now present our main algorithm, {\it Topic-SCORE}, which is a modification of the oracle procedure in Section~\ref{subsec:oracle-method}. 
Input: $D$, $K$, and a vertex hunting (VH) algorithm. Output: $\hat{A}$. 
\begin{itemize} \itemsep -2pt
\item {\it (Pre-SVD normalization)}. Normalize $D$ to $M^{-1/2}D$, where $M$ is as in \eqref{M}. 
\item {\it (SVD)}. Obtain $\hat{\xi}_1,\hat{\xi}_2,\ldots,\hat{\xi}_K$, the left singular vectors of $M^{-1/2}D$. 
\item {\it (Post-SVD normalization)}. Obtain $\hat{R}$ as in \eqref{SCORE-real}. Let $\hat{r}_1,\hat{r}_2,\ldots,\hat{r}_p$ denote its rows.  
\item {\it (Vertex hunting)}. Apply the VH algorithm on $\hat{r}_1,\hat{r}_2,\ldots, \hat{r}_p$ to get $\hat{v}^*_1, \hat{v}^*_2,\ldots, \hat{v}^*_K$. 
\item {\it (Topic matrix estimation)}. For $1 \leq j \leq p$, solve $\hat{\pi}_j^*$ from 
\[
\begin{pmatrix}1 &\ldots& 1\\
\hat{v}_1^* &\ldots & \hat{v}_K^*
\end{pmatrix}\hat{\pi}_j^* = \begin{pmatrix}1\\\hat{r}_j\end{pmatrix}. 
\] 
Obtain $\hat{\pi}_j$ from $\hat{\pi}_j^*$ by first setting the negative entries to $0$ and then renormalizing the vector to have a unit $\ell^1$-norm. \spacingset{1}\footnote{We modify $\hat{\pi}_j^*$ to $\hat{\pi}_j$, to get an eligible weight vector. Note that $\hat{\pi}_j$ differs from $\hat{\pi}_j^*$ only if $\hat{r}_j$ is outside the estimated simplex. The fraction of such $\hat{r}_j$'s is small.}\spacingset{1.45}
Write $\hat{\Pi}=[\hat{\pi}_1,\hat{\pi}_2,\ldots,\hat{\pi}_p]'$.  Obtain $\hat{A}$ from the matrix $M^{1/2}[\diag(\hat{\xi})] \hat{\Pi}$ by re-normalizing each column to have a unit $\ell^1$-norm.  
\end{itemize}

The computational cost of our method mainly comes from the SVD step and the vertex hunting step. For SVD, since we only need to compute a few leading singular vectors, the complexity is $\tilde{O}(np)$. For vertex hunting, if we use SP, the complexity is $O(p^2)$. Therefore, our method is a polynomial time algorithm. 
In Section~\ref{subsec:semi-synthetic}, we report the running time on semi-synthetic data calibrated from real corpora. 
It suggests that our method is much faster than some popular methods. 

\bigskip
\noindent 
{\bf Estimation of $W$}: As a byproduct, once  $\hat{A}$ is obtained, we can estimate $w_i$ by running a regression of $d_i$ on  $\hat{A}_1,\hat{A}_2,\ldots,\hat{A}_K$. We propose a weighted least-squares, where the weights come from the normalizing factors in the pre-SVD normalization and aim to tackle severe frequency heterogeneity: 
\beq \label{w-hat}
\hat{w}^*_i = \mathrm{argmin}_{b}\sum_{j=1}^p \frac{1}{M(j,j)} \Bigl[D(j,i)-\sum_{k=1}^K b(k)\hat{A}_k(j)\Bigr]^2, \qquad 1\leq i\leq n. 
\eeq
We then set the negative entries of $\hat{w}_i^*$ to zero and re-normalize it to have a unit $\ell^1$-norm. This gives $\hat{w}_i$, $1\leq i\leq n$.

{\bf Remark 2} {\it (Our method avoids anchor word selection)}. Although we assume existence of anchor words, our method does not select anchor words explicitly. The anchor word assumption is only needed for the success of vertex hunting, to ensure that there is at least one $\hat{r}_j$ near each true vertex. Our $\hat{A}$ is constructed from the estimated vertices $\hat{v}_1^*,\hat{v}_2^*,\ldots,\hat{v}^*_K$, where each $\hat{v}_k^*$ may not correspond to any particular word (e.g., if we use the SVS algorithm for vertex hunting, each $\hat{v}_k^*$ is a cluster center from the k-means, which is the average of many $\hat{r}_j$'s). In contrast, \cite{Ge, bing2020fast} require to first select a set of anchor words for each topic and then construct $\hat{A}$ using the corresponding rows and columns of $DD'$ (they have to explicitly specify which rows/columns are selected). Our method avoids explicit anchor word selection. This yields several advantages: (i) We need not worry about the errors caused by false selections. (ii)
The error rate for $\hat{A}$ is insensitive to the frequencies of anchor words (c.f., for the method in \cite{Ge}, if a low-frequency anchor word is selected, then the error rate will slow down). (iii) Our method can be extended to settings where the anchor word assumption is not satisfied.  
As long as we plug in a vertex hunting algorithm that estimates the simplex without requiring to have points near each vertex (e.g., \cite{javadi2019nonnegative}), we can drop the anchor word assumption.

{\bf Remark 3} {\it (Connection to LDA)}. The latent Dirichlet allocation (LDA) \citep{blei2003latent} is a popular approach to topic modeling. In the LDA model, $w_i$'s are latent variables from a Dirichlet distribution, and conditioning on $w_1,\ldots,w_n$, the data matrix follows a pLSI model. Therefore, our method still produces a valid estimate of $A$ in the LDA model, provided that $A$ satisfies the anchor-word condition; see Section~\ref{subsec:semi-synthetic} for such numerical experiments. 
The LDA model is thought as having two advantages over the pLSI model: (i) LDA has fewer parameters to estimate, because it treats $w_i$'s as latent variables rather than unknown parameters; (ii) LDA is better in assigning $w$ to a new document, as it takes advantage of the prior information of $w$ learnt from the training corpus. 
However, both arguments are about estimating $w$, not $A$. 
For estimation of $A$, 
the optimal rate is the same, no matter whether we assume a Dirichlet model on $w_i$'s or not; our method is already rate-optimal in many settings (see Section~\ref{sec:Theory}). 
For estimation of $w$, we may benefit from using the LDA model (e.g., \cite{girolami2003equivalence} showed that pLSI is a maximum a posteriori estimated LDA model under a uniform Dirichlet prior, hence, the shortcomings of pLSI on estimating $w$ can be elucidated and resolved within the LDA framework). Our method can be adapted to the LDA framework. Given $\hat{A}$, we can apply \eqref{w-hat} to get $\hat{w}_1,\ldots,\hat{w}_n$, use them to fit a Dirichlet distribution $\mathrm{Dir}(\hat{\alpha})$, and plug $(\hat{A}, \hat{\alpha})$ into the LDA framework to assign $w$ to a training or test document. This approach has the flavor of empirical Bayes. We leave it to future work.

\section{Theoretical properties}  \label{sec:Theory} 
Fix $K\geq 2$ and consider the pLSI model \eqref{pLSI} with $K$ topics. 
Without loss of generality, we assume all documents have the same length $N$.  
Let $a_j'$ denote the $j$th row of $A$ and write $h_j=\|a_j\|_1$, $1\leq j\leq p$. 
These quantities $h_1,h_2,\ldots,h_p$ capture the frequency heterogeneity across words. 
Let $h_{\max}=\max_{1\leq j\leq p}h_j$,  $h_{\min}=\min_{1\leq j\leq p}h_j$, and $\bar{h} = 
\frac{1}{p} \sum_{j = 1}^p h_j$, where by self-normalization of columns of $A$, $\bar{h}=K/p$. 
We assume 
\beq \label{cond-h}
h_{\min} \geq c_1 \bar{h}=c_1K/p, \qquad \mbox{for a constant $c_1 \in (0,1)$}.  
\eeq
This condition on $h_{\min}$ is inspired by the common pre-processing of removing extremely-low-frequency words \citep{blei2003latent}. 
When this condition is not satisfied, we can use the trick suggested by \cite{Ge} to aggregate those extremely-low-frequency words to a `pseudo-word', and the analysis still goes through. Therefore, this is a very mild assumption. We emphasize that we allow for severe word frequency heterogeneity, because $h_{\max}/h_{\min}$ can be as large as $p$ under \eqref{cond-h}.

\begin{defin}
We call $\Sigma_W=n^{-1}WW'$ the topic-topic concurrence matrix and call 
$\Sigma_A=A'H^{-1}A$ the topic-topic overlapping matrix, where $H=\diag(h_1,h_2,\ldots,h_p)$. 
\end{defin}
\noindent
The matrix $\Sigma_W$ is commonly used in the literature \citep{Ge}.  
The matrix $\Sigma_A$ measures the affinity between topics ---
a larger value of $\Sigma_A(k, \ell)$ indicates more overlapping between topics 
 $k$ and $\ell$. 
Both matrices are properly scaled, with all their entries between $0$ and $1$. For a constant $c_2\in (0,1)$, we assume
\beq \label{cond-A}
\lambda_{\min}(\Sigma_W)\geq c_2, \qquad \lambda_{\min}(\Sigma_A)\geq c_2, \qquad \min_{1\leq k,\ell\leq K}\Sigma_A(k,\ell)\geq c_2. 
\eeq
These conditions are mild. Below is a constructive example where \eqref{cond-h}-\eqref{cond-A} are satisfied.

\medskip
\noindent
{\bf Example}. 
Fix $m\geq K$, a positive vector $\alpha\in\mathbb{R}^K$ and a positive matrix $\Gamma = [\eta_1,\eta_2,\ldots,\eta_m]\in \mathbb{R}^{K,m}$ such that $\Gamma$ has a rank $K$ and $\Gamma x={\bf 1}_K$ has at least one non-negative solution $x\in\mathbb{R}^m$ (there exist many such triplets $(\alpha,\Gamma,x)$). Obtain $W$ by drawing $w_i$'s iid from $\mathrm{Dirichlet}(\alpha)$. Let $A^*\in\mathbb{R}^{p\times K}$ be the matrix where its first $K$ rows are  $\{p^{-1}e_k'\}_{1\leq k\leq K}$, and the remaining $(p-K)$ rows are sampled with replacement from $\{p^{-1}\|x\|_1\eta_k\}_{1\leq k\leq m}$ using the probabilities $\{\|x\|^{-1}_1x(k)\}_{1\leq k\leq m}$.  Obtain $A$ by re-normalizing each column of $A^*$ to have a unit $\ell^1$-norm. By straightforward analysis, we can show that \eqref{cond-h}-\eqref{cond-A} hold with high probability.

\medskip

We also need a mild condition on the vertex hunting (VH) algorithm in use:  
\begin{condi}[Efficiency of the VH algorithm] \label{cond:VH}
When the VH algorithm is given a point cloud $X_1,X_2,\ldots,X_p$, where $X_j$ is a proxy to $X_j^*$, and $X_1^*,X_2^*\ldots,X_p^*$ are located in a simplex with $K$ vertices $V^*_1,V^*_2,\ldots, V^*_K$, the algorithm outputs $V_1, V_2,\ldots,V_K$ such that, subject to a label permutation, $\max_{1\leq k\leq K}\|V_k - V^*_k\|\leq C\max_{1\leq j\leq p}\|X_j - X^*_j\|$, for a constant $C>0$. 
\end{condi}

This condition requires that the vertex estimation error is controlled by the maximum noise in the point cloud. 
In Section~\ref{subsec:real-method}, we mentioned two VH algorithms, SP and SVS.  
SP is shown to satisfy Condition~\ref{cond:VH} \citep{gillis2013fast}. SVS is shown to satisfy Condition~\ref{cond:VH} with mild regularity conditions \citep{Mixed-SCORE}. We summarize these results in the supplementary material. 
Our main results below apply to any VH algorithm that satisfies Condition~\ref{cond:VH}.

\subsection{A large-deviation bound for singular vectors} \label{subsec:theory-RMT}
Recall that $\hat{\Xi}=[\hat{\xi}_1,\hat{\xi}_2,\ldots,\hat{\xi}_K]$ contains the first $K$ left singular vectors of $M^{-1/2}D$, where $M=\mathrm{diag}(n^{-1}D{\bf 1}_n)$. We define a population counterpart of $\hat{\Xi}$ as $\Xi=[\xi_1,\xi_2,\ldots,\xi_K]$, where $\xi_k$ is the $k$th singular vector of $M_0^{-1/2}D_0$, with $D_0=\mathbb{E}[D]$ and $M_0=\mathrm{diag}(n^{-1}D_0{\bf 1}_n)$. 
Our key technical tool is the following theorem, which is proved in the supplementary material: 
 
\begin{thm}[Row-wise large-deviation bounds for $\hat{\Xi}$]  \label{thm:noise}
Fix $K\geq 2$ and consider Model \eqref{pLSI} with $N_i= N$. Suppose $\log^2(n) \leq  \min\{p, N\}$, $p \log(n) =o(N n)$, and \eqref{cond-h}-\eqref{cond-A} hold. Define
\[
\beta_n = \begin{cases}
1 + \min\{N^{-1}p, \, N^{-3/2}p^2\}, & \mbox{if } n\geq \max\{N p^2, p^3,N^2p^5\},\cr
1 + N^{-3/2}p^2, & \mbox{if } n< \max\{N p^2, p^3,N^2p^5\}. 
\end{cases}
\]
Let $\hat{\Xi}_j'$ and $\Xi'_j$ denote the $j$-th row of $\hat{\Xi}$ and $\Xi$, respectively.  With probability $1-o(n^{-3})$, there exists a matrix  $\Omega=\mathrm{diag}(\omega,\Omega^*)\in\mathbb{R}^{K\times K}$, where $\omega\in \{\pm 1\}$ and $\Omega^*\in\mathbb{R}^{(K-1)\times (K-1)}$ is an orthogonal matrix, such that 
\[
\|\Omega \hat{\Xi}_j -\Xi_j\| \leq \sqrt{h_j}\cdot C\beta_n\sqrt{\frac{p\log(n)}{Nn}}, \qquad \mbox{for all $1\leq j\leq p$}. 
\]
\end{thm}

In the pLSI model, there is a gap between the 1st and 2nd population singular values (by Perron's theorem \citep{HornJohnson}), so $\xi_1$ can be consistently estimated by $\hat{\xi}_1$, up to a sign flip; however, for $\xi_2,\ldots,\xi_K$, one can only estimate the $(K-1)$-dimensional subspace. This gives rise to the orthogonal matrix $\Omega$ in Theorem~\ref{thm:noise}.
This theorem provides a large-deviation bound for each row of $\hat{\Xi}$ and is useful for analysis of any SVD-based algorithm. 
By Theorem~\ref{thm:noise}, the noise level in different rows of $\hat{\Xi}$ are different: For a higher-frequency word, the corresponding row of $\hat{\Xi}$ has a larger stochastic fluctuation. 

As a consequence of Theorem~\ref{thm:noise}, we can prove a row-wise large-deviation bound for $\hat{R}$, a matrix constructed from $\hat{\Xi}$ by the post-SVD normalization:

\begin{thm}\label{thm:hatR}
Under the conditions of Theorem~\ref{thm:noise}, consider the matrices $\hat{R}$ and $R$ defined in \eqref{SCORE-real} and \eqref{SCORE-oracle}. Let $\hat{r}_j'$ and $r_j'$ denote the $j$th row of $\hat{R}$ and $R$, respectively, $1\leq j\leq p$. With probability $1-o(n^{-3})$, there exists an orthogonal matrix $\Omega^*\in\mathbb{R}^{(K-1)\times (K-1)}$ such that 
\[
\|\Omega^*\hat{r}_j - r_j \|\leq C \beta_n\sqrt{\frac{p\log(n)}{Nn}}, \qquad \mbox{for all $1\leq j\leq p$}.
\] 
\end{thm}

Here, the matrix $\Omega^*$ corresponds to a simultaneous rotation of $\hat{r}_1,\hat{r}_2,\ldots,\hat{r}_p$. Theorem~\ref{thm:hatR} states that each rotated $\hat{r}_j$ is close to $r_j$. Recall that $\hat{v}_1^*,\hat{v}_2^*, \ldots,\hat{v}_K^*$ are the estimated vertices by applying a vertex hunting algorithm on $\hat{r}_1,\hat{r}_2, \ldots,\hat{r}_p$. By Condition~\ref{cond:VH}, each rotated $\hat{v}_k^*$ should be close to the true vertex $v^*_k$. This rotation by $\Omega^*$ is picked by the SVD algorithm as a blackbox and is unknown to users, but it has no effect on the output of Topic-SCORE, because when $\hat{r}_j$ and $\hat{v}_1^*,\hat{v}_2^*, \ldots,\hat{v}_K^*$ are rotated in the same way, the resulting $\hat{\pi}_j^*$ is unchanged, so is $\hat{A}$. Therefore, we can always ``pretend" that $\Omega^*=I_{K-1}$, without loss of generality.

Theorem~\ref{thm:hatR} suggests that the noise levels in different rows of $\hat{R}$ are similar. This is the key that our error rate for $\hat{A}$ (to be presented in Section~\ref{subsec:theory-main}) is insensitive to word frequency heterogeneity. The ``flat" rate in Theorem~\ref{thm:hatR} is not a coincidence: We purposely designed the pre-SVD \& post-SVD normalizations so that they ``coordinate" with each other.

{\bf Remark 4} {\it(The heavy-tail coefficient $\beta_n$)}.  In both Theorems~\ref{thm:noise}-\ref{thm:hatR}, the upper bounds involve a factor $\beta_n$. We call $\beta_n$ the {\it heavy-tail coefficient}. It captures the tail effect of multinomial distributions. Each column of $D$ is a multinomial random vector, whose number of trials is $N$ and dimension is $p$. If $N$ is not large enough compared with $p$, these multinomial entries have heavy tails and will significantly affect the large-deviation bounds for singular vectors. this heavy tail effect  can be partially mitigated if $n$ (the number of documents) is sufficiently large. This is why we have two cases in the definition of $\beta_n$.

{\bf Remark 5} {\it(Proof ideas)}. In the proof of Theorem~\ref{thm:noise}, we introduce two $p\times p$ matrices $
G = M^{-1/2}DD'M^{-1/2}- (N^{-1}n)I_p$ and $G_0=(1-N^{-1})M_0^{-1/2}D_0D_0'M_0^{-1/2}$, and view $\hat{\Xi}$ and $\Xi$ equivalently as containing the eigenvectors of $G$ and $G_0$, respectively. We then provide a non-stochastic perturbation result (Lemma~\ref{lem:EigVecPerturb}) for eigenvectors, which improves the sin-theta theorem \citep{sin-theta} by allocating error to individual rows; this lemma bounds $\|\Omega \hat{\Xi}_j -\Xi_j\|$ in terms of (i) the spectral norm and (ii) the column-wise $\ell^2$-norms of $G-G_0$. The longest part of the proof is to derive a sharp large-deviation bound for the spectral norm of $G-G_0$. We hope to borrow techniques of non-asymptotic random matrix theory in \cite{Vershynin} but face a big challenge: The entries of multinomial random vectors are mutually dependent, and they do not have fast enough tails when $N$ is small or moderately large. We overcome this challenge by a proper way of blending martingale concentration inequalities \citep{freedman1975tail} into non-asymptotic random matrix analysis (see Sections~\ref{sec:high-level-proof} and Sections~\ref{sec:Z-analysis}-\ref{sec:Eigen-perturbation} of the supplementary material).

\subsection{The rates of convergence of Topic-SCORE} \label{subsec:theory-main}
Let $\hat{A}=[\hat{A}_1,\hat{A}_2,\ldots,\hat{A}_K]$ be the estimator by our method Topic-SCORE. 
We measure the performance of $\hat{A}$ by the $\ell^1$-error (subject to a permutation of columns of $\hat{A}$):
\[
\mathcal{L}(\hat{A}, A)  \equiv   \sum_{k=1}^K \|\hat{A}_k  -  A_k \|_1. 
\]
Since each $A_k$ is self-normalized in $\ell^1$-norm, this is a natural loss function. 

The next theorem is our main result. It provides both the error rate for estimating each individual row of $A$ and the rate for the total $\ell^1$-error.

\begin{thm}[Main result]  \label{thm:UB}
Fix $K\geq 2$ and consider Model \eqref{pLSI} with $N_i= N$. As $n\to\infty$, suppose $\log^2(n) \leq  \min\{p, N\}$ and $p \log(n) =o(N n)$. Suppose \eqref{cond-h}-\eqref{cond-A} are satisfied.  
Let $\beta_n$ be the same as in Theorem~\ref{thm:noise}. 
Let $\hat{a}_j'$ and $a'_j$ denote the $j$-th row of $\hat{A}$ and $A$, respectively, $1\leq j\leq p$. With probability $1-o(n^{-3})$, 
\[
\|\hat{a}_j-a_j\|_1 \leq \|a_j\|_1\cdot C\beta_n\sqrt{\frac{p\log(n)}{Nn}}, \qquad \mbox{for all $1\leq j\leq p$}. 
\]
Furthermore, with probability $1-o(n^{-3})$,
\[
\mathcal{L}(\hat{A}, A)\leq C\beta_n\sqrt{\frac{p\log(n)}{Nn}}. 
\]
\end{thm}

By Theorem~\ref{thm:UB}, for estimating the individual rows of  $A$, the error in $\hat{a}_j$ is larger for a higher-frequency word $j$. At the same time, the relative error, measured by $\|\hat{a}_j-a_j\|_1/\|a_j\|_1$, has the same rate for every $1\leq j\leq p$.

For the total $\ell^1$-error, the rate of convergence is primarily governed by $\sqrt{\frac{p\log(n)}{Nn}}$, up to an additional factor captured by the heavy-tail coefficient $\beta_n$ (see Remark 3 in Section~\ref{subsec:theory-RMT}). By plugging in the definition of $\beta_n$, we have the following corollary: 
\begin{cor}\label{cor:rate}
Suppose the conditions of Theorem~\ref{thm:UB} hold. 
We call $N\geq p^{4/3}$, $p\leq N<p^{4/3}$ and $N<p$ the cases of long documents (Case 1), moderately long documents (Case 2) and short documents (Case 3), respectively. In Case 2, if $n\geq \max\{N p^2, p^3,N^2p^5\}$, we call it Case 2a; otherwise, we call it Case 2b. We define Cases 3a-3b similarly. The following holds with probability $1-o(n^{-3})$: 
\begin{itemize} \itemsep -1pt
\item Case 1 (long documents): $\mathcal{L}(\hat{A}, A)\leq C\sqrt{\frac{p\log(n)}{Nn}}$. 
\item Case 2a (moderately long documents): $\mathcal{L}(\hat{A}, A)\leq C\sqrt{\frac{p\log(n)}{Nn}}$. 
\item Case 2b (moderately long documents):  $\mathcal{L}(\hat{A}, A)\leq C \frac{p^2}{N\sqrt{N}} \sqrt{\frac{p\log(n)}{Nn}}$. 
\item Case 3a (short documents): $\mathcal{L}(\hat{A}, A)\leq C\frac{p}{N}\sqrt{\frac{p\log(n)}{Nn}}$.
\item Case 3b (short documents): $\mathcal{L}(\hat{A}, A)\leq C\frac{p^2}{N\sqrt{N}}\sqrt{\frac{p\log(n)}{Nn}}$.
\end{itemize}
\end{cor}

By Corollary~\ref{cor:rate}, our method has two appealing theoretical properties. First, its error rate is {\it insensitive} to word frequency heterogeneity. The bound for ${\cal L}(\hat{A}, A)$ does not depend on $h_{\max}/h_{\min}$ or $h_{\max}/\bar{h}$. 
In real data, the word frequency heterogeneity is usually severe. It is beneficial to have a method whose error rate does not depend on frequency heterogeneity. 
Second, it works for all three case of long, moderately long and short documents. The case of short documents is especially challenging, as the data matrix $D$ will contain many zero's. 
Our SVD-based method can still handle this case.

To assess the optimality of our method, we give a lower bound for the $\ell^1$-error: 
\begin{thm}[Lower bound] \label{thm:LB}
Fix $K\geq 2$ and consider Model \eqref{pLSI} with $N_i= N$. Suppose $\log(n) \leq  \min\{p, N\}$ and $p \log(n) =o(N n)$, as $n\to\infty$. Let $\Phi_{n,N,p}(K,c_1,c_2)$ denote the collection of parameters $(A, W)$ such that \eqref{cond-h}-\eqref{cond-A} hold and that each topic has at least one anchor word. There are constants $C_0>0$ and $\delta_0\in (0,1)$ such that, for sufficiently large $n$, 
\[
\inf_{\hat{A}}\sup_{(A,W)\in\Phi_{n,N,p}(K,c_1,c_2)}\mathbb{P}\biggl( \mathcal{L}(\hat{A},A)\geq C_0\sqrt{\frac{p}{Nn}}\biggr) \geq \delta_0. 
\]
\end{thm}

We compare Theorem~\ref{thm:LB} with Corollary~\ref{cor:rate}. For Cases 1-2, the optimal rate is $\sqrt{\frac{p}{Nn}}$, and our method is rate optimal, up to a logarithm factor. 
For Case 2, we need an additional condition on $n$ to get the sharpest rate. This is likely a technical artifact. Our analysis of singular vectors requires combining martingale tail inequalities \citep{freedman1975tail} with non-asymptotic random matrix theory. We must carefully bound the sum of conditional variances  (SCV) of the martingale constructed in our proof. The SCV is by itself a sum of dependent, heavy-tail random variables, and its own large-deviation bound leads to additional terms. We manage to remove those terms in Case 2a. We conjecture that, with more advanced techniques, these terms can also be removed for Case 2b. 
For Case 3, the optimal rate is unknown, but our rate already improves those in the literature (see the remarks below).

{\bf Remark 6} {\it (Comparison with \cite{Ge})}. Arora et al. \citep{Ge} is among the first who gave explicit error rates of estimating $A$ under the anchor word assumption. They assumed that each topic has an anchor word $j$ such that $\|a_j\|_1\geq C\delta_p$ and showed that, up to a logarithmic factor, the rate of ${\cal L}(\hat{A},A)$ is 
$(Nn)^{-1/2} p\delta_p^{-3}$ (their original result is for $\|\hat{A}-A\|_{\max}$, which we convert to a rate for ${\cal L}(\hat{A}, A)$ by multiplying it by $p$). 
The rate is sensitive to the frequencies of anchor words, captured by $\delta_p$. We note that $\delta_p$ ranges from $p^{-1}$ to $1$. In real applications, it is unlikely that the anchor words are super-frequent. A reasonable case is $\delta_p\asymp \bar{h}\asymp 1/p$. Then, the rate becomes $(Nn)^{-1/2}p^4$, which is slower than our rate in all three cases.

{\bf Remark 7} {\it (Comparison with \cite{bing2020fast})}.
Bing et al. \citep{bing2020fast} proposed a nice method for estimating $A$. In Cases 1-2, when $h_{\max}\leq Ch_{\min}$ (i.e., moderate frequency heterogeneity), their method attains the optimal rate, up to a logarithmic factor.  
However, when $h_{\max}\gg h_{\min}$ (i.e., severe frequency heterogeneity), their error rate has an extra factor of at least $(h_{\max}/h_{\min})^2$ and becomes non-optimal. In comparison, our error rate is unaffected by frequency heterogeneity.  
Furthermore, their result does not cover Case 3 (short documents).

{\bf Remark 8} {\it (Comparison with \cite{bansal2014provable})}. Bansal et al. \citep{bansal2014provable} proposed a method that uses SVD to get a low-rank approximation of $D$. It does not estimate $A$ directly from singular vectors, hence, not the ``SVD-based" method in our sense.  They showed that the rate of ${\cal L}(\hat{A}, A)$ is $
(n\epsilon_n)^{-1/2} \sqrt{p}+ (n\epsilon_n\delta_p)^{-1/2}N$, up to a logarithmic factor,  
where $\delta_p$ is the same as in \cite{Ge} and $\epsilon_n\in (0,1)$ is the fraction of pure or nearly pure documents. 
When $\epsilon_n\asymp 1$ and $\delta_p\asymp 1/p$, the rate is $n^{-1/2}N\sqrt{p}$, which is non-optimal.

\subsection{Estimation of $W$ and $K$, and discussion of misspecified $K$} \label{subsec:W-and-K}

Given $\hat{A}$, we can further obtain an estimator of $w_i$ as in \eqref{w-hat}, for $1\leq i\leq n$. 
The next theorem gives the error rate in $\hat{w}_i$ and is proved in the supplementary material:

\begin{thm}[Estimation of $W$]  \label{thm:W}
Suppose the conditions of Theorem~\ref{thm:UB} hold. For every $1\leq i\leq n$, for any $\delta\in (0,1)$, 
\[
\|\hat{w}_i-w_i\|_1 \leq C \beta_n\sqrt{\frac{p\log(n)}{Nn}} + C\sqrt{\frac{\log(1/\delta)}{N}}, \qquad \mbox{with probability $1-\delta+o(n^{-3})$}. 
\]
\end{thm}

In Theorem~\ref{thm:W}, the two terms come from the error of estimating $A$ and the noise in the $i$th column of $D$, respectively. Usually, the second term dominates. 
Under our assumptions, $Nn\gg p\log(n)$ and $N\to\infty$, so $\hat{w}_i$ is consistent.

In Section~\ref{subsec:theory-main}, we assume $K$ is known. When $K$ is unknown, letting $\hat{\sigma}_1,\ldots,\hat{\sigma}_{n\wedge p}$ be the singular values of $M^{-1/2}D$, and $\beta_n$
be the same as in Theorem~\ref{thm:noise}, we estimate $K$ by 
\beq \label{K-estimate}
\hat{K} = \max\biggl\{1\leq k\leq (n\wedge p):\;  \hat{\sigma}_j^2 > \frac{n}{N} + \beta_n\sqrt{\frac{np\log(n)}{N}}\cdot g_n \biggr\},
\eeq
where $g_n$ is a sequence that converges to $\infty$ slowly (e.g., $g_n=\log(\log(n))$).

\begin{thm}[Estimation of $K$]  \label{thm:K}
Suppose the conditions of Theorem~\ref{thm:UB} hold,  and assume $p \beta_n^2\log(n)=o(Nn)$. Let $g_n$ be any sequence such that $g_n\to\infty$ and $g_n\ll \beta^{-1}_n\sqrt{\frac{Nn}{p\log(n)}}$. With probability $1-o(n^{-3})$, $\hat{K}=K$. 
\end{thm}

In practice, due to weak signals, consistent estimation of $K$ may not hold. We now discuss what happens if $K$ is misspecified. Suppose we apply Topic-SCORE assuming there are $m$ topics. 
We follow  \cite{jin2020estimating} to call $m<K$ the {\it under-fitting case} and $m>K$ the {\it over-fitting case}. 
Let $\hat{A}^{(m)}$ and $A^{(m)}$ be the output of Topic-SCORE and the oracle procedure, respectively. 
Define $\hat{R}^{(m)}$ as in \eqref{SCORE-real} by plugging in $K=m$. Write $(\hat{R}^{(m)})' = [\hat{r}_1^{(m)}, \hat{r}_2^{(m)}, \ldots,\hat{r}_p^{(m)}]$.

In the under-fitting case, we observe a nice property of $\hat{R}^{(m)}$:  Each $\hat{r}_j^{(m)}$ is a sub-vector of $\hat{r}_j$ by restricting it to the first $(m-1)$ coordinates.  
The same argument applies to their population counterparts, $r_j^{(m)}$ and $r_j$. 
Hence, we can apply Theorem~\ref{thm:hatR} directly to obtain a large-deviation bound for $\hat{r}_j^{(m)}$. The only issue comes from the orthogonal matrix $\Omega^*$, because the sub-vector of $\Omega^*\hat{r}_j$ is not necessarily a rotation of the corresponding sub-vector of $\hat{r}_j$. 
To avoid this issue, we assume the gap between each two nested population singular values is comparable with their own magnitude. Under this assumption, it can be shown that $\Omega^*$ is a diagonal matrix with $\pm 1$ in the diagonal.  Let $\Omega^*_m$ be the top left $(m-1)\times (m-1)$ block of $\Omega^*$. 
It follows from Theorem~\ref{thm:hatR} that $\|\Omega^*_m\hat{r}_j^{(m)}-r_j^{(m)}\|\leq \|\Omega^*\hat{r}_j - r_j\|\leq C\beta_n\sqrt{\frac{p\log(n)}{Nn}}$, simultaneously for $1\leq j\leq p$, with probability $1-o(n^{-3})$. Therefore, we can similarly show that ${\cal L}(\hat{A}^{(m)}, A^{(m)})\leq C\sqrt{\frac{p\log(n)}{Nn}}$. The remaining question is to study the oracle output $A^{(m)}$ and understand how it is connected to $A$.

\begin{prop}[The under-fitting case]  \label{prop:under-fitting}
In Model \eqref{pLSI}, suppose each topic has at least one anchor word and the singular values of $M_0^{-1/2}D_0$ are distinct. Fix $2\leq m<K$. Let $R^{(m)}$ be the sub-matrix of $R$ by restricting to the first $(m-1)$ columns, and let $(r_j^{(m)})'$ denote its $j$th row. Recall that $v_1^*, v_2^*, \ldots, v_K^*$ are vertices of the Ideal Simplex in Lemma~\ref{lem:IdealSimplex}. Let $v_k^{(m)}$ be the sub-vector of $v_k^*$ by restricting to the first $(m-1)$ coordinates. Then, the convex hull of $r_1^{(m)},\ldots,r_p^{(m)}$ is a non-degenerate simplex in $\mathbb{R}^{m-1}$ with $K_m$ vertices, where $K_m\leq K$ and the vertices are from $\{v_1^{(m)},\ldots,v_K^{(m)}\}$. 
Without loss of generality, we assume the vertices are $v_1^{(m)},\ldots,v_{K_m}^{(m)}$ and  for $k>K_m$, write $v_k^{(m)}$ as a convex combination of vertices with $\beta_k\in\mathbb{R}^{K_m}$ denoting the combination coefficient vector. Write $B=[\beta_{K_m+1},\beta_{K_m+2},\ldots,\beta_{K}]$. Let $q\in\mathbb{R}^K$ be the same as in Lemma~\ref{lem:Pi-to-A} and $e_1,\ldots,e_K$ be the standard basis of $\mathbb{R}^K$. If $K_m=m$, then 
\[
A_k^{(m)} \;\; \propto \;\; A\cdot \diag(q)\cdot \begin{bmatrix} I_m \\B' \end{bmatrix}e_k, \qquad 1\leq k\leq m. 
\]
\end{prop}

We illustrate Proposition~\ref{prop:under-fitting} using the example in Figure~\ref{fig:simplex}, where $K=3$ and the rows of $R$ are in a triangle. Without loss of generality, we label the top left and top right vertices as $v_1^*$ and $v_2^*$ and the bottom one as $v_3^*$. 
We now consider $m=2$. Each $r_j$ is restricted to its first coordinate to get $r_j^{(m)}$. It is seen in the figure that $r_1^{(m)}, \ldots,r_p^{(m)}$ form a line segment in $\mathbb{R}$, with $v_1^{(m)}$ and $v_2^{(m)}$ as two end points. Suppose $v_3^{(m)}=0.6v_1^{(m)}+0.4v_2^{(m)}$ and $q=(1,1,1)'$. By Proposition~\ref{prop:under-fitting}, $A_1^{(m)}\propto A_1+0.6A_3$ and $A_2^{(m)}\propto A_2+0.4A_3$. 

In the over-fitting case, $m>K$. Each $\hat{r}_j^{(m)}$ is obtained by appending a few noisy coordinates to $\hat{r}_j$. To study this case, we must understand the behavior of the non-leading singular vectors. It is beyond the scope of this paper, which we leave to future work.

\section{Real data applications}  \label{sec:Realdata} 
We apply our method to two real data sets, a corpus of Associated Press news articles (AP) and a corpus of statistical paper abstracts (SLA). Since real data have no ground truth, we evaluate the performance of our method from two perspectives. Perspective 1: the plot of rows of $\hat{R}$. Our theory predicts that the point cloud formed by rows of $\hat{R}$ has the silhouette of a simplex, subject to noise corruption. If we observe this simplex in the plot of $\hat{R}$, it suggests that our model and method fit real data well, especially, our proposed pre-SVD normalization and post-SVD normalization are effective. 
Perspective 2: the interpretation of $\hat{A}$.  Following \cite{PaseII-paper2}, for each word $j$, we define the {\it topic loading vector} $b_j$ by $b_j(k)=A_k(j)/[\sum_{\ell=1}^K A_\ell(j)]$, $1\leq k\leq K$. This vector has a unit $\ell^1$-norm, and word $j$ is an anchor word of topic $k$ if and only if $b_j(k)=1$. Given $\hat{A}$, we compute the empirical topic loading vectors $\hat{b}_j$ by replacing $A$ with $\hat{A}$ in the definition. For each topic $k$, we output a list of words with the largest values of $\hat{b}_j(k)$, as the ``representative words" of this topic, and use them to check whether the topic has a meaningful interpretation. 
Our method has 1 tuning parameter, $\tau$ in \eqref{M}. We set $\tau=0$ for the AP dataset and $\tau=0.1$ for the SLA data set. 
Our method also requires the plug-in of a vertex hunting algorithm. We use SVS (pseudo code is in the supplementary material), which has tuning integers $L$ and $K_0$, and we set $L=10\times K$ and $K_0=\lceil 1.5\times K\rceil$.

\subsection{Associated Press (AP) data} \label{subsec:AP}
The AP data set \citep{harman1overview} consists of $2246$ news articles with a vocabulary of $10473$ words.  In the preprocessing, we first removed  $191$ stop-words. Next, we sorted the remaining words in the descending order of their total counts in the corpus, and we only kept the top $8000$ words. Last, we sorted all documents in the descending order of their lengths (the length only counts those words remaining in the vocabulary) and removed the last $5\%$ of documents; after this operation, some words had zero count in the remaining documents and were removed. We ended up with a corpus with $(n,p)=(2134, 7000)$. 
We need to decide the number of topics. We applied our method for $K\in\{2,3,\ldots,6\}$ and checked goodness-of-fit of the simplex for $\hat{R}$ and interpretability of the ``representative words" for each topic (using the topic loading vectors $\hat{b}_j$ defined above). It suggested that $K=3$ is most appropriate (e.g., the fitting of the simplex is good for $K=3$ but not so for $K=4$; also, as we increased $K$ from 3 to 4, two of the estimated topics had similar interpretations). 
For these reasons, we fix $K=3$.

\spacingset{1} 
 \begin{figure}[tb!]
\centering
\includegraphics[width = .42\textwidth]{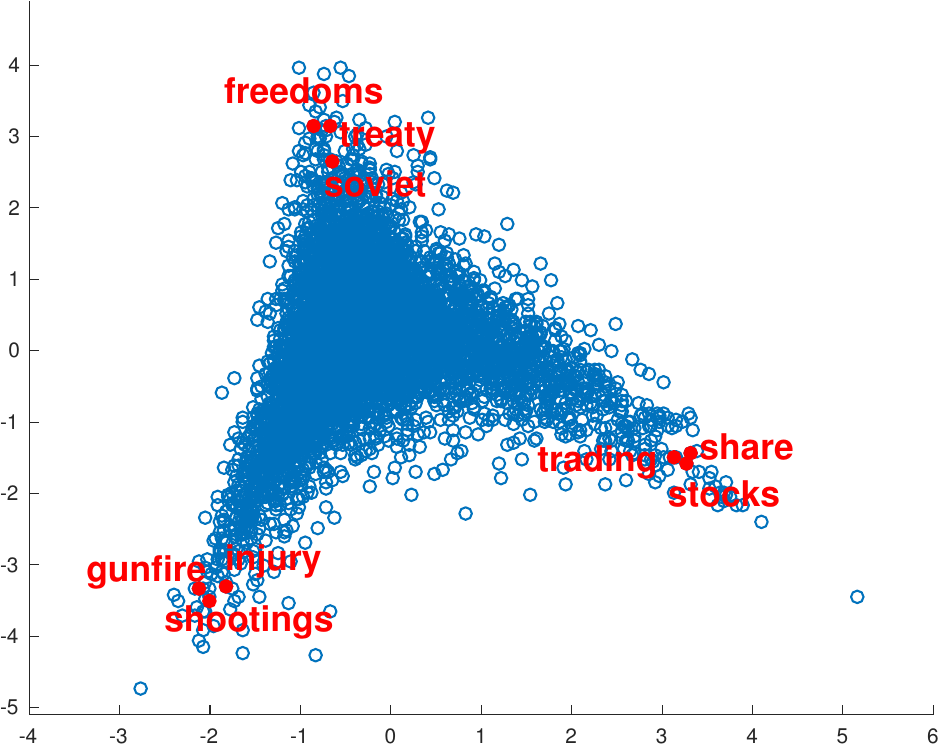}
\includegraphics[width = .42\textwidth]{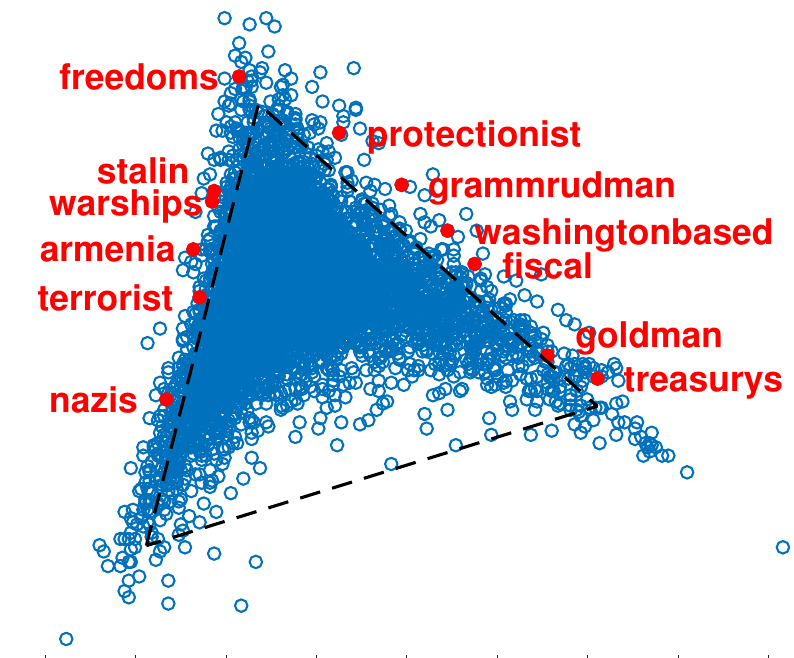}
\caption{The plot of rows of $\hat{R}$ for the AP dataset. Since $K=2$, each row of $\hat{R}$ is a point in $\mathbb{R}^2$. A triangle is visible in the point cloud, where the three vertices represent the topics ``crime", ``politics", and ``finance". For better visualization, we duplicate the plot. In the left panel, we show the representative words near each vertex. In the right plot, we show the estimated triangle by SVS, as well as a few representative words on the edges.}\label{fig:real_ap}
\end{figure}
\spacingset{1.45}


First, we plot the rows of $\hat{R}$ in Figure~\ref{fig:real_ap}. Our theory predicts that the point cloud has approximately the silhouette of a triangle (i.e., a simplex for $K=3$). The plot is a nearly perfect fit! We emphasize that this is the {\it raw plot} of $\hat{R}$. We obtained it by simply applying pre-SVD normalization, SVD and post-SVD normalization, with no additional engineering. This encouraging result suggests that: (i) Singular vectors indeed contain useful information for estimating a topic model (hence, using SVD is a promising direction). (ii) Our pre-SVD and post-SVD normalizations are effective on real data (for example, we can similarly plot the rows of $\hat{\Xi}$ without the post-SVD normalization, but no triangle is visible). 
Each row of $\hat{R}$ is associated with a word in the vocabulary. By our theory, the words near each vertex are the anchor words. In the left panel of Figure~\ref{fig:real_ap}, for each vertex, we show a few nearby words. For the top vertex, the nearby words are {\it freedoms}, {\it treaty} and {\it soviet}, suggesting that this topic is ``Politics". For the bottom left vertex, the nearby words are {\it gunfile}, {\it injury} and {\it shootings}, suggesting that this topic is ``Crime". For the bottom right vertex, the nearby words are {\it trading}, {\it share} and {\it stocks}. We give more representative words for each topic; see Table~\ref{tb:real_ap_k3} below. 
In the right panel of Figure~\ref{fig:real_ap}, we compare the point cloud (formed by rows of $\hat{R}$) with the estimated triangle. A very interesting observation is that there is an ``empty area" near the bottom edge. If a word $j$ is located near the bottom edge, it means in the topic loading matrix $b_j$ (its definition is in the beginning of this section), the weight on ``Politics" is close to 0, but the weights on ``Crime" and ``Finance" are  considerably nonzero. Our results claim that there are almost no such words, which are relevant to both ``Crime" and ``Finance" but irrelevant to ``Politics". This makes sense: These news articles were from early 1990's; at that time, ``Crime (violence)" and ``Finance" seemed to have no direct connection. In contrast, there are many words located near the other two edges. On the edge between ``Finance" and ``Politics", we find words such as {\it treasurys}, {\it goldman}, {\it fiscal}, {\it washingtonbased}, {\it grammrudman} and {\it protectionist}. 
These words are related to both ``Finance" and ``Politics"; the closer to the top vertex, the more connection to ``Politics" and less to ``Finance", and vice versa. 
Similarly, on the edge between ``Crime" and ``Politics", we find words such as {\it nazis}, {\it terrorist}, {\it armenia}, {\it warships} and {\it stalin}. 
We then use the topic loading vectors $\hat{b}_j$ to find a ranked list of ``representative words" for each topic (the higher rank, the more likely it is an anchor word). See Table~\ref{tb:real_ap_k3}. They fit our common sense, especially since these news articles were in early 1990's. For example, {\it bangladesh}, {\it hindus} and {\it dhaka} appear in the list because of the Bangladesh anti-Hindu violence in 1990-1992. 
For a comparison with LDA on this data set, see the supplementary material.

\spacingset{1.1} 
\begin{table}[hbt!]
\caption{Top 15 representative words for each estimated topic in the AP data ($K=3$). In the word list for the ``Finance" topic, {\it rose} is the past tense of {\it rise}.} \label{tb:real_ap_k3}
\scalebox{.85}
{\begin{tabular}{ll}
\toprule
\multirow{2}{*}{``Crime''} & {\it shootings, injury, mafia, detective, bangladesh, dog, hindus, gunfire, aftershocks,}\\
& {\it bears, accidentally, handgun, unfortunate, dhaka, police}\\
\hline
\multirow{2}{*}{``Politics''} & {\it eventual, gorbachevs, openly, soviet, primaries, sununu, yeltsin, cambodia, torture,}\\
&{\it soviets, herbert, gephardt, afghanistan, citizenship, popov }\\
\hline
\multirow{2}{*}{``Finance''} & {\it trading, stock, edged, dow, rose, traders, stocks, indicators, exchange, share,}\\
& {\it guilders, bullion, lire, christies, unleaded} \\
\bottomrule
\end{tabular}}
\end{table}
\spacingset{1.45}

\subsection{Statistical Literature Abstracts (SLA) data} \label{subsec:SLA}
The SLA data set \citep{SCC-JiJin} contains the abstracts of $3193$ papers published in {\it Annals of Statistics}, {\it Biometrika}, {\it Journal of the American Statistical Association}, and {\it Journal of the Royal Statistical Society - Series B}, from 2003 to the first half of 2012. The full vocabulary has $2934$ words. In the pre-processing, we first removed stop words. Since these documents specialize on statistical research, we should remove more stop words than usual (e.g., {\it prove}, {\it propose} and {\it method} are treated as stop words in this corpus, although they are usually not viewed as stop words). We removed a manually selected list of $209$ stop words. Next, we sorted documents in the descending order of length (the length does not count stop words) and removed the last $40\%$ of documents. Those words that had a zero count in the retained documents were also removed from the vocabulary. The pre-processing gave a corpus with $(n, p)=(1916, 2863)$. 
We decided $K$ similarly as before, by running our method for $K=2,3,\ldots, 8$ and checking the interpretability of $\hat{A}$. We also consulted the scree plot. We found that $K = 6$ is the most appropriate choice. Since $K=6$, each row of $\hat{R}$ is a point in $\mathbb{R}^5$. The pairwise coordinate plot suggests that the fitting of the simplex is reasonably good (which is omitted due to space limit).

\spacingset{1.1} 
\begin{table}[hbt]
\caption{Top 15 representative words for each estimated topic in the SLA data ($K=6$).} \label{tb:real_sla_k6}
\scalebox{.85}
{\begin{tabular}{ll}
\toprule
``Multiple & {\it stepup, stepdown, rejections, hochberg, fwer, singlestep, familywise, benjamini,}\\
Testing'' & {\it bonferroni, simes, intersection, false, rejection, positively, kfwer}\\
\hline
\multirow{2}{*}{``Bayes''} & {\it posterior, prior, slice, default, credible, conjugate, priors, improper, wishart,}\\
& {\it admissible, sampler, tractable, probit, normalizing, mode}\\
\hline
``Variable & {\it angle, penalties, zeros, sure, selector, selection, stability, enjoys, penalization,}\\
Selection''& {\it regularization, lasso, tuning, irrelevant, selects, clipped}\\
\hline
``Experimental & {\it aberration, hypercube, latin, nonregular, spacefilling, universally, twofactor,}\\
Design''& {\it blocked, twolevel, designs, crossover, resolution, factorial, toxicity, balanced}\\
\hline
``Spectral & {\it trajectories, amplitude, eigenfunctions, realizations, away, gradient, spectra,}\\
Analysis''& {\it discrimination, functional, auction, nonstationarity, spacetime, slex, curves, jumps}\\
\hline
\multirow{2}{*}{``Application''} & {\it instrument, vaccine, instruments, severity, affects, compliance, infected,}\\
& {\it depression, schools, assignment, participants, causal, warming, rubin, randomized}\\
\bottomrule
\end{tabular}}
\end{table}
\spacingset{1.45}

\begin{figure}[tb!]
\centering
\includegraphics[width=.58\textwidth]{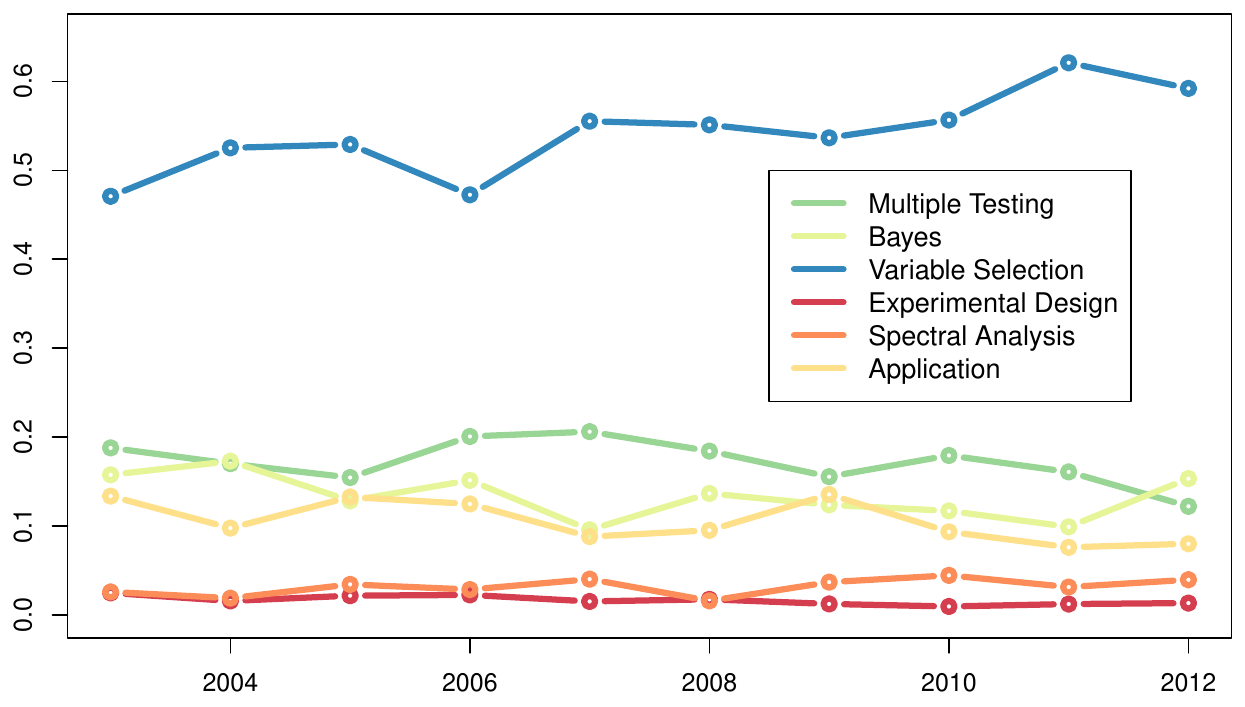}\\
\vspace{10pt}
\includegraphics[width=.26\textwidth]{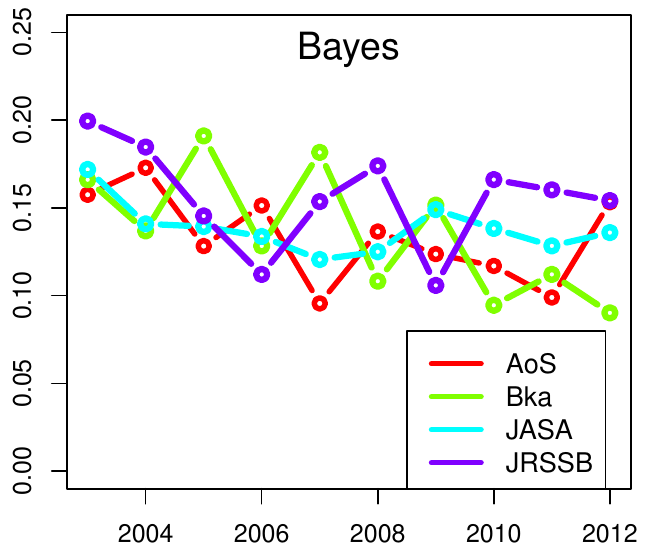}
\includegraphics[width=.26\textwidth]{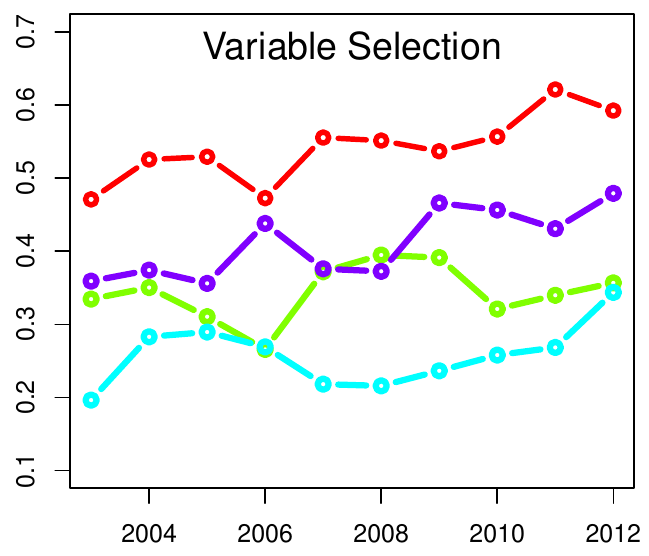}
\includegraphics[width=.26\textwidth]{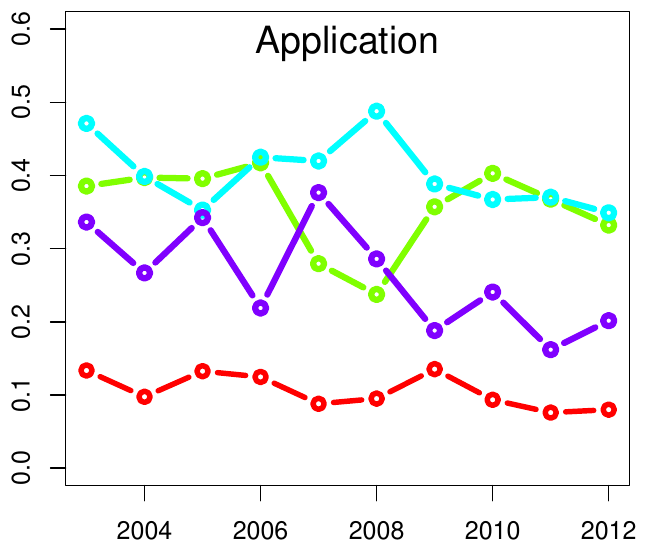}
\caption{Topic trending in the SLA data. Top panel: For each topic $k$, we plot the average of $\hat{w}_i(k)$ over all papers published in year $t$, for $t\in \{2003,2004,\ldots,2012\}$. Bottom: Same as above, except that we restrict separately to papers in each of the four journals.} \label{fig:topic-trends}
\end{figure}

Table~\ref{tb:real_sla_k6} shows the top $15$ representative words in each topic, obtained from $\hat{A}$ by computing the topic loading vectors. Based on these words, we interpret the six topics as ``Multiple Testing", ``Bayes", ``Variable Selection", ``Experimental Design", ``Spectral Analysis", and ``Application".
Given the six estimated topic vectors, we further estimate $\hat{w}_i$ for each document $i$ using the method in \eqref{w-hat}. We then use these $\hat{w}_i$'s to study the topic trending. First, for each year $t$, we compute an average weight vector, $\bar{w}_t\in\mathbb{R}^{K}$, where $\bar{w}_t(k)$ is the average of $\hat{w}_i(k)$ among papers published in year $t$, for $1\leq k\leq K$. In the top panel of Figure~\ref{fig:topic-trends}, we plot the curve $\{\bar{w}_t(k): 2003\leq t\leq 2012\}$ for each of the six estimated topics. We observe that ``Variable Selection" has a much higher average weight than other topics, suggesting that ``Variable Selection" is the most popular topic in these four journals during 2003-2012.  Additionally, ``Multiple Testing", ``Bayes" and ``Application" are moderately popular, and ``Experimental Design" and ``Spectral Analysis" are least popular. 
During this 10-year time period, the average weight of ``Variable Selection" has been steadily increasing (except year 2006). Since the entries of $\bar{w}_t$ sum to 1 for each $t$, the increasing average weight of ``Variable Selection" comes with decreasing average weights of other topics (e.g., ``Multiple Testing"). 
Next, for each journal $J$, we compute the average weight vector $\bar{w}_{t,J}\in\mathbb{R}^{K}$ based on the $\hat{w}_i$'s of papers published in this journal in year $t$. In the bottom panel of Figure~\ref{fig:topic-trends}, we plot the curve of $\{\bar{w}_{t,J}(k): 2003\leq t\leq 2012\}$ for each of the four journals, when the topic $k$ is ``Bayes", ``Variable Selection" and ``Application", respectively. For ``Application", JASA has the highest average weight among all four journals, followed by Biometrika and JRSSB, and AOS has the lowest average weight. For ``Variable Selection", the journal preferences are in the opposite order: AOS has the highest average weight, followed by JRSSB and Biometrika, and JASA has the lowest average weight. For ``Bayes", the average weights of this topic in four journals are quite comparable with each other.

\spacingset{1}
\spacingset{1.45}

\section{Simulations}  \label{sec:Simu}
We compare Topic-SCORE with four other methods: (i) The LDA approach \citep{blei2003latent} (R package {\it lda}, with default Dirichlet priors $\alpha=\beta=0.1$). (ii) The anchor-word recovery (AWR) approach \citep{arora2013practical,Ge} (Python code from \url{http://people.csail.mit.edu/moitra/software.html}). 
(iii) The TSVD approach \citep{bansal2014provable} (Matlab code from \url{http://thetb.github.io/tsvd/}). 
(iv) The EM approach \citep{mei2001note} (Python code from \url{https://github.com/laserwave/plsa}, with the default initialization and maximum number of iterations as 50). 
In Topic-SCORE, we set $\tau=0$ and plug in SVS \citep{Mixed-SCORE} as the vertex hunting algorithm (with default tuning parameters $L=10\times K$ and $K_0=\lceil 1.5\times K\rceil$). 
For all methods, $K$ is given.

\subsection{The calibrated LDA models from real corpora}\label{subsec:semi-synthetic}
We conduct semi-synthetic experiments, where $(A,W)$ are calibrated from real data by LDA. 
Given a real corpus ($n$ documents, vocabulary size $p$), for any pre-specified $(K,N_1,\ldots,N_n)$, we first run LDA by assuming $K$ topics; next, using the posterior mean of $(A,W)$ output by LDA as the true $(A,W)$, we generate $n$ new documents from Model \eqref{pLSI} such that document $i$ has $N_i$ words, $1\leq i\leq n$. We took the AP data set \citep{harman1overview} and the NIPS data set 
\citep{perrone2016poisson} and preprocessed them by removing stop words and keeping the 50\% most frequent words and 95\% longest documents (for AP, the pre-processing is different from in Section~\ref{subsec:AP}, so $(n,p)$ are different). 
For each data set, we conducted two experiments: In the first experiment, $(N_1,\ldots,N_n)$ are the same as in the original data set and $K$ varies in $\{3,5,8,12\}$. In the second experiment, $K=5$ and $N_i=N$, with $N$ varying in $\{100,200,500,1000,2000\}$. 
We measure the performance by $\mathcal{L}(\hat{A},A)=\sum_{k=1}^K \|\hat{A}_k  -  A_k \|_1$, up to a permutation of columns of $\hat{A}$. For each $(K, N_1,\ldots,N_n)$, we generate 20 data sets and report the average $\mathcal{L}(\hat{A}, A)$ for each method. 

\spacingset{1} 
\begin{figure}[tb!]
  \includegraphics[width=.245 \textwidth, trim=17 0 0 0, clip=true]{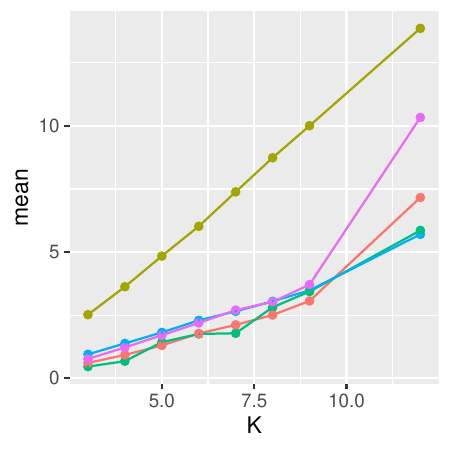}
  \includegraphics[width=.245 \textwidth, trim=17 0 0 0, clip=true]{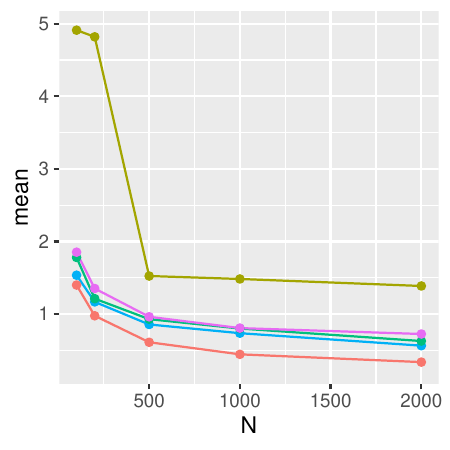}
    \includegraphics[width=.245 \textwidth, trim=17 0 0 0, clip=true]{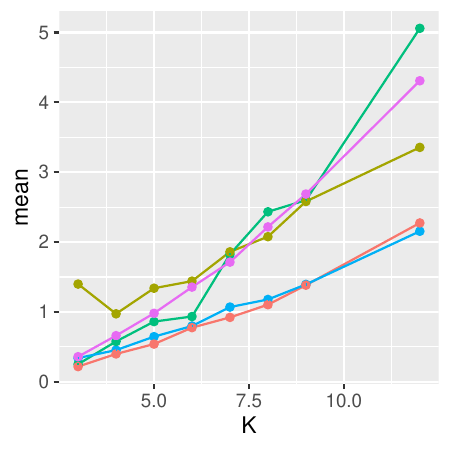}
     \includegraphics[width=.245 \textwidth, trim=17 0 0 0, clip=true]{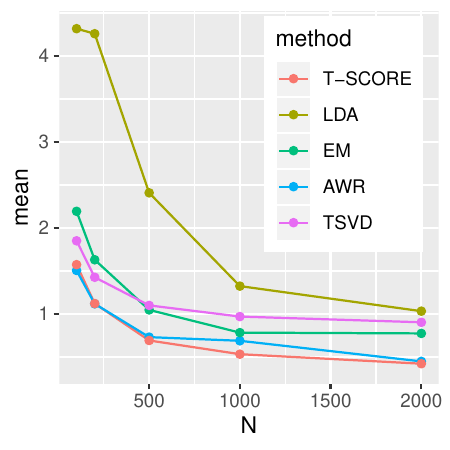}
  \caption{Results of the calibrated LDA models ($y$-axis is average $\mathcal{L}(\hat{A},A)$ over 20 repetitions). Left two panels: calibrated from AP ($n$=$2135$, $p$=$5188$). Right two panels: calibrated from NIPS ($n$=$1417$, $p$=$2508$). In the plots with varying $K$, the document lengths are as in real data. In the plots with varying $N$, all documents have the same length $N$.}\label{fig:semi-synth}
\end{figure}
\spacingset{1.45} 
\spacingset{1} 
\begin{table}[hbt!]
\centering
\caption{Computation time on the semi-synthetic experiments $(N=2000, K=5)$.}
\scalebox{.85}{
\begin{tabular}{ccrr}
  \toprule
  Method			&Software 	& AP data (in second) 	& NIPS data (in second)\\
  \hline
  Topic-SCORE		&R		& $1$	& $< 1$ \\
  LDA					&R		&378 	&395 \\
  AWR				&Python	&113	&37\\
  TSVD					&MATLAB	& $4$	& $2$\\
  EM & Python & 230 & 68 \\
  \bottomrule
\end{tabular}}
\label{table:semi-synthetic}
\end{table}
\spacingset{1.45} 

The results are shown in Figure~\ref{fig:semi-synth}. 
This is an LDA-calibrated model, not the pLSI model with anchor-word condition. However, our method still performs well. In most settings, our method yields the smallest estimation errors. In Table~\ref{table:semi-synthetic}, we report the computing time of different methods, for $(N,K)=(2000,5)$. Our method is much faster than LDA, AWR and EM and comparable with TSVD (note that our method has smaller errors than TSVD).

\subsection{The pLSI models}\label{Subsec Synthetic data}
We simulate data from the pLSI model \eqref{pLSI}, under the anchor-word condition. 
Given parameters $\{p, n, N, K, m_p, \delta_p, m_n\}$, we generate $D$ as follows. First,  we generate the topic matrix $A$.  For $1\leq k\leq K$, let each of the $[(k-1)m_p+1]$-th row to the $(km_p)$-th row equal to $\delta_p e_k'$, where $e_1,\ldots,e_K$ are the standard basis vectors of $\mathbb{R}^K$. For the remaining $(p-Km_p)$ rows, we first generate all entries $iid$ from $Unif(0,1)$, and then normalize each column of the $(p-Km_p)\times K$ sub-matrix to have a sum of $(1-m_p\delta_p)$.
Next, we generate the weight matrix $W$: For $1\leq k\leq K$, let each of the $[(k-1)m_n+1]$-th column to the $(km_n)$-th column equal to $e_k$. For the remaining columns, we first generate all entries $iid$ from $Unif(0,1)$, and then normalize each column to have a sum of $1$.
Last, we generate $D$ from $(A,W)$ using model \eqref{pLSI}. 
Here, $m_p$ is the number of anchor words, $m_n$ is the number of pure documents per topic, and each anchor word satisfies the $\delta_p$-separability condition in \cite{Ge}. 
For each setting, we report the average of $\mathcal{L}(\hat{A}, A)$ over 200 repetitions. 
In Experiments 1-4, we compare our method with LDA (which is not designed for the pLSI model), AWR and TSVD. In Experiment 5, we compare our method with EM.

\spacingset{1} 
\begin{figure}[t]
  \includegraphics[width=.24\textwidth, height=.255\textwidth, trim=42 20 40 70, clip=true]{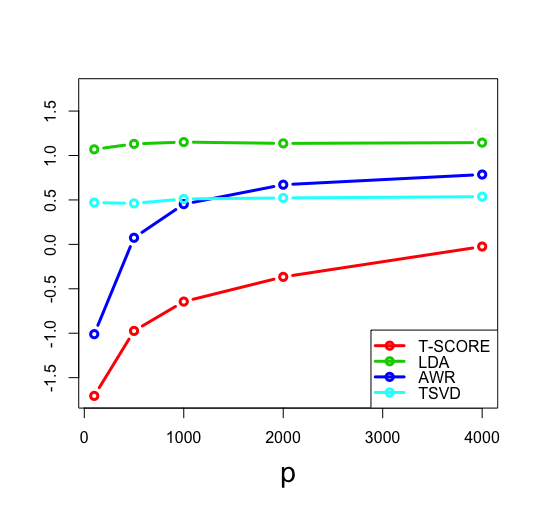}
   \includegraphics[width=.24\textwidth, height=.255\textwidth, scale=0.3, trim=42 20 40 70, clip=true]{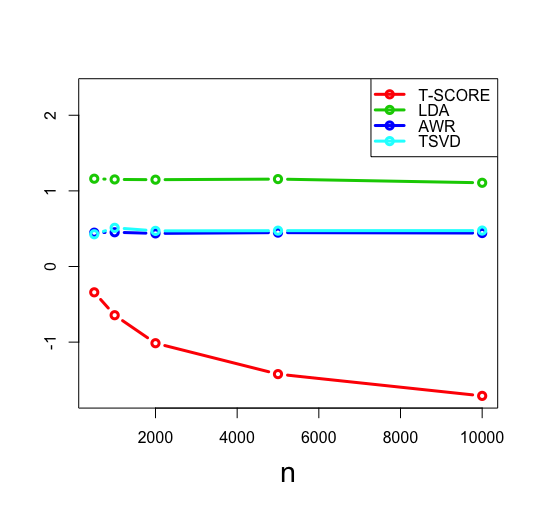}
  \includegraphics[width=.24\textwidth, height=.255\textwidth, trim=42 20 40 70, clip=true]{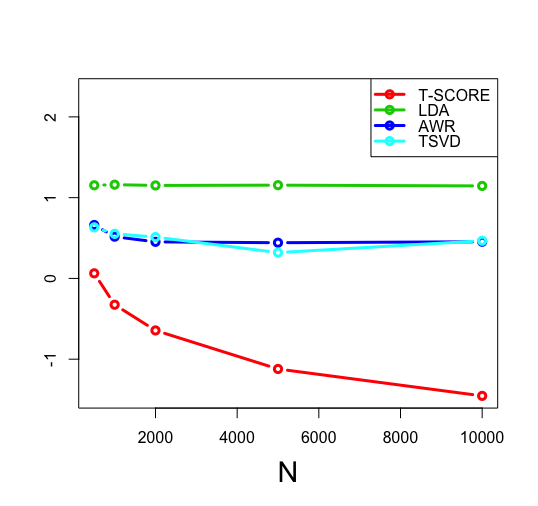}
  \includegraphics[width=.24\textwidth, height=.255\textwidth, trim=42 20 40 70, clip=true]{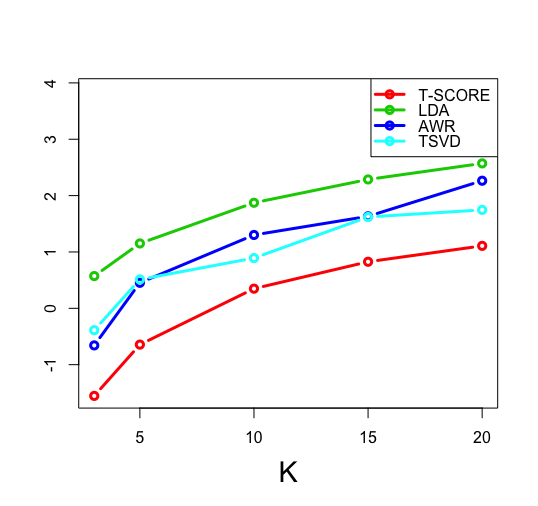}
  \caption{Experiment 1. The $y$-axis is $\log(\mathcal{L}(\hat{A},A))$, and $(p,n,N,K)$ represent the vocabulary size, number of documents, document length, and number of topics, respectively.}\label{fig:synth1_1}
\end{figure}
\spacingset{1.45}

\medskip
\noindent
{\bf Experiment 1: Varying $(p,n,N,K)$.} 
We fix a basic setting where $(p, n, N, K, m_p, \delta_p, m_n)=(1000,1000,2000,5,p/100, 1/p, n/100)$. In each of sub-experiments 1.1-1.4, we vary one of $(p,n,N,K)$ and keep the other parameters the same as in the basic setting. The results are in Figure~\ref{fig:synth1_1}. In all settings, our method yields the smallest error. 
Furthermore, we have the following observations: (i) As $n$ or $N$ increases, our method is the only one whose error has a clear decreasing trend, i.e., our method can take advantage of including {\it more} documents and having {\it longer} documents. (ii) As $K$ increases, the errors of all four methods increase, suggesting that the problem becomes more challenging for larger $K$. (iii) As $p$ increases, the errors of our method and AWR increase, while the errors of LDA and TSVD remain stable; but even for $p$ as large as 4000, our method still outperforms LDA and TSVD.

\spacingset{1} 
\begin{figure}[t]
  \centering
     \includegraphics[width=.28\textwidth, height=.255\textwidth, trim=40 20 35 70, clip=true]{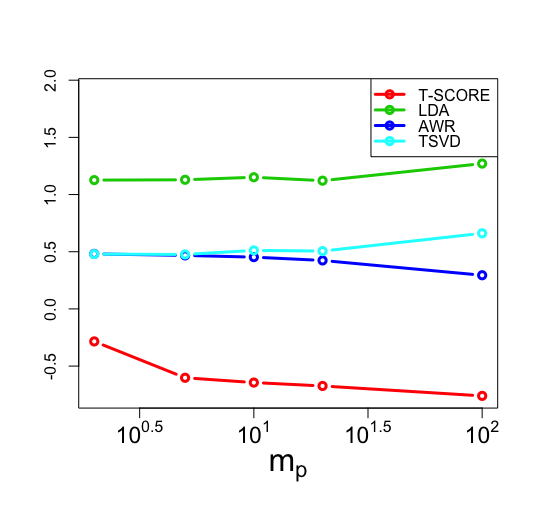}
  \includegraphics[width=.28\textwidth, height=.255\textwidth, trim=40 20 35 70, clip=true]{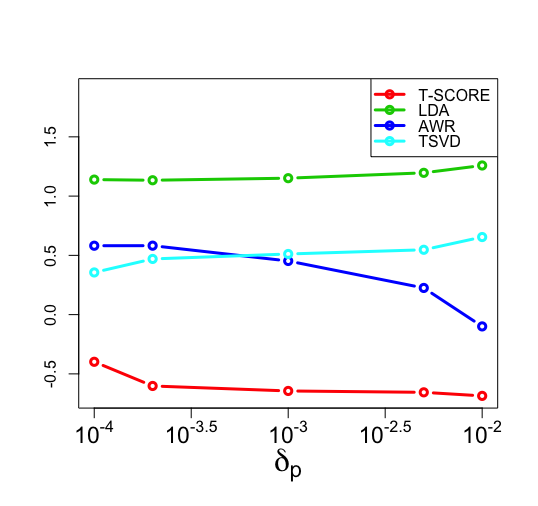}
   \includegraphics[width=.278\textwidth, height=.255\textwidth, trim=40 20 35 70, clip=true]{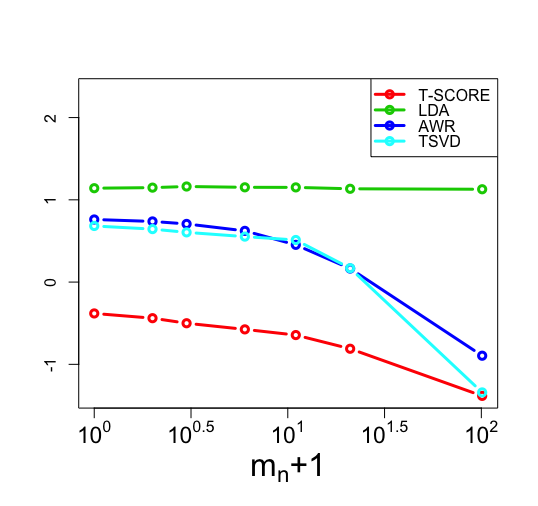}
   \vspace{-1em}
  \caption{Experiment 2. The $y$-axis is $\log(\mathcal{L}(\hat{A},A))$, and $(m_p, \delta_p, m_n)$ represent the number of anchor words, separability of anchor words, and number of pure documents, respectively.}\label{fig:synth1_2}
\end{figure}
\spacingset{1.45}

\medskip
\noindent
{\bf Experiment 2: Anchor words and pure documents.}
We fix the same basic setting as in Experiment 1 and vary one of $(m_p, \delta_p, m_n)$ in each sub-experiment. The results are in Figure~\ref{fig:synth1_2}. First, we look at the effect of anchor words. In the left panel, as $m_p$ (number of anchor words per topic) increases, the error of our method decreases considerably, suggesting that our method can take advantage of having multiple anchor words. Even with $m_p=2$, our method still outperforms the other methods. In the middle panel, as $\delta_p$ (separability of anchor words) increases, the errors of AWR and our method both decrease, and they both outperform LDA and TSVD; moreover, our method outperforms AWR. Furthermore, as long as $\delta_p$ is larger than $2\times 10^{-4}$, our method is relatively insensitive to $\delta_p$; this is consistent with the theory in Section~\ref{sec:Theory}. 
Second, we check the effect of pure documents. In the right panel, as $m_n$ (number of pure documents) increases, the performances of all methods except LDA improves. The improvement on TSVD is especially prominent, because TSVD needs the pure document assumption. When $m_n< 100$, our method has a significant advantage over TSVD; when $m_n=100$, the performance of our method is similar to that of  TSVD.

\begin{figure}[t]
  \centering
  \includegraphics[width=.24\textwidth, height=.255\textwidth, trim=42 20 40 70, clip=true, clip=true]{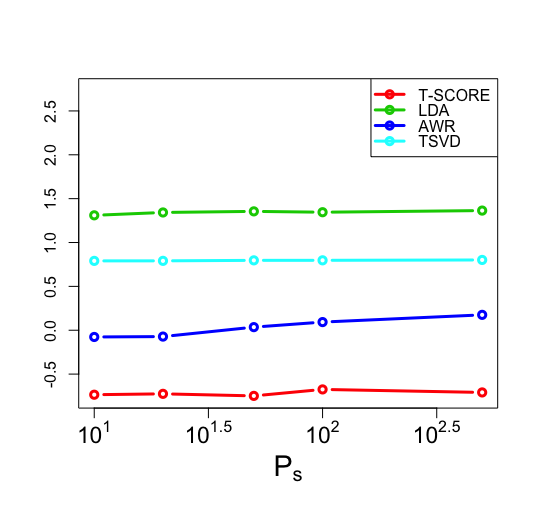}
  \includegraphics[width=.24\textwidth, height=.255\textwidth, trim=42 20 40 70, clip=true]{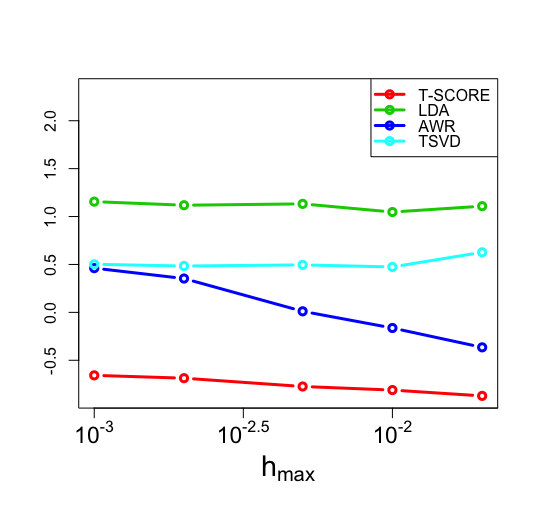}
     \includegraphics[width=.24\textwidth, height=.255\textwidth, trim=42 20 40 70, clip=true]{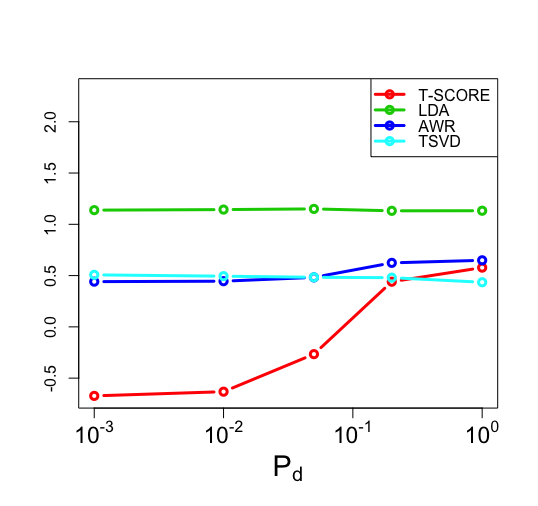}
   \includegraphics[width=.24\textwidth, height=.255\textwidth, trim=42 20 40 70, clip=true]{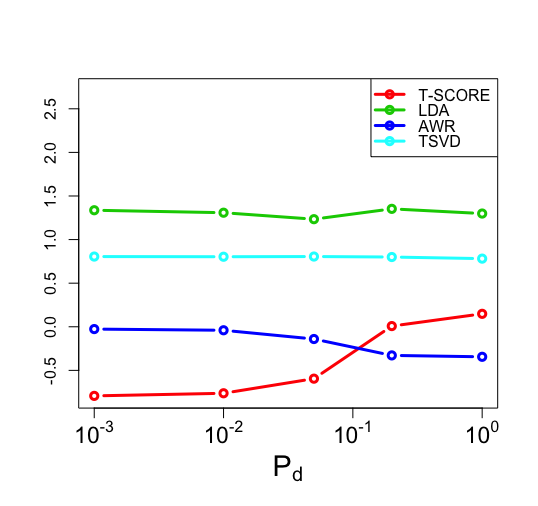}
    \caption{Left two panels: Experiments 3. In the first plot, the word frequency heterogeneity follows the Zipf's law. In the second plot, the word frequency heterogeneity has two scales. In both panels, the frequency heterogeneity increases as either $P_s$ decreases or $h_{\max}$ increases.
    Right two panels: Experiment 4. The first plot is the homogeneous setting, and the second plot is the heterogeneous setting. In both plots, as $P_d$ increases, the almost-anchor words are less anchor-like. The $y$-axis is $\log(\mathcal{L}(\hat{A},A))$.} \label{fig:synth23}
    \end{figure}

\medskip
\noindent
{\bf Experiment 3: Word frequency heterogeneity.}
We consider settings with severe frequency heterogeneity.  
Fix $(p, n, N, K, m_p$, $\delta_p, m_n)=(1000,1000,2000,5,p/100, 1/p, n/100)$. We generate the first $Km_p$ rows of $A$ in the same way as before and generate the remaining $(p-Km_p)$ rows using two different settings below:  
{\it Setting 1: Zipf's law}. Given $P_s>0$, we first generate $A(j,k)$ from the exponential distribution with mean $(P_s+j)^{-1.07}$, independently for all $1\leq k\leq K$, $Km_p+1\leq j\leq p$, and then normalize each column of the $(p-Km_p)\times K$ matrix to have a sum of $(1-m_p\delta_p)$. Under this setting, the word frequencies of each topic roughly follow a Zipf's law with $P_s$ stopping words. A smaller $P_s$ corresponds to larger heterogeneity. 
{\it Setting 2: Two scales}.  Given $h_{\max}\in [1/p,1)$, we generate $\{A(j,k):1\leq k\leq K, Km_p< j\leq Km_p+n_{\max}\}$ $iid$ from $Unif(0, h_{\max})$, where $n_{\max}=\lfloor(1-m_p\delta_p)/(2h_{\max})\rfloor$. Next, for $n_{\min}=p-Km_p-n_{\max}$ and $h_{\min}=(1-m_p\delta_p-h_{\max}n_{\max})/n_{\min}$, we generate $\{A(j,k): 1\leq k\leq K, Km_p+n_{\max}< j\leq p\}$ $iid$ from $Unif(0, h_{\min})$. Last, we normalize each column of the $(p-Km_p)\times K$ matrix to have a sum of $(1-m_p\delta_p)$. Under this setting,  the word frequencies of each topic are in two distinct scales, characterized by $h_{\max}$ and $h_{\min}$, respectively. 
We then generate $(W,D)$ in the same way as before. 
The results are displayed in the left two panels of Figure~\ref{fig:synth23}. Our method always yields the smallest errors. Interestingly, in Setting 2, the performance of AWR improves with increased heterogeneity.

\spacingset{1} 
\spacingset{1.45}

\medskip
\noindent
{\bf Experiment 4: The anchor word assumption only holds approximately.}
We consider settings where we have almost-anchor words but not the exact anchor words as in Definition~\ref{def:anchor}. 
Let $b'_j$ denote the $j$th row of $A$, normalized by its own $\ell^1$-norm. A word $j$ is an anchor word of topic $k$ if and only if $b_j(k)=1$. We define an $\epsilon$-almost-anchor word of topic $k$ if $b_j(k)\geq 1-\epsilon$. Fix $(p, n, N, K, m_p, \delta_p, m_n, P_s) = (1000, 1000,2000,5,p/100, 1/p, n/100, p/20)$. We generate $A$ using two different settings: {\it Setting 1: Homogeneous words}. Given $P_d\in [0,1]$, for each $k$, let all of row $[(k-1)m_p+1]$ to row $(km_p)$ equal to $\delta_p \tilde{e}_k'$, where $\tilde{e}_k(j)=1\{j=k\}+P_d1\{j\neq k\}$, $1\leq j\leq K$. For the remaining $(p-Km_p)$ rows, we first generate entries $iid$ from $Unif(0,1)$, and then normalize each column of the $(p-Km_p)\times K$ sub-matrix to have a sum of $[1-m_p\delta_p-m_p\delta_p(K-1)P_d]$.
{\it Setting 2: Heterogenous words}. Given $P_d\in [0,1]$, first, we generate $A(j,k)$ from the exponential distribution with mean $(P_s+j)^{-1.07}$, independently for all $1\leq k\leq K$, $1\leq j\leq p$; second, for each $1\leq k\leq K$, we randomly select $m_p$ rows from all the rows whose largest entry is the $k$-th entry, and for these selected rows, we keep the $k$-th entry and multiply the other entries by $P_d$; last, we renormalize each column of $A$ to have a sum of $1$. 
We then generate $(W,D)$ in the same way as before. 
In both settings, there are $m_p$ almost-anchor words per topic. The parameter $P_d$ controls the anchorness of these words: a smaller $P_d$ means that the almost-anchor words are more similar to anchor words. In Setting 1, the value of $\epsilon$ for the almost-anchor-words is $\approx \frac{(K-1)P_d}{1+(K-1)P_d}$; when $P_d=0$, these almost-anchor words become exact anchor words. In Setting 2, $P_d$ plays a similar role. 
The results are in the right two panels of Figure~\ref{fig:synth23}. In both settings, our method has the smallest errors in a wide range of $P_d$, suggesting that our method has reasonable performance even without exact anchor words. In Setting 1, when $P_d=1$, TSVD yields the best performance and the performance of our method is slightly worse than that of TSVD. In Setting 2, when $P_d>0.1$, our method is better than LDA and TSVD but is worse than AWR.

\spacingset{1} 
\begin{figure}[t]
  \centering
  \includegraphics[width=.29\textwidth, height=.27\textwidth, trim=17 0 0 0, clip=true]{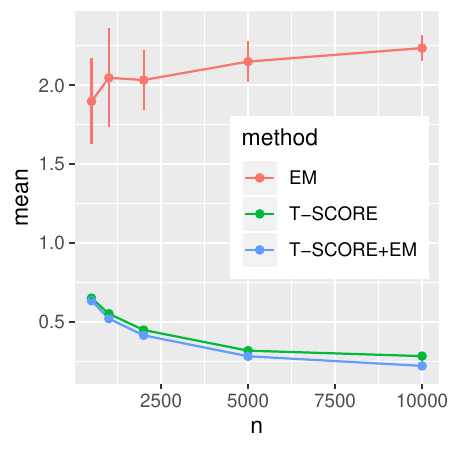} $\quad$
  \includegraphics[width=.29\textwidth, height=.27\textwidth, trim=17 0 0 0, clip=true]{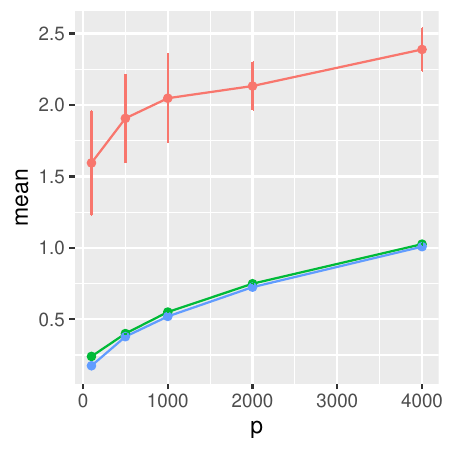}
    \caption{Experiment 5. The EM algorithm is initialized by the default approach (multiple random start). T-SCORE+EM uses T-SCORE to initialize EM.  The dots and vertical bars show the average value and standard deviation of $\mathcal{L}(\hat{A},A)$ over 20 repetitions. For T-SCORE and T-SCORE+EM, the standard deviations are very small, so the vertical bars are invisible.} \label{fig:synth-add}
    \end{figure}
\spacingset{1.45}

\medskip
\noindent
{\bf Experiment 5: Comparison with the EM algorithm.} The EM algorithm is an iterative algorithm to compute a local maximizer of the pLSI likelihood. We find that its performance is sensitive to initialization. 
We use both the default initialization of multiple random starts and the initialization by our method, denoted as EM and T-SCORE-EM, respectively. The maximum number of EM iterations is set to be 50.  
The results are in Figure~\ref{fig:synth-add}, where we plot the average value and standard deviation of ${\cal L}(\hat{A}, A)$ over 20 repetitions. The performance of EM is uniformly worse than our method and has much larger variability across repetitions. The reason is that its performance depends on the quality of initialization, and the default random initialization does not work well. In comparison, if we use T-SCORE to initialize, the performance is much better and becomes more stable. T-SCORE-EM also slightly improves T-SCORE, suggesting that the EM updates can locally improve our estimate. However, this improvement is mild, and the main advantage still comes from the T-SCORE initialization.

\section{Discussion}  \label{sec:Discuss}
We propose an SVD-based method for topic modeling. To our best knowledge, our method is the first that estimates $A$ from only a few leading singular vectors of the corpus matrix. 
Our method combines several non-trivial ideas, including a pre-SVD normalization to tackle severe frequency heterogeneity, a post-SVD normalization to create a low-dimensional word embedding, and a post-SVD procedure that obtains $\hat{A}$ explicitly from the embedded point cloud. 
We give the rate of convergence for our method, and show that it compares favorably with existing results in the literature. Our analysis is based on the row-wise large-deviation bounds for singular vectors, which we spent a significant amount of efforts to derive. 
We apply our method to Associated Press news articles and abstracts of statistical papers.

The Bayesian approaches \citep{wang2021bayesian} and factorization approaches \citep{li2021topic} are also commonly used for topic modeling. For very large corpora, these approaches often need a fast initialization. Our SVD-based method can be potentially used as an initialization for these approaches.

The topic model with sparsity has also attracted recent attentions.  \cite{bing2020optimal} studied the optimal rate for sparse topic modeling, and \cite{bing2021likelihood} derived properties of the likelihood estimation. 
We leave it to future work to generalize our SVD-based method to sparse topic modeling, where we may replace the SVD step by a sparse SVD method \citep{zou2018selective}. 

The current topic model only uses word counts of a document. The multi-gram topic model will also use word orders, and the corpus is stored in a multi-way tensor. We leave it to future work to extend our method to this setting by using tensor factorization \citep{chen2021factor}.

\subsection*{Data and code}

Data and code for reproducing the numerical results of this paper can be found at GitHub (\url{https://github.com/ZhengTracyKe/TopicSCORE}).

\subsection*{Acknowledgements}
The authors thank the Associate Editor and two anonymous referees for helpful comments. The authors thank Jiashun Jin and John Lafferty for reading an early draft of the paper and giving many useful comments. 
The authors thank Pengsheng Ji for sharing the SLA data. Z. Ke thanks Art Owen for useful comments on the real data results. Z. Ke also thanks Rina Barber, Chao Gao and John Lafferty for helpful discussions in the HELIOS reading group, which inspired her to work on topic modeling. The research of Z. Ke is partially supported by the NSF CAREER grant DMS-1943902


\spacingset{.9}
\small

\bibliography{topic}

\begin{thebibliography}{}

\bibitem[\protect\citeauthoryear{Abbe, Fan, Wang, and Zhong}{Abbe
  et~al.}{2020}]{abbe2017entrywise}
Abbe, E., J.~Fan, K.~Wang, and Y.~Zhong (2020).
\newblock Entrywise eigenvector analysis of random matrices with low expected
  rank.
\newblock {\em Ann. Statist.\/}~{\em 48\/}(3), 1452.

\bibitem[\protect\citeauthoryear{Ara{\'u}jo, Saldanha, Galvao, Yoneyama, Chame,
  and Visani}{Ara{\'u}jo et~al.}{2001}]{araujo2001successive}
Ara{\'u}jo, M. C.~U., T.~C.~B. Saldanha, R.~K.~H. Galvao, T.~Yoneyama, H.~C.
  Chame, and V.~Visani (2001).
\newblock The successive projections algorithm for variable selection in
  spectroscopic multicomponent analysis.
\newblock {\em Chemometrics and Intelligent Laboratory Systems\/}~{\em
  57\/}(2), 65--73.

\bibitem[\protect\citeauthoryear{Arora, Ge, Halpern, Mimno, Moitra, Sontag, Wu,
  and Zhu}{Arora et~al.}{2013}]{arora2013practical}
Arora, S., R.~Ge, Y.~Halpern, D.~Mimno, A.~Moitra, D.~Sontag, Y.~Wu, and M.~Zhu
  (2013).
\newblock A practical algorithm for topic modeling with provable guarantees.
\newblock In {\em International Conference on Machine Learning (ICML)}, pp.\
  280--288.

\bibitem[\protect\citeauthoryear{Arora, Ge, and Moitra}{Arora
  et~al.}{2012}]{Ge}
Arora, S., R.~Ge, and A.~Moitra (2012).
\newblock Learning topic models--going beyond {SVD}.
\newblock In {\em Foundations of Computer Science (FOCS)}, pp.\  1--10.

\bibitem[\protect\citeauthoryear{Bansal, Bhattacharyya, and Kannan}{Bansal
  et~al.}{2014}]{bansal2014provable}
Bansal, T., C.~Bhattacharyya, and R.~Kannan (2014).
\newblock A provable {SVD}-based algorithm for learning topics in dominant
  admixture corpus.
\newblock In {\em Adv. Neural Inf. Process. Syst.}, pp.\  1997--2005.

\bibitem[\protect\citeauthoryear{Bing, Bunea, Strimas-Mackey, and Wegkamp}{Bing
  et~al.}{2021}]{bing2021likelihood}
Bing, X., F.~Bunea, S.~Strimas-Mackey, and M.~Wegkamp (2021).
\newblock Likelihood estimation of sparse topic distributions in topic models
  and its applications to wasserstein document distance calculations.
\newblock {\em arXiv:2107.05766\/}.

\bibitem[\protect\citeauthoryear{Bing, Bunea, and Wegkamp}{Bing
  et~al.}{2020a}]{bing2020fast}
Bing, X., F.~Bunea, and M.~Wegkamp (2020a).
\newblock A fast algorithm with minimax optimal guarantees for topic models
  with an unknown number of topics.
\newblock {\em Bernoulli\/}~{\em 26\/}(3), 1765--1796.

\bibitem[\protect\citeauthoryear{Bing, Bunea, and Wegkamp}{Bing
  et~al.}{2020b}]{bing2020optimal}
Bing, X., F.~Bunea, and M.~Wegkamp (2020b).
\newblock Optimal estimation of sparse topic models.
\newblock {\em J. Mach. Learn. Res.\/}~{\em 21\/}(177), 1--45.

\bibitem[\protect\citeauthoryear{Bioucas-Dias, Plaza, Dobigeon, Parente, Du,
  Gader, and Chanussot}{Bioucas-Dias et~al.}{2012}]{bioucas2012hyperspectral}
Bioucas-Dias, J.~M., A.~Plaza, N.~Dobigeon, M.~Parente, Q.~Du, P.~Gader, and
  J.~Chanussot (2012).
\newblock Hyperspectral unmixing overview: Geometrical, statistical, and sparse
  regression-based approaches.
\newblock {\em IEEE journal of selected topics in applied earth observations
  and remote sensing\/}~{\em 5\/}(2), 354--379.

\bibitem[\protect\citeauthoryear{Blei}{Blei}{2012}]{blei2012probabilistic}
Blei, D. (2012).
\newblock Probabilistic topic models.
\newblock {\em Commun. ACM\/}~{\em 55\/}(4), 77--84.

\bibitem[\protect\citeauthoryear{Blei, Ng, and Jordan}{Blei
  et~al.}{2003}]{blei2003latent}
Blei, D., A.~Ng, and M.~Jordan (2003).
\newblock Latent dirichlet allocation.
\newblock {\em J. Mach. Learn. Res.\/}~{\em 3}, 993--1022.

\bibitem[\protect\citeauthoryear{Cai, Zhang, et~al.}{Cai
  et~al.}{2018}]{cai2018rate}
Cai, T.~T., A.~Zhang, et~al. (2018).
\newblock Rate-optimal perturbation bounds for singular subspaces with
  applications to high-dimensional statistics.
\newblock {\em Ann. Statist.\/}~{\em 46\/}(1), 60--89.

\bibitem[\protect\citeauthoryear{Chen, Jin, and Ke}{Chen
  et~al.}{2021}]{DYChen2}
Chen, D., J.~Jin, and Z.~T. Ke (2021).
\newblock A new approach to vertex hunting by k-nearest neighborhood denoising.
\newblock {\em Manuscript\/}.

\bibitem[\protect\citeauthoryear{Chen, Ke, and Zhang}{Chen
  et~al.}{2021}]{VALISE}
Chen, D., Z.~T. Ke, and S.~Zhang (2021).
\newblock Valise: A robust vertex hunting algorithm with theoretical
  guarantees.
\newblock {\em Manuscript\/}.

\bibitem[\protect\citeauthoryear{Chen, Yang, and Zhang}{Chen
  et~al.}{2022}]{chen2021factor}
Chen, R., D.~Yang, and C.-H. Zhang (2022).
\newblock Factor models for high-dimensional tensor time series (with
  discussions).
\newblock {\em J. Amer. Statist. Soc.\/}~{\em 117\/}(537), 94--116.

\bibitem[\protect\citeauthoryear{Davis and Kahan}{Davis and
  Kahan}{1970}]{sin-theta}
Davis, C. and W.~M. Kahan (1970).
\newblock The rotation of eigenvectors by a perturbation. iii.
\newblock {\em SIAM J. Numer. Anal.\/}~{\em 7\/}(1), 1--46.

\bibitem[\protect\citeauthoryear{Deerwester, Dumais, Furnas, Landauer, and
  Harshman}{Deerwester et~al.}{1990}]{deerwester1990indexing}
Deerwester, S., S.~T. Dumais, G.~W. Furnas, T.~K. Landauer, and R.~Harshman
  (1990).
\newblock Indexing by latent semantic analysis.
\newblock {\em J. Amer. Statist. Soc.\/}~{\em 41\/}(6), 391.

\bibitem[\protect\citeauthoryear{Dey, Hsiao, and Stephens}{Dey
  et~al.}{2017}]{dey2017visualizing}
Dey, K.~K., C.~J. Hsiao, and M.~Stephens (2017).
\newblock Visualizing the structure of {RNA}-seq expression data using grade of
  membership models.
\newblock {\em PLoS genetics\/}~{\em 13\/}(3), e1006599.

\bibitem[\protect\citeauthoryear{Ding, Li, and Peng}{Ding
  et~al.}{2008}]{ding2008equivalence}
Ding, C., T.~Li, and W.~Peng (2008).
\newblock On the equivalence between non-negative matrix factorization and
  probabilistic latent semantic indexing.
\newblock {\em Comput. Statist. Data Anal.\/}~{\em 52\/}(8), 3913--3927.

\bibitem[\protect\citeauthoryear{Donoho and Stodden}{Donoho and
  Stodden}{2004}]{donoho2003does}
Donoho, D. and V.~Stodden (2004).
\newblock When does non-negative matrix factorization give a correct
  decomposition into parts?
\newblock In {\em Adv. Neural Inf. Process. Syst.}, pp.\  1141--1148.

\bibitem[\protect\citeauthoryear{Fan, Fan, Han, and Lv}{Fan
  et~al.}{2022}]{fan2020asymptotic}
Fan, J., Y.~Fan, X.~Han, and J.~Lv (2022).
\newblock Asymptotic theory of eigenvectors for random matrices with diverging
  spikes.
\newblock {\em J. Amer. Statist. Soc.\/}~{\em 117\/}(538), 996--1009.

\bibitem[\protect\citeauthoryear{Fan, Xue, and Zhou}{Fan
  et~al.}{2021}]{fan2021much}
Fan, J., L.~Xue, and Y.~Zhou (2021).
\newblock How much can machines learn finance from chinese text data?
\newblock {\em Available at SSRN\/}.

\bibitem[\protect\citeauthoryear{Freedman}{Freedman}{1975}]{freedman1975tail}
Freedman, D.~A. (1975).
\newblock On tail probabilities for martingales.
\newblock {\em Ann. Probab.\/}~{\em 3\/}(1), 100--118.

\bibitem[\protect\citeauthoryear{Gillis and Vavasis}{Gillis and
  Vavasis}{2013}]{gillis2013fast}
Gillis, N. and S.~A. Vavasis (2013).
\newblock Fast and robust recursive algorithmsfor separable nonnegative matrix
  factorization.
\newblock {\em IEEE transactions on pattern analysis and machine
  intelligence\/}~{\em 36\/}(4), 698--714.

\bibitem[\protect\citeauthoryear{Girolami and Kab{\'a}n}{Girolami and
  Kab{\'a}n}{2003}]{girolami2003equivalence}
Girolami, M. and A.~Kab{\'a}n (2003).
\newblock On an equivalence between {PLSI} and {LDA}.
\newblock In {\em Proceedings of the 26th annual international ACM SIGIR
  conference on Research and development in informaion retrieval}, pp.\
  433--434.

\bibitem[\protect\citeauthoryear{Harman}{Harman}{1993}]{harman1overview}
Harman, D. (1993).
\newblock Overview of the first text retrieval conference (trec-1).
\newblock In {\em Proceedings of the first Text REtrieval Conference (TREC-1)},
  pp.\  1--20.

\bibitem[\protect\citeauthoryear{Hofmann}{Hofmann}{1999}]{hofmann1999}
Hofmann, T. (1999).
\newblock Probabilistic latent semantic indexing.
\newblock In {\em International ACM SIGIR conference}, pp.\  50--57.

\bibitem[\protect\citeauthoryear{Horn and Johnson}{Horn and
  Johnson}{1985}]{HornJohnson}
Horn, R. and C.~Johnson (1985).
\newblock {\em Matrix Analysis}.
\newblock Cambridge University Press.

\bibitem[\protect\citeauthoryear{Javadi and Montanari}{Javadi and
  Montanari}{2020}]{javadi2019nonnegative}
Javadi, H. and A.~Montanari (2020).
\newblock Nonnegative matrix factorization via archetypal analysis.
\newblock {\em J. Amer. Statist. Soc.\/}~{\em 115\/}(530), 896--907.

\bibitem[\protect\citeauthoryear{Ji and Jin}{Ji and Jin}{2016}]{SCC-JiJin}
Ji, P. and J.~Jin (2016).
\newblock Coauthorship and citation networks for statisticians.
\newblock {\em Ann. Appl. Statist.\/}~{\em 10\/}(4), 1779--1812.

\bibitem[\protect\citeauthoryear{Ji, Jin, Ke, and Li}{Ji
  et~al.}{2021}]{PaseII-paper2}
Ji, P., J.~Jin, Z.~T. Ke, and W.~Li (2021).
\newblock Meta-analysis on citations for statisticians.
\newblock {\em Manuscript\/}.

\bibitem[\protect\citeauthoryear{Jin}{Jin}{2015}]{SCORE}
Jin, J. (2015).
\newblock Fast community detection by {SCORE}.
\newblock {\em Ann. Statist.\/}~{\em 43\/}(1), 57--89.

\bibitem[\protect\citeauthoryear{Jin, Ke, and Luo}{Jin
  et~al.}{2017}]{Mixed-SCORE}
Jin, J., Z.~T. Ke, and S.~Luo (2017).
\newblock Estimating network memberships by simplex vertex hunting.
\newblock {\em arXiv:1708.07852\/}.

\bibitem[\protect\citeauthoryear{Jin, Ke, Luo, and Wang}{Jin
  et~al.}{2022}]{jin2020estimating}
Jin, J., Z.~T. Ke, S.~Luo, and M.~Wang (2022).
\newblock Optimal estimation of the number of network communities.
\newblock {\em J. Amer. Statist. Soc.\/}, 1--16.

\bibitem[\protect\citeauthoryear{Li, Zhu, Qu, Ye, and Sun}{Li
  et~al.}{2021}]{li2021topic}
Li, Y., R.~Zhu, A.~Qu, H.~Ye, and Z.~Sun (2021).
\newblock Topic modeling on triage notes with semiorthogonal nonnegative matrix
  factorization.
\newblock {\em J. Amer. Statist. Soc.\/}, 1--16.

\bibitem[\protect\citeauthoryear{Mei and Zhai}{Mei and
  Zhai}{2001}]{mei2001note}
Mei, Q. and C.~Zhai (2001).
\newblock A note on {EM} algorithm for probabilistic latent semantic analysis.
\newblock In {\em International Conference on Information and Knowledge
  Management}.

\bibitem[\protect\citeauthoryear{Perrone, Jenkins, Spano, and Teh}{Perrone
  et~al.}{2017}]{perrone2016poisson}
Perrone, V., P.~A. Jenkins, D.~Spano, and Y.~W. Teh (2017).
\newblock Poisson random fields for dynamic feature models.
\newblock {\em J. Mach. Learn. Res.\/}~{\em 18}, 1--45.

\bibitem[\protect\citeauthoryear{Tsybakov}{Tsybakov}{2009}]{tsybakov2009introduction}
Tsybakov, A.~B. (2009).
\newblock Introduction to nonparametric estimation. revised and extended from
  the 2004 french original. translated by vladimir zaiats.

\bibitem[\protect\citeauthoryear{Vershynin}{Vershynin}{2012}]{Vershynin}
Vershynin, R. (2012).
\newblock Introduction to the non-asymptotic analysis of random matrices.
\newblock In {\em Compressed Sensing: Theory and Applications}, pp.\  210--268.
  Cambridge Univ. Press.

\bibitem[\protect\citeauthoryear{Wang, Zhang, Li, Deng, and Liu}{Wang
  et~al.}{2021}]{wang2021bayesian}
Wang, F., J.~L. Zhang, Y.~Li, K.~Deng, and J.~S. Liu (2021).
\newblock Bayesian text classification and summarization via a class-specified
  topic model.
\newblock {\em J. Mach. Learn. Res.\/}~{\em 22\/}(89), 1--48.

\bibitem[\protect\citeauthoryear{Zou and Xue}{Zou and
  Xue}{2018}]{zou2018selective}
Zou, H. and L.~Xue (2018).
\newblock A selective overview of sparse principal component analysis.
\newblock {\em Proceedings of the IEEE\/}~{\em 106\/}(8), 1311--1320.

\end{thebibliography}
\bibliographystyle{chicago}

\newpage

\spacingset{1.45}

\addtocontents{toc}{\protect\setcounter{tocdepth}{3}}

\tableofcontents

\appendix

\section{Two vertex hunting algorithms}
Our main algorithm contains a step, that is, to estimate the $K$ vertices of the Ideal Simplex from the low-dimensional point cloud $\hat{r}_1,\hat{r}_2,\ldots,\hat{r}_p\in\mathbb{R}^{K-1}$. This is conducted by a vertex hunting algorithm. Below, we give the detailed code of these two algorithms. 

\subsection{Successive projection}

The successive projection \cite{araujo2001successive} is a greedy algorithm. It successively projects the data points onto the orthogonal space of previously determined vertices and decides the next vertex by identifying the extreme point after projection. 

\bigskip
\noindent
{\it Successive projection}. Input: $K$, $\hat{r}_1,\ldots,\hat{r}_p$. Output: $\hat{v}_1^*,\ldots,\hat{v}_K^*$. 
\begin{itemize}
\item Initialize $Y_i=(1, \hat{r}'_i)'\in\mathbb{R}^K$, for $1\leq i\leq p$.
\item At iteration $k=1,2,\ldots,K$: Find $i_k=\mathrm{argmax}_{1\leq i\leq n}\|Y_i\|$ and let $u_k=Y_{i_k}/\|Y_{i_k}\|$. Set the $k$-th estimated vertex as $\hat{v}_k=\hat{r}_{i_k}$. Project all data points by updating $Y_i$ to $(1-u_ku_k')Y_i$, for $1\leq i\leq p$. 
\item Output $\hat{v}_1,\hat{v}_2,\ldots,\hat{v}_K$. 
\end{itemize}
 
We now show that SP satisfies Condition~\ref{cond:VH} in our setting. By \cite{Ge, gillis2013fast}, the statement in Condition~\ref{cond:VH} holds if (i) for each true vertex $v_k^*$, there is at least one $r_j$ that is located on $v_k^*$, and (ii) the volume of this simplex is lower bounded by a constant. 
In our setting, the requirement (i) is guaranteed by the anchor-word assumption, and the requirement (ii) is proved in Lemma~\ref{lem:V}. 
Therefore, if we plug in SP as the vertex hunting algorithm, all the theoretical results in Section~\ref{sec:Theory} hold.



\subsection{Sketched vertex search}
The sketched vertex search (SVS) \cite{Mixed-SCORE} is another vertex hunting algorithm. Its main idea is to first apply a de-noise step (Step VH-1 below) to reduce the noise level and then search for the best-fit simplex on the post-de-noise point cloud (Step VH-2 below).  
 
\bigskip
\noindent
{\it Sketched vertex search}. Input: $K$, $\hat{r}_1,\cdots,\hat{r}_p$, and tuning integers $L =10\times K$ and $K_0=\lceil 1.5\times K\rceil$.  
Output: $\hat{v}^*_1,\cdots,\hat{v}^*_K$.     
\begin{enumerate}
\item[VH-1.] Cluster by applying the classical $k$-means to $\hat{r}_1,\cdots,\hat{r}_p$,  assuming there are $L$ clusters.   Let $\hat{\theta}_1,\cdots,\hat{\theta}_L$ be the Euclidean centers of the clusters. 
\item[VH-2.] Let $\bar{\theta}=L^{-1}\sum_{j=1}^L\hat{\theta}_j$. Sort $\hat{\theta}_1,\hat{\theta}_2,\ldots, \hat{\theta}_L$ in the decreasing order of $\|\hat{\theta}_j-\bar{\theta}\|$ and retain only the first $K_0$ of them. Let $B$ denote the index set of retained cluster centers (note: $|B|=K_0$). Select $K$ distinct indices $\hat{j}_1, \hat{j}_2, \ldots, \hat{j}_K$ from $B$ such that $\hat{\theta}_{\hat{j}_1},\cdots,\hat{\theta}_{\hat{j}_K}$ are affinely independent and minimize 
\beq \label{SVS-obj}
\max_{1\leq j\leq L} \big\{\mathrm{distance}\big(\hat{\theta}_j, \; {\cal S}(\hat{\theta}_{j_1},\cdots,\hat{\theta}_{j_K})  \big)\big\},
\eeq
where ${\cal S}(\hat{\theta}_{j_1},\cdots,\hat{\theta}_{j_K})$ is the simplex with $\hat{\theta}_{j_1},\cdots,\hat{\theta}_{j_K}$ as vertices, and $\mathrm{distance}(\cdot,\cdot)$ is the Euclidean distance. 
Output $\hat{v}^*_k=\hat{\theta}_{\hat{j}_k}$, $1\leq k\leq K$. 
\end{enumerate}

In Step VH-2,  for arbitrary $(u,v_1,\ldots,v_K)$, the Euclidean distance from $u$ to ${\cal S}(v_1,\ldots,v_K)$ is computed as follows. We re-formulate it as to minimize $\|u-\sum_{k=1}^K \alpha_k v_k\|^2$ over $(\alpha_1,\ldots,\alpha_K)$, subject to the constraints that $0\leq \alpha_k\leq 1$ and $\sum_{k=1}^K\alpha_k=1$. This is a standard quadratic programming and can be computed easily. Therefore, for each given $\{j_1,j_2,\ldots,j_K\}\subset B$, we can compute \eqref{SVS-obj} directly. The optimization reduces to searching over $j_1,j_2,\ldots,j_K$ among the $K_0$ indices in $B$. 
For the default choice of $K_0$, this search is computationally manageable. In Table~\ref{table:semi-synthetic} of the main article, we report the computing time of Topic-SCORE with SVS as the vertex hunting algorithm, using calibrated data from real corpora. 
It only takes only 1.04 second on the AP data and 0.29 second on the NIPS data. 

We have observed appealing numerical performance of SVS. The topic-SCORE by plugging in SVS has lower errors than the topic-SCORE by plugging in SP. For this reason, we use SVS in all numerical experiments. 


In theory, \cite{Mixed-SCORE} showed that SVS satisfies Condition~\ref{cond:VH} when there are multiple points of $r_j$ located on each vertex and the remaining $r_j$'s are continuously distributed in an open set in the interior of the simplex. They also found settings where the vertex estimation error by SVS is strictly faster than that of SP (this supports our numerical observations). Later, \cite{DYChen2, VALISE} provided modifications of SVS and showed that Condition~\ref{cond:VH} can be satisfied provided that there is at least one $r_j$ located on the vertex of each simplex.

\section{Comparison with LDA on the AP data set}

Table~\ref{tb:real_ap_k3} shows the results of Topic-SCORE on the AP data set. We now report the results of LDA \cite{blei2003latent} on the same data set. We use the R package {\it lda} with $K=3$ and default algorithm parameters. Given $\hat{A}$ from LDA, we similarly compute $\hat{b}_j$ by $\hat{b}_j(k)=\hat{A}_k(j)/[\sum_{\ell=1}^K \hat{A}_\ell(j)]$, for $1\leq j\leq p$ and $1\leq k\leq K$. For each topic $k$, we obtain a list of ``representative words" corresponding to those $j$ with largest values of $\hat{b}_j(k)$. The output of LDA varies with the random seed. 
What we report below is with respect to setting the random seed as $0$. We also tried other random seeds, and the results are more or less similar.

\begin{table}[h]
\centering
\caption{Results of LDA on the AP data ($K=3$). For each topic, we report the top 15 representative words. }
\scalebox{.9}
{\begin{tabular}{ll}
\toprule
\multirow{2}{*}{Topic 1} & {\it waste, ready, nasa, requires, coast, july, tuesday, half, turn, health, }\\ 
& {\it speaker,  decide, virus, head, gallon}\\
\hline
\multirow{2}{*}{Topic 2} & {\it dan, source, sandinistas, gulf, diplomatic, crude, standards, control,}\\
&{\it rising, high,  feet, ability, prepared, armed, relatives}\\
\hline
\multirow{2}{*}{Topic 3} & {\it keep, scene, past, secretary, pacific, think, tv, saw, impact, sales, }\\
& {\it activists, fish, express, cuba, ago} \\
\bottomrule
\end{tabular}}
\end{table}

\section{A high-level description of the proof ideas} \label{sec:high-level-proof}
To study the rate of convergence of Topic-SCORE, we start from an equivalent description of the algorithm in matrix operation. Recall that $\hat{\xi}_1,\hat{\xi}_2,\ldots,\hat{\xi}_K$ are the first $K$ left singular vectors of $M^{-1/2}D$. 
The matrix $\hat{R}$ can be re-written as 
\[
\hat{R} = [\diag(\hat{\xi}_1)]^{-1}[\hat{\xi}_2,\ldots,\hat{\xi}_K], \qquad\mbox{or}\qquad  [{\bf 1}_p, \, \hat{R}] = [\diag(\hat{\xi}_1)]^{-1}\hat{\Xi}. 
\] 
Recall that $\hat{v}^*_1,\hat{v}^*_2,\ldots,\hat{v}^*_K$ are the vertices estimated by the vertex hunting step. Given these vertices, 
we first solve $\hat{\pi}_j^*$ from the linear equations $\sum_{k=1}^K\hat{\pi}_j^*(k)=1$ and $\sum_{k=1}^K\hat{\pi}_j^*(k)\hat{v}_k^*=\hat{r}_j$.
Introduce a $K\times K$ matrix
\[
\hat{Q} = \begin{pmatrix}
1 & \ldots & 1\\
\hat{v}_1^* & \ldots & \hat{v}_K^*
\end{pmatrix}. 
\]
We can write $\hat{\pi}_j^*$ explicitly as $\hat{\pi}_j^*=\hat{Q}^{-1}(1,\hat{r}'_j)'$. Write $\hat{\Pi}^*=[\hat{\pi}_1^*,\hat{\pi}_2^*,\ldots,\hat{\pi}_p^*]'$. It follows that
\[
\hat{\Pi}^*= [{\bf 1}_p, \, \hat{R}](\hat{Q}')^{-1}.  
\]
Our algorithm sets the negative entries in $\hat{\pi}_j^*$ to 0 and re-normalizes the vector to have a unit $\ell^1$-norm; this gives $\hat{\pi}_j$. Write $\hat{\Pi}=[\hat{\pi}_1,\hat{\pi}_2,\ldots,\hat{\pi}_p]'$. 
The estimate $\hat{A}$ is obtained by re-normalizing each column of the matrix $M^{1/2}[\diag(\hat{\xi}_1)] \hat{\Pi}$ to have a unit $\ell^1$-norm. 
To express $\hat{A}$ in terms of $\hat{\Pi}^*$, we define two operators, ${\cal P}_{\mathrm{round}}$ and ${\cal N}_{\mathrm{col}}$: for a given matrix $B$, ${\cal P}_{\mathrm{round}}(B)$ is the matrix by setting all negative entries in $B$ to 0 and renormalizing each row to have a unit $\ell^1$-norm, and ${\cal N}_{\mathrm{col}}(B)$ is the matrix by re-normalizing each column of $B$ by its own $\ell^1$-norm. It follows that
\[
\hat{A} = {\cal N}_{\mathrm{col}}\Bigl( M^{1/2} \diag(\hat{\xi}_1)\cdot {\cal P}_{\mathrm{round}}(\hat{\Pi}^*) \Bigr). 
\]
Combining the above, we can express the Topic-SCORE algorithm in matrix form:
\beq \label{Explain-1}
\hat{A}  = {\cal N}_{\mathrm{col}}\Bigl( M^{1/2}\diag(\hat{\xi}_1)\cdot {\cal P}_{\mathrm{round}}\Bigl( [\diag(\hat{\xi}_1)]^{-1}\hat{\Xi}(\hat{Q}')^{-1} \Bigr) \Bigr). 
\eeq

First, in Section~\ref{sec:supp-oracle}, we study the oracle case, where every quantity on the right hand side of \eqref{Explain-1} is replaced by its population counterpart. Let $\Xi = [\xi_1,\xi_2, \ldots,\xi_K]$ contain the left singular vectors of $M_0^{-1/2}AW$, where $M_0$ is the population counterpart of $M$. Define
\[
Q =\begin{pmatrix}
1 & \ldots & 1\\
v_1^* & \ldots & v_K^*
\end{pmatrix}, 
\]
where $v_1^*, v_2^*,\ldots,v_K^*$ are the vertices of the Ideal Simplex. Let $\hat{A}^{\text{oracle}}$ denote the output of applying Topic-SCORE on the population singular vectors. By \eqref{Explain-1}, 
\beq\label{Explain-2}
\hat{A}^{\mathrm{oracle}} := {\cal N}_{\mathrm{col}}\Bigl( M_0^{1/2}\diag(\xi_1)\cdot {\cal P}_{\mathrm{round}}\Bigl( [\diag(\xi_1)]^{-1}\Xi (Q')^{-1} \Bigr) \Bigr).  
\eeq
In Section~\ref{subsec:proof-Section2}, we show that there exists a positive vector $q\in\mathbb{R}^K$ such that 
\[
\Xi = M_0^{-1/2}A\cdot\diag(q)\cdot Q'. 
\]
We plug it into \eqref{Explain-2} to get 
\[
\hat{A}^{\text{oracle}} = {\cal N}_{\mathrm{col}}\Bigl( M_0^{1/2}\diag(\xi_1)\cdot {\cal P}_{\mathrm{round}}\Bigl( [\diag(\xi_1)]^{-1} M_0^{-1/2}A\cdot \diag(q)\Bigr) \Bigr).
\]
Recall that each $r_j$ is a convex combination of $v_1^*,v_2^*, \ldots,v_K^*$, and $\pi_j$ is the vector of combination coefficients. 
Write $\Pi=[\pi_1,\pi_2,\ldots,\pi_p]'$.  In Section~\ref{subsec:proof-Section2}, we also show that 
\[
[\diag(\xi_1)]^{-1} M_0^{-1/2}A\cdot \diag(q) = \Pi. 
\]
In particular, each row of $[\diag(\xi_1)]^{-1} M_0^{-1/2}A\cdot \diag(q)$ is already a nonnegative vector with unit $\ell^1$-norm. Hence, the operator ${\cal P}_{\mathrm{round}}$ has no effect on this matrix. It follows that
\begin{align*}
\hat{A}^{\text{oracle}} &= {\cal N}_{\mathrm{col}}\Bigl( M_0^{1/2}\diag(\xi_1)\cdot  [\diag(\xi_1)]^{-1} M_0^{-1/2}A\cdot \diag(q) \Bigr)\cr
&={\cal N}_{\mathrm{col}}\bigl( A\cdot\diag(q)\bigr) = A. 
\end{align*}
In other words, the Topic-SCORE outputs $A$ exactly in the oracle case. 

Next, in Section~\ref{sec:UBproof}, we study the real case and bound ${\cal L}(\hat{A}, A)={\cal L}(\hat{A}, \hat{A}^{\text{oracle}})$. Comparing \eqref{Explain-1} and \eqref{Explain-2}, the key is to control the noise accumulation in every step. In \eqref{Explain-1}, $\hat{A}$ is obtained from $M$, $\hat{\Xi}$ and $\hat{Q}$, where $\hat{Q}$ is constructed from the estimated vertices. Hence, there are three sources of noise: (i) noise in the diagonal matrix $M$, (ii) noise in the singular vectors $\hat{\xi}_1,\hat{\xi}_2,\ldots,\hat{\xi}_K$, and (iii) noise in the estimated vertices $\hat{v}_1^*,\ldots,\hat{v}_K^*$. We now introduce three quantities, $\Delta_1$, $\Delta_2$ and $Err_{VH}$, to measure three sources of noise, respectively.  
For noise source (i), $M$ is a diagonal matrix, whose population counterpart is $M_0$. Recall that $h_j=\|a_j\|_1$ captures the overall frequency of word $j$. 
Accounting for the potentially severe frequency heterogeneity, a proper measure of noise in $M$ is
\beq \label{Explain-4}
\Delta_1 = \max_{1\leq j\leq p}\bigl\{ h_j^{-1}|M(j,j)-M_0(j,j)|\bigr\}. 
\eeq
We then consider noise source (ii). 
Denote by $\hat{\Xi}_j'$ and $\Xi_j'$ the $j$th row of $\hat{\Xi}$ and $\Xi$, respectively. At first glance, it seems natural to measure the noise in $\hat{\Xi}$ by the maximum of $\|\hat{\Xi}_j-\Xi_j\|$ over $1\leq j\leq p$. However, this is incorrect, because $\hat{\Xi}$ is not necessarily close to $\Xi$. By sin-theta theorem \citep{sin-theta}, $\hat{\Xi}$ is close to $\Xi$ only if there is a properly large gap between every two nested singular values of $M_0^{-1/2}AW$. 
In our setting, there is only an appropriately large gap between the first and second singular values (this is because $M_0^{-1/2}AW$ is an irreducible nonnegative matrix; by Perron's theorem, its first singular value is always apart from the remaining singular values; furthermore, our regularity condition \eqref{cond-A} ensures that this gap is properly large; see Lemma~\ref{lem:PopEigVal}). Therefore, $\hat{\xi}_1$ is close to $\xi_1$. However, the other singular values may have zero or very small gaps between each other. Write $\hat{\Xi}^*=[\hat{\xi}_2,\ldots,\hat{\xi}_K]$ and $\Xi^*=[\xi_2,\ldots,\xi_K]$. By Perron's theorem, $\hat{\Xi}^*$ is close to $\Xi^*$ only up to a rotation of the  $(K-1)$ columns; i.e., there exists an (unknown) orthogonal matrix $\Omega^*\in\mathbb{R}^{(K-1)\times (K-1)}$ such that $\hat{\Xi}^*\approx \Xi^*\Omega^*$. Additionally, each singular vector is determined up to a sign flip (this sign flip is arbitrarily chosen by the SVD algorithm; for $\hat{\xi}_2,\ldots,\hat{\xi}_K$, such sign flips are already absorbed into the orthogonal matrix $\Omega^*$, so we only consider the sign flip of $\hat{\xi}_1$). It follows that $\hat{\Xi}=[\hat{\xi}_1, \hat{\Xi}^*]\approx [\omega \xi_1, \Xi^*\Omega^*]=\Xi \Omega$, where $\omega\in\{\pm 1\}$ and $\Omega = \diag(\omega, \Omega^*)$. Note that $\Omega$ is a $K\times K$ orthogonal matrix. It further implies that $(\Omega\hat{\Xi}_j)'=e_j'\hat{\Xi}\Omega'=e'_j\hat{\Xi}\Omega^{-1}\approx e_j'\Xi=(\Xi_j)'$. 
In light of this, let $\mathcal{O}_K$ denote the set of all matrices of the form $\Omega = \mathrm{diag}(\omega, \Omega^*)\in\mathbb{R}^{K,K}$, where $\omega\in \{\pm 1\}$ and $\Omega^*$ is an orthogonal matrix. A proper measure of noise in $\hat{\Xi}$ is 
\beq \label{Explain-5}
\Delta_2= \min_{\Omega\in\mathcal{O}_K}\max_{1\leq j\leq p} \bigl\{ h_j^{-1/2}\|\Omega\hat{\Xi}_j -\Xi_j\|\bigr\}. 
\eeq
For noise resource (iii), we also need to take into account this rotation. The vertex hunting is conducted on $\hat{r}_1,\hat{r}_2,\ldots,\hat{r}_p$, where by definition, $\hat{r}_j= [\hat{\xi}_1(j)]^{-1}\hat{\Xi}^*_j$, and $(\hat{\Xi}_j^*)'$ is the $j$th row of $\hat{\Xi}^*$. It gives
$\Omega^*\hat{r}_j=[\hat{\xi}_1(j)]^{-1}\Omega^*\hat{\Xi}^*_j \approx  [\omega\xi_1(j)]^{-1}\Xi^*_j= \omega^{-1}r_j$. Without loss of generality, we assume the signs of $\hat{\xi}_1$ and $\xi_1$ are picked such that $\omega=1$. Then,  $\Omega^*\hat{r}_j\approx r_j$, for $1\leq j\leq p$. We thus expect to see $\Omega^*\hat{v}_k^*\approx v^*_k$, where   $v_1^*,v_2^*, \ldots,v_K^*$ are the vertices of the Ideal Simplex, and $\hat{v}_1^*,\hat{v}_2^*, \ldots,\hat{v}_K^*$ are the estimated ones. A proper measure of noise in vertex hunting is
\beq \label{Explain-6}
Err_{VH}\equiv \min_{\substack{\kappa: \text{ a permutation}\\\text{ on }\{1,\ldots,K\}}}\Bigl\{ \max_{1\leq k\leq K}\|\Omega^*\hat{v}_k^*-v^*_{\kappa(k)}\| \Bigr\}.
\eeq
Here, the permutation comes from that we can re-label the estimated vertices in an arbitrary order. 
After defining the three quantities that capture three noise sources, in Sections~\ref{subsec:supp-UB-proof}-\ref{subsec:proof-lem-method}, 
we study how the estimation errors in $\hat{A}$ are affected by these quantities. Lemma~\ref{lem:method} gives the key result: Recall that $\hat{a}_j'$ and $a_j$ denote the $j$th row of $\hat{A}$ and $A$, respectively. Up to a permutation of columns of $\hat{A}$, 
\beq \label{Explain-7}
\max_{1\leq j\leq p}\biggl\{\frac{\|\hat{a}_j-a_j\|_1}{\|a_j\|_1}\biggr\} \leq C(\Delta_1 +\Delta_2+ Err_{VH}). 
\eeq
The proof of \eqref{Explain-7} is in Section~\ref{subsec:proof-lem-method}. Given the expressions of $\hat{A}$ and $A$ in \eqref{Explain-1}-\eqref{Explain-2}, the proof is easy to digest. We now briefly explain why $\Omega^*$ does not cause a trouble. Recall that after vertex hunting, our algorithm solves  $\hat{\pi}_j^*$ using $\hat{r}_j$ and the estimated vertices. When all the $\hat{r}_j$'s are rotated by $\Omega^*$, the estimated vertices are also rotated by the same matrix $\Omega^*$, so the solution $\hat{\pi}_j^*$ remains unchanged.  Mathematically, we have (recall that $\Omega=\diag(\omega,\Omega^*)$ and we have assumed $\omega=1$ without loss of generality)
\[
\hat{\Pi}^*=[{\bf 1}_p, \hat{R}]\Omega'\Omega(\hat{Q}')^{-1}=  [{\bf 1}_p, \hat{R}(\Omega^*)'] \cdot [(\Omega\hat{Q})']^{-1}, \quad\mbox{where}\quad\Omega\hat{Q} = \begin{pmatrix}
1 & \ldots & 1\\
\Omega^*\hat{v}_1^* & \ldots & \Omega^*\hat{v}_K^*
\end{pmatrix}.
\]
This shows that $\hat{\Pi}^*$ is indeed invariant of the rotation $\Omega^*$. After obtaining $\hat{\Pi}^*$, we no longer need the $\hat{r}_j$'s and $\hat{v}_k^*$'s in the remaining steps, so $\hat{A}$ is not affected by rotation. See the proof of Lemma~\ref{lem:method} for more details. 

Comparing \eqref{Explain-7} with the claims in Theorem~\ref{thm:UB}, the remaining work is to derive tight large-deviation bounds for $\Delta_1$, $\Delta_2$ and $Err_{VH}$. 
By Condition~\ref{cond:VH}, $Err_{VH}$ is controlled by $\max_{1\leq j\leq p}\|\Omega^*\hat{r}_j-r_j\|$. With a few lines of proofs (see \eqref{proof-UB-temp2}), we can show $\|\Omega^*\hat{r}_j-r_j\|\leq Ch_j^{-1/2}\|\Omega\hat{\Xi}_j-\Xi_j\|$. Hence, $Err_{VH}$ is controlled by $\Delta_2$. We only need to bound $\Delta_1$ and $\Delta_2$. The analysis of $\Delta_1$ is comparably easier. By definition, $M(j,j)=\frac{1}{n}\sum_{i=1}^n D(j,i)$. 
Fixing $j$, $D(j,i)$'s are independent Binomial random variables. A Binomial variable with $N$ trials is a sum of $N$ independent Bernoulli variables. Therefore, we write each $M(j,j)$ as a sum of $Nn$ independent Bernoulli variables and apply the Martingale Bernstein inequality. 
This is contained in Lemma~\ref{lem:M}. The analysis of $\Delta_2$ is much more sophisticated. It is related to the entry-wise eigenvector analysis we present in Section~\ref{subsec:theory-RMT} of the main paper, which proof is explained below. 

The entry-wise eigenvector analysis is contained in Section~\ref{sec:Eigen-perturbation}. Intuitively, by sin-theta theorem \citep{sin-theta}, if we view $\hat{\xi}_k$'s and $\xi_k$'s as the respective eigenvectors of two symmetric matrices $G$ and $G_0$ such that $G\approx G_0$, then we expect that $\hat{\xi}_k$'s are close to $\xi_k$'s. 
Since $\hat{\xi}_k$'s are singular vectors of $M^{-1/2}D$ and $\xi_k$'s are singular vectors of $M_0^{-1/2}D_0$, it seems natural to use
\[
\widetilde{G} = M^{-1/2}DD'M^{-1/2}, \qquad \widetilde{G}_0=M_0^{-1/2}D_0D_0'M^{-1/2}. 
\]
Unfortunately, $\widetilde{G}$ and $\widetilde{G}_0$ are not close enough. To see where the issue comes, let $Z=D-D_0$ denote the `noise' matrix. 
It is seen that 
\beq \label{Explain-8}
\widetilde{G} \approx M_0^{-1/2}DD'M_0^{-1/2} = \widetilde{G}_0 +  M_0^{-1/2}(D_0Z'+ZD_0')M_0^{-1/2} + M_0^{-1/2}ZZ'M_0^{-1/2}. 
\eeq
The last term, $M_0^{-1/2}ZZ'M_0^{-1/2}$, is a random matrix with nonzero mean. Its spectral norm is much larger than the desirable bound.  
To resolve this issue, we calculate the mean of this random matrix. Note that $Z_i=d_i - \mathbb{E}[d_i]$, where $Nd_i\sim \mathrm{Multinomial}(N, d_i^0)$. By properties of multinomial random vectors, $\mathbb{E}[Z_iZ_i'] = N^{-1}[\diag(d_i^0)-d_i^0(d_i^0)']$. It follows that
\begin{align*}
\mathbb{E}\bigl[ M_0^{-1/2}ZZ'M_0^{-1/2}\bigr]  & = M_0^{-1/2}\biggl\{ \sum_{i=1}^n  \frac{1}{N} [\diag(d_i^0)-d_i^0(d_i^0)'] \biggr\}M_0^{-1/2} \cr
&=  M_0^{-1/2}\biggl[ \frac{n}{N}\diag\Bigl( \frac{1}{n}D_0{\bf 1}_n\Bigr) - \frac{1}{N}D_0D_0'\biggr] M_0^{-1/2}\cr
&= M_0^{-1/2}\Bigl( \frac{n}{N}M_0 - \frac{1}{N}D_0D_0'\Bigr) M_0^{-1/2}\cr
&= \frac{n}{N}I_p - \frac{1}{N}\widetilde{G}_0.   
\end{align*}
Here, the first term is large in spectral norm. However, since it is proportional to an identity matrix, subtracting this matrix from $\widetilde{G}$ only changes eigenvalues but not eigenvectors! We will absorb it into $\widetilde{G}$. The second term is proportional to $\widetilde{G}_0$, hence, we will absorb it into $\widetilde{G}_0$, which does not change the eigenvectors either. 
We plug it into \eqref{Explain-8} to get
\begin{align*}
\widetilde{G} - \frac{n}{N}I_p  & \approx M_0^{-1/2}DD'M_0^{-1/2} - \frac{n}{N}I_p \cr
&= (1-\frac{1}{N})\widetilde{G}_0 + M_0^{-1/2}(D_0Z'+ZD_0'+ZZ-\mathbb{E}[ZZ'])M_0^{-1/2}. 
\end{align*} 
Now, the last term is a zero-mean random matrix, whose spectral norm can be controlled. This motivates us to define
\begin{align} \label{Explain-9}
G = \widetilde{G}-\frac{n}{N}I_p &:= M^{-1/2}DD'M^{-1/2}-\frac{n}{N}I_p, \cr
G_0 =(1-\frac{1}{N})\widetilde{G}_0 &:= (1-\frac{1}{N})M_0^{-1/2}D_0D_0'M_0^{-1/2}.   
\end{align}
It is easy to see that $\hat{\xi}_k$'s are indeed the eigenvectors of $G$, and $\xi_k$'s are indeed the eigenvectors of $G_0$. To obtain the entry-wise large-deviation bounds, we need a technical lemma, Lemma~\ref{lem:EigVecPerturb}, which implies that there exists an orthogonal matrix $\Omega$ such that, simultaneously for all $1\leq j\leq p$, 
\beq \label{Explain-10}
\|\Omega \hat{\Xi}_j - \Xi_j\|\leq \frac{C}{\|G_0\|} \bigl(\|G-G_0\|\|\Xi_j\| + \|(G-G_0)e_j\|\bigr).   
\eeq
Using this lemma, we reduce the study of entries of empirical eigenvectors to the study of the spectral norm and row-wise $\ell^2$-norms of the matrix $G-G_0$. 
Given \eqref{Explain-10}, it remains to derive a lower bound for $\|G_0\|$, an upper bound for $\|\Xi_j\|$, and  large-deviation upper bounds for $\|G-G_0\|$ and $\|(G-G_0)e_j\|$. These are given in Lemmas~\ref{lem:PopEigVal}-\ref{lem:Enorm}.

The proofs of Lemmas~\ref{lem:PopEigVal}-\ref{lem:Enorm} combine several techniques in probability. Here we give a brief explanation. Write $J=M^{-1/2}M_0^{1/2}$ and 
\[
B_1 = M_0^{-1/2}(D_0Z'+ZD_0')M_0^{-1/2}, \qquad B_2 = M_0^{-1/2}(ZZ'-\mathbb{E}[ZZ'])M_0^{-1/2}. 
\]
By \eqref{Explain-8}-\eqref{Explain-9}, $J^{-1}(G+\frac{n}{N}I_p)J^{-1}=M_0^{-1/2}DD'M_0^{-1/2}=G_0 + B_1+B_2$. It follows that
\[
G-G_0 = (JG_0J-G_0) + JB_1J+JB_2J + \frac{n}{N}(JJ'-I_p). 
\]
The analysis of $\Delta_1$ (see Lemma~\ref{lem:M}) already yields $J\approx I_p$. 
To bound the spectral norm and row-wise $\ell^2$ norms of $G-G_0$, the key is to study the two matrices $B_1$ and $B_2$. 
This is contained in Section~\ref{sec:Z-analysis}, where Lemma~\ref{lem:cross-term} is for the analysis of $B_1$, and Lemmas~\ref{lem:ZZ'-2}-\ref{lem:new-ZZ'} are for the analysis of $B_2$. Take the analysis of $\|B_2\|$ for example. 
Using the techniques of non-asymptotic random matrix analysis \citep{Vershynin}, we consider an $\alpha$-net $\mathcal{M}_\alpha$ on the unit sphere $\mathcal{S}^{p-1}$, satisfying $|{\cal M}_\alpha|\leq (1+2/\alpha)^p$. It is known that 
\[
\|B_2\| \leq (1 - 2 \alpha)^{-1} \sup_{u \in \mathcal{M}_\alpha} \{ |u'B_2 u| \}.
\]
It suffices to bound $|u'B_2u|$ for every $u\in {\cal M}_\alpha$. 
By definition, $u'B_2u = \sum_{i=1}^n (u'M_0^{-1/2}z_i)^2-\sum_{i=1}^n \mathbb{E}[(u'M_0^{-1/2}z_i)^2]$, where $z_i=d_i-d_i^0$ and $Nd_i\sim \mathrm{Multinomial}(N, d_i^0)$. In distribution, $\mathrm{Multinomial}(N, d_i^0)$ is  the sum of $N$ independent random vectors $T_{im}\sim  \mathrm{Multinomial}(1, d_i^0)$. Let $\widetilde{T}_{im}=T_{im}-\mathbb{E}[T_{im}]$. It follows that $z_i=N^{-1}\sum_{i=1}^n \widetilde{T}_{im}$. 
We then have 
\begin{align*}
u'B_2u &= \sum_{i=1}^n\biggl( \frac{1}{N} \sum_{m=1}^T u'M_0^{-1/2}\widetilde{T}_{im}\biggr)^2 - \sum_{i=1}^n\mathbb{E}\biggl(\frac{1}{N} \sum_{m=1}^T u'M_0^{-1/2}\widetilde{T}_{im}\biggr)^2 \cr
&=\frac{1}{N^2} \sum_{i=1}^n\sum_{m, s=1}^N u'M_0^{-1/2}\bigl(\widetilde{T}_{im}\widetilde{T}_{is}'- \mathbb{E}[\widetilde{T}_{im}\widetilde{T}_{is}']\bigr)M_0^{-1/2}u. 
\end{align*}
The random vectors $\{\widetilde{T}_{im}\}_{1\leq i\leq n,1\leq m\leq N}$ are independent and have zero means. Hence, $u'B_2u$ is a V-statistic. We bound it using Martingale large-deviation inequalities. The challenging case is  $N\ll p$, where $T_{im}$ has many zero entries and $u'M_0^{-1/2}\widetilde{T}_{im}$ have heavier tails than subGaussian variables. We tackle these challenges in the proof of Lemma~\ref{lem:new-ZZ'}. 

In summary, our analysis of Topic-SCORE can be divided into three major parts: 
\begin{itemize}
\item Part 1: Analysis of the oracle case, where we show that the output of the algorithm is exactly $A$. This part is contained in Section~\ref{sec:supp-oracle}. 
\item Part 2: Analysis of the real case, where we show that the estimation errors of $\hat{A}$ come from three sources: (i) noise in $M$, (ii) noise in $\hat{\xi}_1,\hat{\xi}_2,\ldots,\hat{\xi}_K$, and (iii) vertex hunting errors. We define  $\Delta_1$, $\Delta_2$ and $Err_{VH}$ to measure each noise source and express the estimation errors in terms of these quantities. This part is contained in Section~\ref{sec:UBproof}. 
\item Part 3: Derivation of the large-deviation bounds for $\Delta_1$, $\Delta_2$ and $Err_{VH}$. This part requires careful study of the noise matrix $Z$ and entry-wise eigenvector analysis, which are contained in Section~\ref{sec:Z-analysis} and Section~\ref{sec:Eigen-perturbation}, respectively. 
\end{itemize}

\section{Analysis of the oracle case} \label{sec:supp-oracle}
We first prove Lemmas~\ref{lem:simplicial-cone}-\ref{lem:Pi-to-A}. These lemmas give the rationale of the oracle Topic-SCORE algorithm. We then give a few lemmas about properties of the matrices $(M_0, V^*, R)$. These lemmas will be used frequently in the proofs of our main theorems.

\subsection{Proofs of Lemmas~\ref{lem:simplicial-cone}-\ref{lem:Pi-to-A}} \label{subsec:proof-Section2}
In Section~\ref{subsec:proof-lem-V}, we state and prove a useful lemma, Lemma~\ref{lem:V}. Using the first bullet point of that lemma,  there exists a unique non-singular matrix $V\in\mathbb{R}^{K,K}$ such that 
\beq \label{oracle-key-1}
\Xi=M_0^{-1/2}AV.
\eeq

First, we prove Lemma~\ref{lem:simplicial-cone}. Denote by $u_k'$ the $k$th row of $V$, for $1\leq k\leq K$. Recall that $x_j'$ and $a_j'$ denote the $j$th row of $\Xi$ and $A$, respectively. By \eqref{oracle-key-1},
\[
x_j = \sum_{k=1}^K \frac{a_j(k)}{\sqrt{M_0(j,j)}}u_k, \qquad 1\leq j\leq p. 
\]
By comparing it with the definition of simplicial cones, we immediately see that each $x_j$ is contained in the simplicial cone spanned by $u_1,u_2,\ldots,u_K$. Furthermore, if $j$ is an anchor word of topic $k$, then $a_j(k)\neq 0$ and $a_j(\ell)=0$ for all $\ell\neq j$. It follows that $x_j=\frac{a_j(k)}{\sqrt{M_0(j,j)}}u_k$. This means $x_j$ is located on the supporting ray defined by $u_k$. 

Next, we prove Lemma~\ref{lem:IdealSimplex}. In Section~\ref{subsec:proof-lem-R}, we state and prove a lemma, Lemma~\ref{lem:R}. Using the first bullet point of that lemma, $\xi_1$ is a strictly positive vector, so $R$ is well-defined. Let $1_p$ be the $p$-dimensional vector of $1$'s. By the definition of $R$, 
\beq \label{oracle-key-2}
[1_p, R] = [\mathrm{diag}(\xi_1)]^{-1}\Xi. 
\eeq 
Let $V$ be the same as in \eqref{oracle-key-1}. Write $V=[V_1, V_2, \ldots, V_K]$. Using the second bullet point of Lemma~\ref{lem:V}, $V_1$ is a strictly positive vector. Define a matrix $V^*\in \mathbb{R}^{K\times (K-1)}$ by
\[
V^*(\ell,k) = V_{k+1}(\ell)/V_1(\ell), \qquad 1\leq \ell\leq K, 1\leq k\leq K-1. 
\]
Let $1_K$ be the $K$-dimensional vector of $1$'s. The above definition implies 
\beq \label{oracle-key-3}
V=\mathrm{diag}(V_1)\cdot [1_K, V^*]. 
\eeq
We plug \eqref{oracle-key-1} into \eqref{oracle-key-2}, and then use the expression of $V$ in \eqref{oracle-key-3}. It follows that 
\[
[1_p, R] = [\mathrm{diag}(\xi_1)]^{-1}M_0^{-1/2}A \cdot \mathrm{diag}(V_1)\cdot [1_K, V^*]. 
\]
The above equality can be equivalently written as
\beq \label{oracle-key-4}
1_p = \Pi\cdot 1_K, \qquad R = \Pi \cdot V^*, \qquad \mbox{with}\quad \Pi=[\mathrm{diag}(\xi_1)]^{-1}M_0^{-1/2}A \cdot \mathrm{diag}(V_1). 
\eeq
Write $\Pi=[\pi_1,\pi_2,\ldots,\pi_p]'$ and $V^*=[v_1^*, v_2^*,\ldots,v_K^*]'$. It follows from \eqref{oracle-key-4} that
\[
1 = \sum_{k=1}^K \pi_j(k), \qquad r_j = \sum_{k=1}^K \pi_j(k)v_k^*, \qquad\mbox{for all }1\leq j\leq p. 
\]
Note that $\xi_1$ and $V_1$ are strictly positive vectors, and $M_0$ is a diagonal matrix with positive diagonals. Then, $\Pi$ must be a nonnegative matrix. Therefore, the above implies that each $r_j$ is a convex combination of $v_1^*,v_2^*,\ldots,v_K^*$. This proves that the point cloud $r_1,r_2,\ldots,r_p$ are contained in a simplex ${\cal S}_K^*$, whose vertices are $v_1^*,v_2^*,\ldots,v_K^*$. Furthermore, by definition of $\Pi$ in \eqref{oracle-key-4},  
\[
\pi_j(k) = \frac{V_1(k)}{\xi_1(j)\sqrt{M_0(j,j)}}\cdot a_j(k), \qquad 1\leq k\leq K. 
\] 
Therefore, $\pi_j(k)\neq 0$ if and only if $a_j(k)\neq 0$. It follows that, for an anchor word $j$ of topic $k$, $\pi_j(\ell)=0$ for all $\ell\neq k$. Then, $\pi_j$ can only equal to $e_k$, the $k$th standard basis of $\mathbb{R}^K$. It implies that $r_j=v_k^*$, i.e., $r_j$ is located exactly on the vertex $v_k^*$. 

Last, we prove Lemma~\ref{lem:Pi-to-A}. It suffices to check the uniqueness of the convex combination coefficient vector $\pi_j$ for each $1\leq j\leq p$. Then, the claim of this lemma follows immediately from the definition of $\Pi$ in \eqref{oracle-key-4}. We now show the uniqueness of $\pi_j$. Note that $\pi_j$ is the solution of 
\[
\begin{pmatrix}
1 & \cdots & 1\\
v_1^* & \cdots & v_K^* 
\end{pmatrix}\pi_j = \begin{pmatrix}1 \\ r_j\end{pmatrix}. 
\]
The solution is unique if and only if the $K\times K$ matrix $[1_K, V^*]$ is non-singular. By \eqref{oracle-key-3}, this matrix is equal to $[\diag(V_1)]^{-1}V$. Since $V$ is non-singular and $V_1$ is a strictly positive vector, the matrix $[\diag(V_1)]^{-1}V$ is non-singular. 
\qed

\subsection{A useful lemma about $M_0$}

Although the oracle Topic-SCORE works for an arbitrary positive diagonal matrix $M_0$, one specific choice of interest is 
\[
M_0 = \diag(n^{-1}D_0 1_n). 
\]
The next lemma gives its properties (recall that $h_j=\|a_j\|_1$, where $a_j'$ is the $j$th row of $A$).  
\begin{lem} \label{lem:M0}
Consider $D_0=AW$ and $M_0=\diag(n^{-1}D_0 1_n)$, where the regularity condition \eqref{cond-A} holds. Then, 
\[
c_2h_j\leq M_0(j,j)\leq h_j,\qquad\mbox{for all } 1\leq j\leq p.
\] 
Here, $c_2$ is the same constant as in \eqref{cond-A}.
\end{lem}

\bigskip
\noindent 
{\it Proof of Lemma~\ref{lem:M0}}: Recall that $\Sigma_W=n^{-1}WW'$. By \eqref{cond-A}, $\lambda_{\min}(\Sigma_W)\geq c_2$. We write 
\[
M_0(j,j)=\frac{1}{n}\sum_{i=1}^n \Bigl[\sum_{k=1}^K A_k(j)w_i(k)\Bigr]=\sum_{k=1}^K A_k(j)\Bigl[\frac{1}{n}\sum_{i=1}^n w_i(k)\Bigr].
\]
Since $w_i(k)\leq 1$, we have $M_0(j,j)\leq \sum_{k=1}^K A_k(j)=h_j$. At the same time, $\frac{1}{n}\sum_{i=1}^n w_i(k)\geq \frac{1}{n}\sum_{i=1}^nw^2_i(k)=\Sigma_W(k,k)\geq \lambda_{\min}(\Sigma_W)$; consequently, $M_0(j,j)\geq c_2\sum_{k=1}^K A_k(j)=c_2h_j$. \qed

\subsection{A useful lemma about $V$ and $V^*$} \label{subsec:proof-lem-V}

In Section~\ref{subsec:proof-Section2}, we have defined a matrix $V$ through $\Xi=AV$ (if it exists). We have also defined $V^*$ by $V^*(\ell,k)=V_{k+1}(\ell)/V_1(\ell)$, for $1\leq\ell\leq K$, $1\leq k\leq K-1$ (if it exists). Write $V^*=[v_1^*,\ldots,v_K^*]'$. The next lemma confirms that these two matrices are well-defined and have some nice properties. 

We must note that $\Xi$ and $V$ are not uniquely defined. They are up to the sign flips and rotations of eigenvectors. The following lemma applies to any eligible choice of $\Xi$:

\begin{lem} \label{lem:V}
Consider $D_0=AW$ and an arbitrary positive diagonal matrix $M_0$. Suppose the regularity conditions \eqref{cond-h}-\eqref{cond-A} hold. The following statements are true:
\begin{itemize}
\item For any eligible choice of $\Xi$, there exists a unique non-singular matrix $V\in\mathbb{R}^{K,K}$ such that $\Xi=M_0^{-1/2}AV$; moreover, $(VV')^{-1}=A'M_0^{-1}A$. 
\item All entries of $V_1$ have the same sign. 
\item $\mathcal{S}_K^*=\mathcal{S}(v_1^*,\ldots,v_K^*)$ is a non-degenerate simplex.
\end{itemize}
Furthermore, if $M_0=\diag(n^{-1}D_01_n)$, then the following statements are true: 
\begin{itemize}
\item $C_1^{-1}\leq |V_1(k)|\leq C_1$ for all $1\leq k\leq K$. 
\item The volume of ${\cal S}^*_K$ is lower bounded by $C_2^{-1}$ and upper bounded by $C_2$.
\item $\max_{1\leq k\leq K} \|v_k^*\|\leq C_3$. 
\item $C_4^{-1}\leq \|v^*_k-v^*_\ell\|\leq C_4$ for all $1\leq k\neq \ell \leq K$.
\end{itemize}
Here, $C_1$-$C_4$ are positive constants satisfying that $C_1,C_2,C_4>1$. 
\end{lem}

\bigskip
\noindent
{\it Proof of Lemma~\ref{lem:V}}: Consider the first claim. Note that $M_0^{-1/2}D_0$ has a full column rank $K$. We write the SVD of $M_0^{-1/2}D_0$ by 
\[
M_0^{-1/2}D_0=\Xi\Lambda B', 
\]
where $\Lambda =\mathrm{diag}(\lambda_1,\ldots, \lambda_K)$ contains the singular values and $B\in\mathbb{R}^{n,K}$ contains the right singular vectors; note that $\Xi'\Xi=B'B=I_K$. It is seen that
\[
\Xi = (\Xi \Lambda B')B \Lambda^{-1} = M_0^{-1/2}D_0B \Lambda^{-1} =  M_0^{-1/2}A(WB\Lambda^{-1}).  
\]
By letting $V=WB\Lambda^{-1}$, we have $\Xi=AV$; i.e., such a $V$ exists. Furthermore, for any $V$ such that $\Xi=M_0^{-1/2}AV$, we have $\Xi'M_0^{-1/2}AV=\Xi'\Xi=I_K$. This implies that $V$ is the inverse of $(\Xi'M_0^{-1/2}A)$, so $V$ is unique and non-singular. Last, we plug $\Xi=M_0^{-1/2}AV$ into $\Xi'\Xi=I_K$; it yields $I_K = V'A'M_0^{-1}AV$. Multiplying both sides of this equation by $V$ from the left and by $V'$ from the right, we obtain:
\[
VV' = (VV')A'M_0^{-1}A(VV').    
\] 
This proves that $VV'=(A'M_0^{-1}A)^{-1}$.

Consider the second claim. 
Let $\lambda_1,\ldots,\lambda_K$ be the singular values of $M_0^{-1/2}D_0$. Then, 
\[
M_0^{-1/2}D_0D_0'M_0^{-1/2}\xi_k=\lambda^2_k\xi_k,
\]
where $D_0=AW$ and $\xi_k=M_0^{-1/2}AV_k$. Combining these facts gives 
\[
(M_0^{-1/2}AWW'A'M_0^{-1/2})(M_0^{-1/2}AV_k) = \lambda^2_k(M_0^{-1/2}AV_k).
\] 
Multiplying both sides by $(A'M_0^{-1}A)^{-1}A'M_0^{-1/2}$ from the left, we have
\[
(WW'A'M_0^{-1}A)V_k = \lambda^2_k V_k. 
\]
Recall that $\Sigma_W=n^{-1}WW'$. We immediately have 
\beq \label{lem-Vprop-temp}
\bigl[\Sigma_W(A'M_0^{-1}A)\bigr] V_k = (n^{-1}\lambda^2_k) V_k. 
\eeq
Therefore, for each $1\leq k\leq K$,  $V_k$ is a right eigenvector of $\Sigma_W(A'M_0^{-1}A)$ associated with the eigenvalue $n^{-1}\lambda^2_k$ (these eigenvectors are not necessarily orthogonal with each other). 

By Perron's theorem \cite{HornJohnson}, the leading eigenvector of a strictly positive matrix must be a strictly positive vector. Therefore, to show that $V_1$ is a strictly positive vector, it suffices to show that $\Theta \equiv \Sigma_W(A'M_0^{-1}A)$ is a strictly positive matrix. We note that $\sum_{s=1}^K\Sigma_W(k,s)\geq \Sigma_W(k,k)\geq \lambda_{\min}(\Sigma_{W})\geq c_2$. It follows that 
\begin{align*}
\Theta(k,\ell) & = \sum_{s=1}^K\Sigma_W(k,s)\cdot (A'M_0^{-1}A)(s, \ell)\cr
&\geq \min_{s,t}\bigl\{(A'M_0^{-1}A)(s, \ell)\bigr\}\cdot \sum_{s=1}^K \Sigma_W (k,s)\cr
&\geq c_2\cdot \min_{s,t}\bigl\{(A'M_0^{-1}A)(s, \ell)\bigr\}.  
\end{align*}
It suffices to show that $A'M_0^{-1}A$ is a strictly positive matrix. Write $M_{0,\max}=\max_{1\leq j\leq p}M_0(j,j)$. Recall that by the condition \eqref{cond-A}, $\Sigma_A=A'H^{-1}A$ is a strictly positive matrix. We have
\begin{align*}
\Sigma_A(k,\ell) & =\sum_{j=1}^p h_j^{-1}a_j(k)a_j(\ell) \leq h_{\min}^{-1}\sum_{j=1}^p a_j(k)a_j(\ell) \cr
&\leq h_{\min}^{-1}M_{0,\max} \sum_{j=1}^p [M_0(j,j)]^{-1}a_j(k)a_j(\ell) = h_{\min}^{-1}M_{0,\max}  (A'M_0^{-1}A)(k,\ell). 
\end{align*}
Therefore, $A'M_0^{-1}A$ must be a strictly positive matrix. The second claim follows.

Consider the third claim. The simplex ${\cal S}_K^*$ is not degenerate if and only if $v_1^*,v_2^*,\ldots,v_K^*$ are affinely independent, which holds if and only if the following matrix is non-singular: 
\beq \label{mat-Q}
Q\equiv \begin{pmatrix}
1  & \ldots & 1\\
v_1^* & \ldots & v_K^*
\end{pmatrix}. 
\eeq
By \eqref{oracle-key-3}, $Q'=[\diag(V_1)]^{-1}V$. Since $V$ is non-singular and $V_1$ is a strictly positive vector, we know that $Q$ is non-singular. This implies that ${\cal S}_K^*$ is a non-degenerate simplex.

The above claims hold for an arbitrary choice of $M_0$. The remaining four claims are for the particular choice of $M_0=\diag(n^{-1}D_0 1_n)$. 

Consider the fourth claim. We first show that 
\beq   \label{lem-Vprop-1}
|V_1(k)|\leq C, \qquad \mbox{for }1\leq k\leq K. 
\eeq 
By Lemma~\ref{lem:M0}, $c_2 h_j\leq M_0(j,j)\leq h_j$, for every $j$. Then, $A'(M_0^{-1}-H^{-1})A$ is a positive semi-definite matrix. It follows that $\lambda_{\min}(A'M^{-1}_0A)\geq \lambda_{\min}(A'H^{-1}A)$. 
Similarly, $A'(c_2^{-1}H^{-1}-M_0^{-1})A$ is a positive semi-definite matrix, and we get $\lambda_{\max}(A'M_0^{-1}A)\leq c_2^{-1}\lambda_{\max}(A'H^{-1}A)$. Note that $A'H^{-1}A=\Sigma_A$. The condition \eqref{cond-A} gives $\lambda_{\min}(\Sigma_A)\geq c_2$; also, using the fact that the column sums of $A$ are all equal to $1$, we have $\lambda_{\max}(\Sigma_A)\leq \|\Sigma_A\|_1=1$. Combining the above gives 
\beq \label{AM0A}
c_2\leq \lambda_{\min}(A'M_0^{-1}A)\leq \lambda_{\max}(A'M_0^{-1}A)\leq c_2^{-1}. 
\eeq
In the first claim, we have seen that $VV'=(A'M_0^{-1}A)^{-1}$. So, \eqref{AM0A} yields:
\beq \label{VV'}
c_2 \leq \lambda_{\min}(VV')\leq \lambda_{\max}(VV')\leq c_2^{-1}. 
\eeq
Observing that $\sum_{\ell=1}^K V_\ell^2(k)$ is the $k$-th diagonal of $VV'$, we obtain \eqref{lem-Vprop-1}. 

Next, we show that for a constant $c>0$, up to a multiple of $\pm 1$ on $V_1$, 
\beq \label{lem-Vprop-2}
V_1(k)\geq c, \qquad\mbox{for }1\leq k\leq K. 
\eeq
Since $\|V_1\|^2$ is the first diagonal of $V'V$, we have $\|V_1\|^2\geq \lambda_{\min}(V'V)=\lambda_{\min}(VV')\geq c_2$, where the last inequality is due to \eqref{VV'}. Therefore, to show \eqref{lem-Vprop-2}, it suffices to show that
\beq \label{lem-Vprop-3}
\liminf_{n\to\infty} \min_{1\leq k\leq K} \{\eta_1(k)\}\geq c, \qquad\mbox{with}\quad \eta_1=\mathrm{sign}(V_1(1))\cdot \|V_1\|^{-1}V_1. 
\eeq
By \eqref{lem-Vprop-temp}, $V_1$ is the leading right singular vector of $\Theta=\Sigma_W(A'M_0^{-1}A)$, i.e., 
\[
\mbox{$\eta_1$ is the unit-norm leading eigenvector of $\Theta=\Sigma_W(A'M_0^{-1}A)$}. 
\]
Write $\eta_1=\eta_1^{(n)}$ to indicate its dependence on $n$; similar for other quantities. Suppose \eqref{lem-Vprop-3} is not true. Then, there exists $k$ and a subsequence $\{n_m\}_{m=1}^\infty$ such that $\lim_{m\to\infty}\eta_1^{(n_m)}(k)= 0$. Furthermore, the spectral norm of $\Sigma_W$ is bounded (because each column of $W$ has a unit $\ell^1$-norm), and the spectral norm of $A'M_0^{-1}A$ is also bounded (by \eqref{AM0A}). Therefore, there exists a subsequence of $\{n_m\}_{m=1}^\infty$ such that $\Theta$ tends to a fixed matrix $\Theta_0$; without loss of generality, we assume this subsequence is $\{n_m\}_{m=1}^\infty$ itself. The above implies 
\[
\lim_{m\to\infty} \eta^{(n_m)}_1(k)=0, \qquad \lim_{m\to\infty}\Theta^{(n_m)} = \Theta_0. 
\]
In the proof of Lemma~\ref{lem:PopEigVal}, we show that the eigengap of $\Theta$ is bounded below by a positive constant; see \eqref{lem-eigenval-0}. Using the sine-theta theorem \citep{sin-theta}, when $\Theta^{(n_m)} \to\Theta_0$, up to a multiple of $\pm 1$ on $\eta_1^{(n_m)}$, 
\[
\eta^{(n_m)}_1 \to q_0, \qquad \mbox{$q_0$ is the unit-norm leading eigenvector of $\Theta_0$}. 
\]
Combining the above gives
\beq \label{lem-Vprop-7}
q_0(k)=0. 
\eeq
We then study $\Theta_0$. Write $\Theta=\Theta_1+\Theta_2$, where $\Theta_1=\Sigma_W(A'H^{-1}A)$ and $\Theta_2=\Sigma_WA'(M_0^{-1}-H^{-1})A$. By Lemma~\ref{lem:M0}, $M_0(j,j)\leq h_j$, so all the entries of $\Theta_2$ are non-negative. Moreover, the assumption \eqref{cond-A} yields that all entries of $\Sigma_A=A'H^{-1}A$ are lower bounded by a constant $c_2>0$; as a result, all entries of $\Theta_1$ are lower bounded by a positive constant. Combining the above, all entries of $\Theta$ are lower bounded by a positive constant, which implies:
\beq \label{lem-Vprop-8}
\mbox{$\Theta_0$ is a strictly positive matrix}. 
\eeq 
By Perron's theorem \citep{HornJohnson}, the leading unit-norm eigenvector (up to $\pm 1$) of a positive matrix has all positive entries. So \eqref{lem-Vprop-7} and \eqref{lem-Vprop-8} are contradicting with each other. This proves \eqref{lem-Vprop-3}; then, \eqref{lem-Vprop-2} follows. The fourth claim follows by combining \eqref{lem-Vprop-1} and \eqref{lem-Vprop-2}. 

Consider the fifth claim. Let $Q$ be the same as in \eqref{mat-Q}. The volume of ${\cal S}_K^*$ is equal to 
\[
\frac{1}{(K-1)!}\det([v_2^*-v^*_1,\ldots,v^*_K- v^*_1]) = \frac{1}{(K-1)!}\det(Q). 
\]
We have seen $Q'=[\mathrm{diag}(V_1)]^{-1}\cdot V$. It follows that 
\[
Q'Q = [\mathrm{diag}(V_1)]^{-1}VV'[\mathrm{diag}(V_1)]^{-1}. 
\] 
We plug in \eqref{lem-Vprop-1}, \eqref{lem-Vprop-2} and \eqref{VV'} to get
\beq \label{Vprop-8}
C^{-1}\leq \lambda_{\min}(Q'Q)\leq \lambda_{\max}(Q'Q)\leq C. 
\eeq
Therefore, all singular values of $Q$ are upper/lower bounded by constants. It follows that $\det(Q)$ is upper/lower bounded by constants, so is the volume of ${\cal S}_K^*$. 

Consider the sixth and seventh claims. Note that
\[
\begin{pmatrix} 1\\ v_k^* \end{pmatrix}
 = Qe_k,  \qquad \mbox{$e_k$: the $k$-th standard basis of $\mathbb{R}^K$}. 
\]
Therefore, $\|v_k^*\|\leq \|Q\|\leq C$, $\|v_k^*-v^*_\ell\|\leq \|Q\|\cdot \|e_k-e_\ell\| \leq \sqrt{2}\|Q\|\leq C$, and $\|v_k^*-v^*_\ell\|^2\geq \|e_k-e_\ell\|^2\cdot \lambda_{\min}(Q'Q)\geq C^{-1}$. The last two claims follow immediately. 
\qed

\subsection{A useful lemma about $R$} \label{subsec:proof-lem-R}

We present a lemma about the matrix $R$. For $1\leq j\leq p$, we recall that $a_j'$ denotes the $j$-th row of $A$, and $\tilde{a}_j=h_j^{-1}a_j$, where $h_j=\|a_j\|_1$. Write $R=[r_1,\ldots,r_p]'$. 

\begin{lem} \label{lem:R}
Consider $D_0=AW$ and $M_0=\diag(n^{-1}D_0 1_n)$, where the regularity condition \eqref{cond-A} holds. The following statements are true:
\begin{itemize}
\item We can choose the sign of $\xi_1$ such that all the entries are positive and that $C_5^{-1}\sqrt{h_j}\leq \xi_1(j)\leq C_5\sqrt{h_j}$ for all $1\leq j\leq p$. 
\item $\max_{1\leq j\leq p}\|r_j\|\leq C_6$. 
\item $C_7^{-1}\|\tilde{a}_i-\tilde{a}_j\|\leq \|r_i-r_j\|\leq C_7\|\tilde{a}_i-\tilde{a}_j\|$, for all $1\leq i,j\leq p$. 
\end{itemize}
Here, $C_5$-$C_7$ are positive constants satisfying that $C_5,C_7>1$. 
\end{lem}

\bigskip
\noindent
{\it Proof of Lemma~\ref{lem:R}}: Consider the first claim. From $\Xi=M_0^{-1/2}AV$, we have $\xi_1(j)=[M_0(j,j)]^{-1/2}a_j'V_1$ for $1\leq j\leq p$. Note that $a_j$ is a non-negative vector with $\|a_j\|_1\neq 0$ and that all entries of $V_1$ are either all positive or all negative; so the entries of $a_j'V_1$ all have the same sign. Consequently, the entries of $\xi_1$ also have the same sign; this means we can choose the sign of $\xi_1$ so that all the entries are positive. 

Assuming all entries of $\xi_1$ and $V_1$ are positive, we now give lower/upper bound of $\xi_1(j)$, for $1\leq j\leq p$. Since $\xi_1(j)=[M_0(j,j)]^{-1/2}a_j'V_1$, 
\[
\xi_1(j)\geq [M_0(j,j)]^{-1/2} \|a_j\|_1 \min_{1\leq k\leq K}V_1(k). 
\]
By definition, $\|a_j\|_1=h_j$. By Lemma~\eqref{lem:M0}, $M_0(j,j)\leq h_j$. 
By Lemma~\ref{lem:V}, $V_1(k)\geq C^{-1}$ for all $1\leq k\leq K$. Combining the above gives
\[
\xi_1(j)\geq C^{-1}\sqrt{h_j}. 
\]
Similarly, we can prove that $\xi_1(j)\leq C\sqrt{h_j}$.

Consider the second claim. Since each $r_j$ is in the simplex $\mathcal{S}_K^*$, it follows that $\|r_j\|\leq \max_{1\leq k\leq K}\|v_k^*\|$; by Lemma~\ref{lem:V}, $\max_{1\leq k\leq K}\|v_k^*\|\leq C$. The claim then follows. 

Consider the third claim. By Lemma~\ref{lem:IdealSimplex}, each $r_j$ is a convex combination of $v_1^*,\ldots, v_K^*$, where the weight vector $\pi_j$ is the $j$-th row of $\Pi=[\mathrm{\diag}(\xi_1)]^{-1}\cdot M_0^{-1/2}A\cdot\mathrm{diag}(V_1)$. So
\[
\begin{pmatrix}0\\ r_i - r_j\end{pmatrix} = Q(\pi_i-\pi_j), \qquad \mbox{where } Q = \begin{pmatrix}
1  & \ldots & 1\\
v_1^* & \ldots & v_K^*
\end{pmatrix}. 
\]
In \eqref{Vprop-8}, we have seen that $C^{-1}\leq \lambda_{\min}(Q'Q)\leq \lambda_{\max}(Q'Q)\leq C$. So,
\[
C^{-1}\|\pi_i-\pi_j\|\leq \|r_i-r_j\|\leq C\|\pi_i-\pi_j\|. 
\] 
To show the claim, it suffices to prove that 
\beq \label{Rprop-2}
C^{-1}\|\tilde{a}_i - \tilde{a}_j\| \leq \|\pi_i-\pi_j\|\leq C\|\tilde{a}_i - \tilde{a}_j\|. 
\eeq

We now show \eqref{Rprop-2}. We assume the sign of $\xi_1$ is chosen such that all entries of $\xi_1$ and $V_1$ are positive. Since $\Pi=[\mathrm{\diag}(\xi_1)]^{-1}\cdot M_0^{-1/2}A\cdot\mathrm{diag}(V_1)$, 
\begin{align} \label{Rprop-3}
\pi_j & = [\xi_1(j)]^{-1}[M_0(j,j)]^{-1/2}\cdot \mathrm{diag}(V_1) a_j\cr
& = [\xi_1(j)]^{-1}[M_0(j,j)]^{-1/2}h_j\cdot \mathrm{diag}(V_1) \tilde{a}_j\cr
& \propto (V_1\circ \tilde{a}_j), 
\end{align}
where $\circ$ denotes the entry-wise product of two vectors. 
Noting that both $\pi_j$ and $\tilde{a}_j$ are weight vectors, we have $\pi_j=(V_1\circ \tilde{a}_j)/\|V_1\circ\tilde{a}_j\|_1$. Therefore, 
\[
\pi_i - \pi_j = \frac{(V_1\circ \tilde{a}_i)}{\|V_1\circ \tilde{a}_i\|_1} - \frac{(V_1\circ \tilde{a}_j)}{\|V_1\circ \tilde{a}_j\|_1}=\frac{V_1\circ (\tilde{a}_i-\tilde{a}_j)}{\|V_1\circ \tilde{a}_i\|_1} + \frac{\|V_1\circ \tilde{a}_j\|_1-\|V_1\circ \tilde{a}_i\|_1}{\|V_1\circ \tilde{a}_i\|_1}\pi_j. 
\]
By the triangle inequality, $|\|V_1\circ \tilde{a}_j\|_1-\|V_1\circ \tilde{a}_i\|_1|\leq \|(V_1\circ\tilde{a}_j)-(V_1\circ \tilde{a}_i)\|_1=\|V_1\circ (\tilde{a}_i-\tilde{a}_j)\|_1$. Moreover, $\|\pi_j\|_1=1$. It follows that
\[
\|\pi_i-\pi_j\|_1 \leq 2\frac{\|V_1\circ (\tilde{a}_i-\tilde{a}_j)\|_1}{\|V_1\circ \tilde{a}_i\|_1}. 
\]
By Lemma~\ref{lem:V}, $C^{-1}\leq V_1(k)\leq C$ for all $k$. So $
\|V_1\circ (\tilde{a}_i-\tilde{a}_j)\|_1\leq C\|\tilde{a}_i-\tilde{a}_j\|_1$, and $ \|V_1\circ \tilde{a}_i\|_1\geq C^{-1}$. 
It follows that 
\[  
\|\pi_i-\pi_j\|_1\leq C\|\tilde{a}_i-\tilde{a}_j\|_1. 
\]
Using the Cauchy-Schwarz inequality, $\|\tilde{a}_i-\tilde{a}_j\|_1\leq \sqrt{K}\|\tilde{a}_i-\tilde{a}_j\|$. Moreover, since $\|\pi_i-\pi_j\|_\infty\leq 1$, we have $\|\pi_i-\pi_j\|\leq \|\pi_i-\pi_j\|_1$. It follows that
\beq \label{Rprop-4}
\|\pi_i-\pi_j\|\leq C\|\tilde{a}_i-\tilde{a}_j\|. 
\eeq
This gives the second inequality in \eqref{Rprop-2}. 

To get the first inequality in \eqref{Rprop-2}, introduce a vector $b\in\mathbb{R}^K$ with $b(k)=1/V_1(k)$. Then \eqref{Rprop-3} implies $\tilde{a}_j\propto(b\circ \pi_j)$ for all $1\leq j\leq p$. Since both $\tilde{a}_j$ and $\pi_j$ are weight vectors, we have $\tilde{a}_j=\frac{b\circ\pi_j}{\|b\circ\pi_j\|_1}$. Note that $C^{-1}\leq \min_kV_1(k)\leq \max_k V_1(k)\leq C$ implies $C^{-1}\leq \min_k b(k)\leq\max_k b(k)\leq C$. By replacing $V_1$ with $b$ in the proof of \eqref{Rprop-4}, we immediately obtain
\[
\|\tilde{a}_i -\tilde{a}_j\| \leq C\|\pi_i - \pi_j\|. 
\]
This gives the second inequality in \eqref{Rprop-2}. \qed

\section{Properties of the noise matrix $Z=D-D_0$} \label{sec:Z-analysis}

Write $Z=[z_1,z_2,\ldots,z_n]=[Z_1,Z_2,\ldots,Z_p]'$. We state a few lemmas about this matrix.

First, let $M=\diag(n^{-1}D1_n)$ and $M_0=\diag(n^{-1}D_0 1_n)$. The next lemma characterizes the diagonal matrix $M-M_0=n^{-1}\mathrm{diag}(Z1_n)$. 
\begin{lem}\label{lem:M}
Consider model \eqref{pLSI}, where $K$ is fixed, $N_i= N$, and the condition \eqref{cond-A} holds. As $n\to\infty$, suppose $Nnh_{\min}/\log(n)\to\infty$. With probability $1-o(n^{-3})$, 
\[
|M(j,j) - M_0(j,j)| \leq C(Nn)^{-1/2}\sqrt{h_j\log(n)}, \qquad \mbox{for all $1\leq j\leq p$}. 
\]
\end{lem}

Second, we give a lemma about the $p$-dimensional vector $M_0^{-1/2}ZW_k$, where $W'_k$ denotes the $k$-th row of $W$, for $1\leq k\leq K$. 
\begin{lem}\label{lem:cross-term}
Consider model \eqref{pLSI}, where $K$ is fixed, $N_i= N$, and the condition \eqref{cond-A} holds.  As $n\to\infty$, suppose $Nnh_{\min}/\log(n)\to\infty$. With probability $1-o(n^{-3})$, for all $1\leq k\leq K$,
\begin{align*}
& |Z_j'W_k|\leq CN^{-1/2}\sqrt{nh_j\log(n)}, \qquad \mbox{for all $1\leq j\leq p$}, \cr
&\|M_0^{-1/2}ZW_k\| \leq CN^{-1/2}\sqrt{np\log(n)}. 
\end{align*}
\end{lem}

Next, we give two lemmas that characterize the entries of the matrix $ZZ'$.  Lemma~\ref{lem:ZZ'-2} is for the general case, and Lemma~\ref{lem:new-ZZ'-2} improves the bound in Lemma~\ref{lem:ZZ'-2} when $n$ satisfies an additional requirement.  

\begin{lem}\label{lem:ZZ'-2}
Consider model \eqref{pLSI}, where $K$ is fixed, $N_i= N$, and \eqref{cond-A} is satisfied. As $n\to\infty$, suppose $\log(n)=O(\min\{N,p\})$. With probability $1-o(n^{-3})$, for all $1\leq j,\ell\leq p$, 
\[
|Z_j'Z_\ell - E[Z_j'Z_\ell]|\leq C\biggl(\frac{1}{N}+\frac{\log(n)}{N^2h_{\min}}\biggr)\sqrt{nh_jh_\ell\log(n)}. 
\]
\end{lem}

\begin{lem} \label{lem:new-ZZ'-2}
Under the assumptions of Lemma~\ref{lem:ZZ'-2}, if additionally $n\geq\frac{p}{h^2_{\min}}(1+\frac{p^2}{N^2}+Nh_{\min})$, then with probability $1-o(n^{-3})$, simultaneously for all $1\leq j,\ell\leq p$, 
\[
|Z_j'Z_\ell - E[Z_j'Z_\ell]|\leq C\Bigl(\frac{1}{N}+\frac{1}{N\sqrt{Nh_{\min}}}  \Bigr)\sqrt{nh_jh_\ell\log(n)}. 
\]
\end{lem}

Last, we derive large-deviation bounds for the matrix
\[
M_0^{-1/2}(ZZ'-E[ZZ'])M_0^{-1/2}. 
\]
Below, Lemma~\ref{lem:ZZ'} is for the general case, and Lemma~\ref{lem:new-ZZ'} improves the bound in Lemma~\ref{lem:ZZ'} when $n$ satisfies an additional requirement.

\begin{lem}\label{lem:ZZ'}
Consider model \eqref{pLSI}, where $K$ is fixed, $N_i= N$, and \eqref{cond-A} is satisfied. As $n\to\infty$, suppose $\log(n+N)=O(\min\{N,p\})$ and $p=O(n)$. With probability $1-o(n^{-3})$, 
\[
\|M_0^{-1/2}(ZZ'-E[ZZ'])M_0^{-1/2}\|\leq C\Bigl(\frac{1}{N}+\frac{p}{N^2h_{\min}}\Bigr)\sqrt{np}.  
\] 
\end{lem}

\begin{lem} \label{lem:new-ZZ'}
Under the assumptions of Lemma~\ref{lem:ZZ'}, if additionally $n\geq\frac{p}{h^2_{\min}}(1+\frac{p^2}{N^2}+Nh_{\min})$, then with probability $1-o(n^{-3})$, 
\[
\|M_0^{-1/2}(ZZ'-E[ZZ'])M_0^{-1/2}\|\leq C\frac{\sqrt{np}}{N}\Bigl(1 + \frac{1}{\sqrt{Nh_{\min}}}\Bigr). 
\]
\end{lem}

The above lemmas are proved in Sections~\ref{subsec:proof-lem-M}-\ref{subsec:proof-lem-ZZ'-new} below. The proofs of Lemmas~\ref{lem:ZZ'}-\ref{lem:new-ZZ'} are especially sophisticated, where we combine non-asymptotic random matrix theory with martingale tail inequalities.

\subsection{Proof of Lemma~\ref{lem:M}} \label{subsec:proof-lem-M}
Recall that $Z=D-D_0=[z_1,z_2,\ldots,z_n]$. 
Introduce a set of $p$-dimensional random vectors $\{T_{im}:1\leq i\leq n, 1\leq m\leq N\}$ such that they are independent of each other and that $T_{im}\sim \mathrm{Multinomial}(1,d_i^0)$. From the definition of multinomial distributions, 
  \beq \label{prop-binomial-0}
    z_i\overset{(d)}{=}\frac{1}{N}\sum_{m=1}^{N}(T_{im}-E[T_{im}]), \qquad 1\leq i\leq n.
  \eeq
It follows that
\[
M(j,j)- M_0(j,j)=\frac{1}{n}\sum_{i=1}^n z_i(j) \overset{(d)}{=}\frac{1}{Nn}\sum_{i=1}^n\sum_{m=1}^N \{T_{im}(j)-E[T_{im}(j)] \}.
\]
Fix $j$ and write $X_{im}=T_{im}(j)-E[T_{im}(j)]$. Then, $\{X_{im}:1\leq i\leq n,1\leq m\leq N\}$ are independent of each other. Moreover, since $T_{im}(j)\sim \mathrm{Bernoulli}(d_i^0(j))$, we have $|X_{im}|\leq 2$ and $\mathrm{Var}(X_{im})\leq d_i^0(j)=\sum_{k=1}^KA_k(j)w_i(k)\leq \sum_{k=1}^K A_k(j)=h_j$. 
We now apply the Bernstein inequality: 

\begin{lem}[Bernstein inequality] \label{lem:Bennett}
Suppose $X_1,\cdots, X_n$ are independent random variables such that $EX_i=0$, $|X_i|\leq b$ and $\mathrm{Var}(X_i)\leq \sigma^2_i$ for all $i$. Let $\sigma^2 = n^{-1}\sum_{i=1}^n\sigma^2_i$. 
Then, for any $t> 0$, 
\[
P\Big( n^{-1}|\sum_{i=1}^n X_i|\geq t \Big) \leq 2 \exp\left(  -\frac{nt^2/2}{\sigma^2 + bt/3} \right). 
\]
\end{lem}

Using Lemma~\ref{lem:Bennett}, we obtain
\[
P\bigl( |M(j,j)-M_0(j,j)|\geq t \bigr) \leq 2 \exp\left(  -\frac{Nnt^2/2}{h_j + 2t/3} \right). 
\]
Let $t=(Nn)^{-1/2}\sqrt{10h_j\log(n)}$. Since $h_j\geq h_{\min}\gg(Nn)^{-1}\log(n)$, we have $t\ll h_j$; therefore, in the denominator of the exponent, the term $h_j$ is dominating. It follows that, with probability $1-o(n^{-4})$,
\[
|M(j,j)-M_0(j,j)|\leq (Nn)^{-1/2}\sqrt{10h_j\log(n)}. 
\]
According to the probability union bound, the above holds simultaneously for all $1\leq j\leq p$ with probability $1-o(pn^{-4})=1-o(n^{-3})$. Here, we have assumed $n\geq \max\{N,p\}$ without loss of generality. If $n<\max\{N,p\}$, the result continues to hold with $\log(n)$ replaced by $\log(\max\{n,N,p\})$.
\qed

\subsection{Proof of Lemma~\ref{lem:cross-term}}
Consider the first claim. Fix $k$. Let $\{T_{im}:1\leq i\leq n, 1\leq m\leq N\}$ be as in \eqref{prop-binomial-0}. It follows that
\[
Z_j'W_k =\sum_{i=1}^n z_i(j)w_i(k) \overset{(d)}{=}\frac{1}{Nn}\sum_{i=1}^n\sum_{m=1}^N nw_i(k) \bigl\{T_{im}(j)-E[T_{im}(j)]\bigr\}.
\]
Write $X_{im}=nw_i(k) \{T_{im}(j)-E[T_{im}(j)]\}$. Since $T_{im}(j)\sim \mathrm{Bernoulli}(d_i^0(j))$, we find that $\mathrm{Var}(X_{im})\leq n^2w_i^2(k)d_i^0(j)\leq n^2h_j$ and $|X_{im}|\leq 2nw_i(k)\leq 2n$. We now apply Lemma~\ref{lem:Bennett} with $\sigma^2=n^2h_j$ and $b=2n$. It yields that
\[
P(|Z_j'W_k|>t)\leq 2\exp\left( \frac{Nnt^2/2}{n^2h_j + 2nt/3}\right).
\]
Set $t=C\sqrt{N^{-1}nh_j\log(n)}$ for a constant $C>0$ to be decided. For such $t$, since $h_j\geq h_{\min}\gg (Nn)^{-1}\log(n)$, the term $n^2h_j$ is the dominating term in the denominator of the exponent. Therefore, when $C$ is properly large, the right hand side is $o(n^{-4})$. In other words, with probability $1-o(n^{-4})$,
\beq \label{lem-cross-1}
|Z_j'W_k|\leq CN^{-1/2}\sqrt{nh_j\log(n)}. 
\eeq
Combing this with the probability union bound gives the claim. 

Consider the second claim. Write 
\[
\|M_0^{-1/2}ZW_k\|^2=\sum_{j=1}^p \frac{1}{M_0(j,j)}|Z_j'W_k|^2. 
\]
We have obtained the upper bound \eqref{lem-cross-1}, which holds simultaneously for all $1\leq j\leq p$, with probability $1-o(n^{-3})$. Moreover, 
from Lemma~\ref{lem:M0}, $M_0(j,j)\geq c_1h_j$. As a result, with probability $1-o(n^{-3})$,
\[
\|M_0^{-1/2}ZW_k\|^2\leq \sum_{j=1}^p\frac{1}{c_1h_j}\frac{Cnh_j\log(n)}{N} = \frac{Cnp\log(n)}{c_1N}.
\]
This proves the claim. \qed

\subsection{Proof of Lemma~\ref{lem:ZZ'-2}} \label{subsec:proof-lem-ZZ'-2}
We aim to show that, for any given $1\leq j,\ell\leq p$, with probability $1-o(n^{-5})$, 
\beq\label{lem-ZjZell-0}
\frac{1}{\sqrt{h_jh_\ell}}|Z_j'Z_\ell - E[Z_j'Z_\ell]|\leq C\biggl(\frac{1}{N}+\frac{\log(n)}{N^2h_{\min}}\biggr)\sqrt{n\log(n)}. 
\eeq
Once \eqref{lem-ZjZell-0} is true, the claim follows from the probability union bound. 

Below, we show \eqref{lem-ZjZell-0}. 
Fix $(j,\ell)$. 
Write $Z=[z_1,\ldots,z_n]$, and let $H=\mathrm{diag}(h_1,\ldots,h_p)$. Using the equality $xy=\frac{1}{4}(x+y)^2-\frac{1}{4}(x-y)^2$, we find that
\begin{align*}
\frac{Z_j'Z_\ell}{\sqrt{h_jh_\ell}} &= \sum_{i=1}^n \frac{z_i(j)}{\sqrt{h_j}}\cdot\frac{z_i(\ell)}{\sqrt{h_\ell}}\cr
& =\sum_{i=1}^n \left(  \frac{z_i(j)}{2\sqrt{h_j}} +  \frac{z_i(\ell)}{2\sqrt{h_\ell}}\right)^2 -  \sum_{i=1}^n \left(  \frac{z_i(j)}{2\sqrt{h_j}} -  \frac{z_i(\ell)}{\sqrt{2h_\ell}}\right)^2\cr
&= \sum_{i=1}^n (u_1'H^{-1/2}z_i)^2 - \sum_{i=1}^n (u_2'H^{-1/2}z_i)^2, \quad \;\; u_1\equiv\frac{e_j+e_\ell}{2}, u_2 \equiv \frac{e_j-e_\ell}{2}; 
\end{align*}
here $e_1,\ldots,e_p$ denote the standard basis vectors of $\mathbb{R}^p$. Taking the expectation on both sides, we find that $E[Z_j'Z_\ell]$ has a similar decomposition. As a result, 
\begin{align} \label{lem-ZjZell-1}
\frac{Z_j'Z_\ell - E[Z_j'Z_\ell]}{\sqrt{h_jh_\ell}} &= \sum_{i=1}^n \bigl\{(u_1'H^{-1/2}z_i)^2-E[(u_1'H^{-1/2}z_i)^2]\bigr\}\cr
& -  \sum_{i=1}^n \bigl\{(u_2'H^{-1/2}z_i)^2-E[(u_2'H^{-1/2}z_i)^2]\bigr\}\cr
&\equiv I + II. 
\end{align}
Below, we focus on deriving an upper bound for $I$. In the end of the proof, we explain how to bound $II$ in a similar way. 

We start from studying $u_1'H^{-1/2}z_i$. Let $\{T_{im}:1\leq i\leq n,1\leq m\leq N\}$ be the same as in \eqref{prop-binomial-0}. It follows that
\[
u_1'H^{-1/2}z_i\overset{(d)}{=} \frac{1}{N}\sum_{m=1}^N u_1'H^{-1/2}(T_{im}-E[T_{im}]). 
\]
Write $Y_{im}=u_1'H^{-1/2}(T_{im}-E[T_{im}])$. Since $T_{im}\sim \mathrm{Multinomial}(1, d_i^0)$, the covariance matrix of $T_{im}$ equals to $\mathrm{diag}(d_i^0)-d_i^0(d_i^0)'$. It follows that 
$\mathrm{Var}(Y_{im})\leq u_1'H^{-1/2}\mathrm{diag}(d_i^0)H^{-1/2}u_1=\frac{1}{4}(\frac{\sqrt{d_i^0(j)}}{\sqrt{h_j}}+\frac{\sqrt{d_i^0(\ell)}}{\sqrt{h_\ell}})^2\leq 1$, where the last inequality is because  $d_i^0(j)\leq h_j$. Furthermore, $|Y_{im}|\leq 1/\sqrt{h}_j+1/\sqrt{h}_\ell\leq 2/\sqrt{h_{\min}}$. We now apply the Bernstein inequality, Lemma~\ref{lem:Bennett}, with $\sigma^2=1$, $b=2/\sqrt{h_{\min}}$. It gives
\beq \label{lem-ZjZell-2}
P\bigl( |u_1'H^{-1/2}z_i|>t \bigr)\leq 2\exp\left( -\frac{Nt^2/2}{1+2t/(3\sqrt{h_{\min}})}\right), \qquad \mbox{for all }t>0. 
\eeq
As a result, with probability $1-o(n^{-5})$,
\[
|u_1'H^{-1/2}z_i| \leq C\max\biggl\{ \frac{\sqrt{\log(n)}}{\sqrt{N}},\;\;\frac{\log(n)}{N\sqrt{h_{\min}}} \biggr\}. 
\]
It motivates us to consider two different cases: (a) $Nh_{\min}\geq \log(n)$, and (b) $Nh_{\min}<\log(n)$.

Consider case (a).  
Let $t_0=\tilde{C}N^{-1/2}\sqrt{\log(n)}$ for a properly large $\tilde{C}>0$ to be decided. For all $0<t\leq t_0$, the right hand side of \eqref{lem-ZjZell-2} is bounded by $2e^{-CNt^2/4}$. 
Define 
\[
X_i=(u_1'H^{-1/2}z_i)\cdot 1\bigl\{|u_1'H^{-1/2}z_i|\leq t_0\bigr\}.
\]
For any fixed $\beta>0$, when $\tilde{C}=\tilde{C}(\beta)$ is chosen properly large, we have the following results: 
\begin{itemize}
\item[(i)] $X_i=u_1'H^{-1/2}z_i$ with probability $1-o(n^{-6})$. 
\item[(ii)] $X_i$ is a sub-Gaussian random variable with the sub-Gaussian norm $\|X_i\|_{\psi_2}=O(1/\sqrt{N})$. 
\item[(iii)] $|E[(u'H^{-1/2}z_i)^2]-E[X_i^2]|=o(n^{-\beta})$.
\end{itemize}
Here (i) is because $P(X_i\neq u_1'H^{-1/2}z_i)=P(|u_1'H^{-1/2}z_i|>t_0)\leq 2e^{-CNt_0^2/4}=O(n^{-C\tilde{C}^2/4})$; (ii) is because: for $0<t\leq t_0$, $P(|X_i|>t)\leq P(|u_1'H^{-1/2}z_i|>t)\leq 2e^{-CNt^2/4}$, and for $t>t_0$, $P(|X_i|>t)=0$; (iii) is because $|E[(u'H^{-1/2}z_i)^2]-E[X_i^2]|\leq (2/\sqrt{h_{\min}})^2\cdot P(|u'H^{-1/2}z_i|>t_0)=o(N)\cdot O(n^{-C\tilde{C}^2/4})$. We choose $\beta$ large enough such that $N^{-1}\sqrt{n\log(n)}\geq n^{-\beta}$. Using (i)-(iii) above, with probability $1-o(n^{-5})$, 
\beq \label{lem-ZjZell-3}
I = \sum_{i=1}^n (X_i^2-E[(u_1'H^{-1/2}z_i)]\bigr) = \sum_{i=1}^n (X_i^2-E[X_i^2]) + o\Bigl(\frac{\sqrt{n\log(n)}}{N}\Bigr). 
\eeq
Since each $X_i$ is sub-Gaussian, $X_i^2-E[X_i^2]$ is a sub-exponential random variable with the sub-exponential norm $\|X_i^2-E[X_i^2]\|_{\psi_1}\leq 2\|X_i\|^2_{\psi_2}=O(1/N)$ \citep[Lemma 5.14, Remark 5.18]{Vershynin}. We apply the Bernstein's inequality for sub-exponential variables \citep[Corollary 5.17]{Vershynin}:
\begin{lem}[Bernstein's inequality for sub-exponential variables] \label{lem:Bernstein2}
Suppose $X_1,\cdots, X_n$ are independent random variables such that $EX_i=0$ and $\max_{1\leq i\leq n}\|X\|_{\psi_1}\leq \kappa$.  
Then, for any $t> 0$, 
\[
P\Big( |\sum_{i=1}^n X_i|>nt \Big) \leq 2 \exp\left(  -cn\min\left\{\frac{t^2}{\kappa^2}, \frac{t}{\kappa}   \right\}\right),
\]
where $c>0$ is a universal constant. 
\end{lem}

\noindent
We apply Lemma~\ref{lem:Bernstein2} with $\kappa=C_1/N$ and $t=C_2\kappa\sqrt{n^{-1}\log(n)}$ for $C_1, C_2>0$ that are large enough. It follows that with probability $1-o(n^{-5})$, 
\[
|\sum_{i=1}^n (X_i^2-E[X_i^2])| \leq CN^{-1}\sqrt{n\log(n)}.
\]
Combining it with \eqref{lem-ZjZell-3} gives: with probability $1-o(n^{-5})$, 
\beq \label{lem-ZjZell-4}
|I| \leq CN^{-1}\sqrt{n\log(n)}. 
\eeq

Consider case (b). In this case, let $\delta_n = C_3\log(n)/(N\sqrt{h_{\min}})$ for a large enough constant $C_3$ to be decided. It follows from \eqref{lem-ZjZell-2} that
\[
P\bigl( |u_1'H^{-1/2}z_i|>t \bigr)\leq
\begin{cases}
2\exp\bigl( - Nt^2/[2 + 4C_3 \frac{\log(n)}{Nh_{\min}}] \bigr), & 0<t\leq \delta_n,\\
2\exp\bigl(- \frac{3}{6C_3^{-1}+4}\frac{N}{\sqrt{h_{\min}}} t\bigr), & t> \delta_n. 
\end{cases}
\]
Define 
\[
\tilde{X}_i = u_1'H^{-1/2}z_i\cdot 1\bigl\{ |u_1'H^{-1/2}z_i | \leq \delta_n\bigr \}. 
\]
Therefore, for each fixed $\beta>0$, by choosing $C_3=C_3(\beta)$ appropriately large, we conclude that
\begin{itemize}
\item[(i)] $\tilde{X}_i=u_1'H^{-1/2}z_i$ with probability $1-o(n^{-6})$. 
\item[(ii)] $\tilde{X}_i$ is a sub-Gaussian random variable with the sub-Gaussian norm $\|\tilde{X}_i\|_{\psi_2}=O\bigl(\sqrt{\log(n)/(N^2h_{\min})}\bigr)$. 
\item[(iii)] $|E[(u'H^{-1/2}z_i)^2]-E[X_i^2]|=o(n^{-\beta})$.
\end{itemize}
We choose $\beta$ large enough such that $\frac{\log(n)}{N^2h_{\min}}\sqrt{n\log(n)}\geq n^{-\beta}$. It follows that with probability $1-o(n^{-5})$, 
\[
I = \sum_{i=1}^n (X_i^2-E[(u_1'H^{-1/2}z_i)]\bigr) = \sum_{i=1}^n (X_i^2-E[X_i^2]) + o\biggl(\frac{\log(n)}{N^2h_{\min}}\sqrt{n\log(n)}\biggr). 
\]
Each $X_i^2-E[X_i^2]$ is a sub-exponential random variable with the sub-exponential norm $\|X_i^2-E[X_i^2]\|_{\psi_1}=O(\log(n)/(N^2h_{\min}))$. We then apply Lemma~\ref{lem:Bernstein2} with $\kappa=C_4\log(n)/(N^2h_{\min})$ and $t=C_5\kappa\sqrt{n^{-1}\log(n)}$, with $C_4,C_5$ being large enough constants. It follows that with probability $1-o(n^{-5})$,
\[
|\sum_{i=1}^n (X_i^2-E[X_i^2])| \leq nt \leq \frac{C\log(n)}{N^2h_{\min}}\sqrt{n\log(n)}. 
\] 
It follows that 
\beq \label{lem-ZjZell-5}
|I| \leq C\frac{\log(n)}{N^2h_{\min}}\sqrt{n\log(n)}. 
\eeq
Combining \eqref{lem-ZjZell-4}-\eqref{lem-ZjZell-5} gives that
\beq  \label{lem-ZjZell-6}
|I|\leq C\biggl(\frac{1}{N}+\frac{\log(n)}{N^2h_{\min}}\biggr)\sqrt{n\log(n)}. 
\eeq

We then bound $II$. When $j=\ell$, $II$ is exactly equal to $0$. When $j\neq \ell$, we can similarly write $u_2'H^{-1/2}z_i=N^{-1}\sum_{m=1}^N Y_{im}$, with $Y_{im}=u_2'H^{-1/2}(T_{im}-E[T_{im}])$. Then, $|Y_{im}|\leq \max\{1/\sqrt{h_j},1/\sqrt{h_\ell}\}\leq 1/\sqrt{h_{\min}}$, and $\mathrm{Var}(Y_{im})\leq u_2'H^{-1}\mathrm{diag}(d_i^0)H^{-1/2}u_2\leq \frac{1}{4}(\frac{\sqrt{d_i^0(j)}}{\sqrt{h_j}}-\frac{\sqrt{d_i^0(\ell)}}{\sqrt{h_\ell}})^2\leq \frac{1}{4}$. We again apply Lemma~\ref{lem:Bennett} to bound the tail probability of $u_2'H^{-1/2}z_i$, and then apply Lemma~\ref{lem:Bernstein2} to bound $II$. Similarly, we find that, with probability $1-o(n^{-5})$, 
\beq  \label{lem-ZjZell-7}
|II|\leq C\biggl(\frac{1}{N}+\frac{\log(n)}{N^2h_{\min}}\biggr)\sqrt{n\log(n)}. 
\eeq
Then, \eqref{lem-ZjZell-0} follows from plugging \eqref{lem-ZjZell-6}-\eqref{lem-ZjZell-7} into \eqref{lem-ZjZell-1}. \qed

\subsection{Proof of Lemma~\ref{lem:new-ZZ'-2}}
Following the lines in the proof of Lemma~\ref{lem:ZZ'-2} until \eqref{lem-ZjZell-1}, we know that the key is to get upper bounds for 
\begin{align*}
X_1 &= \sum_{i=1}^n \{ (u_1'H^{-1/2}z_i)^2 - E[(u_1'H^{-1/2}z_i)^2]\},\cr
X_2 &= \sum_{i=1}^n \{ (u_2'H^{-1/2}z_i)^2 - E[(u_2'H^{-1/2}z_i)^2]\},
\end{align*}
where $u_1$ and $u_2$ are as in \eqref{lem-ZjZell-1}. We will analyze these terms in the proof of Lemma~\ref{lem:new-ZZ'}. For this reason, we no longer repeat the proof but quote the results from the proof of Lemma~\ref{lem:new-ZZ'}. 

We can bound $X_1$ and $X_2$ similarly as in the proof of \eqref{thm-large-0}, except that we only need the bounds hold with probability $1-o(n^{-5})$ but in \eqref{thm-large-0} we need the bound to hold with probability $1-o(9^{-p}n^{-3})$. So, we simply replace $p$ in \eqref{thm-large-0} by $\sqrt{\log(n)}$. This proves Lemma~\ref{lem:new-ZZ'-2}.

\subsection{Proof of Lemma~\ref{lem:ZZ'}} \label{subsec:proof-lem-ZZ'}
Let $H=\mathrm{diag}(h_1,\ldots,h_p)$. By Lemma~\ref{lem:M0}, $M_0(j,j)\geq c_1h_j$ for all $1\leq j\leq p$. It follows that $\|M_0^{-1/2}H^{1/2}\|\leq c_1^{-1/2}$. As a result, 
\begin{align*}
& \|M_0^{-1/2}(ZZ'-E[ZZ'])M_0^{-1/2}\|\cr
=& \|M_0^{-1/2}H^{1/2}\|\cdot \|H^{-1/2}(ZZ'-E[ZZ'])H^{-1/2}\|\cdot \|H^{1/2}M_0^{-1/2}\|\cr
\leq & c_1^{-1} \|H^{-1/2}(ZZ'-E[ZZ'])H^{-1/2}\|. 
\end{align*}
Therefore, to show the claim, it suffices to show that 
\beq \label{lem-ZZ'-0}
\|H^{-1/2}(ZZ'-E[ZZ'])H^{-1/2}\|\leq C\Big(\frac{1}{N}+\frac{p}{N^2h_{\min}}\Big)\sqrt{np}. 
\eeq

To show \eqref{lem-ZZ'-0}, we need some existing results on $\alpha$-nets. 
For any $\alpha>0$, a subset $\mathcal{M}$ of the unit sphere $\mathcal{S}^{p-1}$ is called an $\alpha$-net if $\sup_{x\in\mathcal{S}^{p-1}}\inf_{y\in\mathcal{M}}\|x-y\|\leq \alpha$. The following lemma combines Lemmas 5.2-5.3 in \cite{Vershynin}. 
\begin{lem}[$\alpha$-net]\label{lem:alpha-net}
Fix $\alpha\in (0,1/2)$. There exists an $\alpha$-net $\mathcal{M}_\alpha$ of $\mathcal{S}^{p-1}$ such that  $|\mathcal{M}_\alpha |\leq (1+2/\alpha)^p$. Moreover, for any symmetric $p\times p$ matrix $B$,
$\|B\| \leq (1 - 2 \alpha)^{-1} \sup_{u \in \mathcal{M}_\alpha} \{ |u'B u| \}$. 
\end{lem}
\noindent 
By Lemma~\ref{lem:alpha-net}, there exists a $(1/4)$-net $\mathcal{M}_{1/4}$, such that $|\mathcal{M}_{1/4}|\leq 9^p$ and 
\[
\|H^{-1/2}(ZZ'-E[ZZ'])H^{-1/2}\|\leq 2\max_{u\in\mathcal{M}_{1/4}}\{|u'H^{-1/2}(ZZ'- E[ZZ'])H^{-1/2}u|\}. 
\]
Therefore, 
to show \eqref{lem-ZZ'-0}, it is sufficient to show that, for any fixed $u\in\mathcal{S}^{p-1}$, with probability $1-o(9^{-p}n^{-3})$,
\beq \label{lem-ZZ'-1}
|u'H^{-1/2}(ZZ'- E[ZZ'])H^{-1/2}u|\leq C \Big(\frac{1}{N}+\frac{p}{N^2h_{\min}}\Big)\sqrt{np}. 
\eeq

Below, we show \eqref{lem-ZZ'-1}. Write $Z=[z_1,\ldots,z_n]$. 
For any $u\in\mathcal{S}^{p-1}$, 
\begin{align} \label{lem-ZZ'-decompose}
& u'H^{-1/2}(ZZ'- E[ZZ'])H^{-1/2}u \cr
=& \sum_{i=1}^n \{ (u'H^{-1/2}z_i)^2 - E[(u'H^{-1/2}z_i)^2]\}. 
\end{align}
Our plan is to first get a tail bound for $u'H^{-1/2}z_i$, which is similar to \eqref{lem-ZjZell-2}. We then consider two separate cases, $Nh_{\min}\geq p$ and $Nh_{\min}<p$: for each case, we use the tail bound of $u'H^{-1/2}z_i$ to prove \eqref{lem-ZZ'-1}.

First, we study $u'H^{-1/2}z_i$. Let $\{T_{im}: 1\leq i\leq n,1\leq m\leq N\}$ be the set of random variables as in \eqref{prop-binomial-0}. Write
\beq \label{Yim}
u'H^{-1/2}z_i\overset{(d)}=\frac{1}{N}\sum_{m=1}^NY_{im}, \qquad\mbox{with }Y_{im}=u'H^{-1/2}(T_{im}-E[T_{im}]). 
\eeq
Since $T_{im}$ follows a distribution of $\mathrm{Multinomial}(1,d_i^0)$, 
it is easy to see that $|Y_{im}|\leq 2/\sqrt{h_{\min}}$ and $\mathrm{var}(Y_{im})\leq u'H^{-1/2}\mathrm{diag}(d_i^0)H^{-1/2}u\leq \|u\|^2\leq 1$ (note that $d_i^0(j)= \sum_{k=1}^K A_k(j)w_i(k)\leq \sum_{k=1}^K A_k(j)=h_j$). We apply the Bernstein's inequality, Lemma~\ref{lem:Bennett}, and obtain that, for any $t>0$, 
\beq \label{lem-ZZ'-2}
P( |u'H^{-1/2}z_i|>t )\leq 2\exp\left( -\frac{Nt^2/2}{1+2t/(3\sqrt{h_{\min}})}\right), \qquad \mbox{for all }t>0. 
\eeq

Next, we prove \eqref{lem-ZZ'-1} for two cases separately: $Nh_{\min}\geq p$ and $Nh_{\min}<p$. In the first case, 
for a constant $C_1>0$ to be decided, let $\delta_{n1}= C_1\sqrt{p/N}$. Since $Nh_{\min}\geq p$, we have 
\beq  \label{lem-ZZ'-3}
P(|u'H^{-1/2}z_i|>t)\leq 2\exp\left( -\frac{Nt^2/2}{1+2C_1/3} \right), \qquad\mbox{for all }0< t\leq \delta_{n1}. 
\eeq
We then define a truncated version of $u'H^{-1/2}z_i$: 
\[
X_i \equiv u'H^{-1/2}z_i \cdot 1\bigl\{|u'H^{-1/2}z_i|\leq \delta_{n1}\bigr\}, \qquad 1\leq i\leq n. 
\]
We claim that
\begin{itemize}
\item[(i)] $X_i=u'H^{-1/2}z_i$ with probability $1-o(9^{-p}n^{-4})$. 
\item[(ii)] $X_i$ is a sub-Gaussian random variable with the sub-Gaussian norm $\|X_i\|_{\psi_2}=O(1/\sqrt{N})$. 
\item[(iii)] $|E[(u'H^{-1/2}z_i)^2]-E[X_i^2]|$ is negligible compared with the right hand side of \eqref{lem-ZZ'-1}.
\end{itemize}
Here (ii) is a direct result of \eqref{lem-ZZ'-3}. To see (i), note that by \eqref{lem-ZZ'-3}, $P(|u'H^{-1/2}z_i|>\delta_{n1})\leq 2\exp(-\tfrac{C_1^2/2}{1+2C_1/3}p)$; since $p\geq C \log(n)$, with an appropriately large $C_1$, this probability is $o(9.1^{-p})= o(9^{-p}n^{-4})$. To see (iii), note that $|u'H^{-1/2}z_i|\leq 2/\sqrt{h_{\min}}\leq 2\sqrt{N/p}$; so, $|E[(u'H^{-1/2}z_i)^2]-E[X_i^2]|\leq (4N/p)\cdot P(|u'H^{-1/2}z_i|>\delta_{n1})\leq (8N/p)\cdot\exp(-\tfrac{C_1^2/2}{1+2C_1/3}p)$. Since $p\geq C\log(N+n)$, when $C_1$ is large enough, this quantity is $o(N^{-1}\sqrt{np})$.
Combining (i)-(iii) with \eqref{lem-ZZ'-decompose}, with probability $1-o(9^{-p}n^{-3})$, 
\beq \label{lem-ZZ'-4}
|u'H^{-1/2}(ZZ'-E[ZZ'])H^{-1/2}u|\leq |\sum_{i=1}^n (X^2_i - E[X^2_i])| + o(N^{-1}\sqrt{np}). 
\eeq
Since each $X_i$ is sub-Gaussian, $X_i^2-E[X_i^2]$ is a sub-exponential random variable with the sub-exponential norm $\|X_i^2-E[X_i^2]\|_{\psi_1}\leq 2\|X_i\|^2_{\psi_2}=O(1/N)$ \citep[Lemma 5.14, Remark 5.18]{Vershynin}. We then apply Lemma~\ref{lem:Bernstein2} with $\kappa=O(1/N)$ and $t=C\kappa\cdot \sqrt{p/n}$. When the constant $C$ is large enough, with probability $1-o(9^{-p}n^{-3})$,
\beq \label{lem-ZZ'-5}
| \sum_{i=1}^n (X_i^2 - E[X_i^2])|\leq nt\leq C N^{-1}\sqrt{np}. 
\eeq
Combining \eqref{lem-ZZ'-4}-\eqref{lem-ZZ'-5} gives \eqref{lem-ZZ'-1} in the first case. 

In the second case, let $\delta_{n2}=C_2p/(N\sqrt{h_{\min}})$ for a constant $C_2>0$ to be determined. We study the right hand of \eqref{lem-ZZ'-2}. Note that $Nh_{\min}<p$. For $t\leq \delta_{n2}$, we have $1+2t/(3\sqrt{h_{\min}}) \leq p/(Nh_{\min})+2\delta_{n2}/(3\sqrt{h_{\min}})=(1+2C_2/3)\cdot p/(Nh_{\min})$; for $t>\delta_{n2}$, we have $1+2t/(3\sqrt{h_{\min}})\leq \delta_{n2}/(C_2\sqrt{h_{\min}}) +2t/(3\sqrt{h_{\min}}) =(C_2^{-1}+2/3)\cdot t/\sqrt{h_{\min}}$. Plugging them into \eqref{lem-ZZ'-2} gives
\beq  \label{lem-ZZ'-6}
P(|u'H^{-1/2}z_i|>t)\leq 2
\begin{cases}
\exp\Big( -\frac{1/2}{1+2C_2/3}\cdot p^{-1}N^2h_{\min}\cdot t^2\Big), &\mbox{for }0< t\leq \delta_{n2},\\
\exp\Big( -\frac{1/2}{C_2^{-1}+2/3}\cdot N\sqrt{h_{\min}}\cdot t\Big), &\mbox{for }t > \delta_{n2}.
\end{cases} 
\eeq
In particular, $P(|u'H^{-1/2}z_i|>\delta_{n2})\leq 2e^{-\frac{3C^2_2}{6+4C_2}p}$. In light of this, we introduce a truncated version of $u'H^{-1/2}z_i$: 
\[
\tilde{X}_i \equiv u'H^{-1/2}z_i \cdot 1\bigl\{|u'H^{-1/2}z_i|\leq \delta_{n2}\bigr\}, \qquad 1\leq i\leq n. 
\]
We have the following observations, whose proofs are similar to the (i)-(iii) in the first case and are omitted. 
\begin{itemize}
\item[(i)] $\tilde{X}_i=u'H^{-1/2}z_i$ with probability $1-o(9^{-p}n^{-4})$. 
\item[(ii)] $\tilde{X}_i$ is a sub-Gaussian random variable with the sub-Gaussian norm $\|\tilde{X}_i\|_{\psi_2}=O(\sqrt{p/(N^2h_{\min})})$. 
\item[(iii)] $|E[(u'H^{-1/2}z_i)^2]-E[\tilde{X}_i^2]|$ is negligible compared with the right hand side of \eqref{lem-ZZ'-1}.
\end{itemize} 
From (ii), $\tilde{X}_i^2-E[\tilde{X}^2_i]$ is a sub-exponential random variable with the sub-exponential norm $\|\tilde{X}_i^2-E[\tilde{X}^2_i]\|_{\psi_1}=O(p/(N^2h_{\min}))$. We apply Lemma~\ref{lem:Bernstein2} with $\kappa=O(p/(N^2h_{\min}))$ and $t=O(\kappa\sqrt{p/n})$. Combining the result with (i) and (iii), we find that, with probability $1-o(9^{-p}n^{-3})$,
\begin{align}  \label{lem-ZZ'-7}
 |u'H^{-1/2}&(ZZ'-E[ZZ'])H^{-1/2}u| \leq |\sum_{i=1}^n (\tilde{X}^2_i - E[\tilde{X}^2_i])| +  o\big(\frac{p\sqrt{np}}{N^2h_{\min}}\big)\cr
&\leq Cn\kappa \sqrt{p/n} + o\big(\frac{p\sqrt{np}}{N^2h_{\min}}\big) \leq \frac{Cp\sqrt{np}}{N^2h_{\min}}.  
\end{align}
This proves \eqref{lem-ZZ'-1} in the second case. \qed

\subsection{Proof of Lemma~\ref{lem:new-ZZ'}} \label{subsec:proof-lem-ZZ'-new}

Following the lines of proof of Lemma~\ref{lem:ZZ'} until equation \eqref{lem-ZZ'-decompose}, we find out that it suffices to prove: for any fixed unit-norm vector $u$, with probability $1-o(9^{-p}n^{-3})$, 
\beq \label{thm-large-0}
\sum_{i=1}^n \{ (u'H^{-1/2}z_i)^2 - E[(u'H^{-1/2}z_i)^2]\}\leq C\frac{\sqrt{np}}{N}\Bigl(1 + \frac{1}{\sqrt{Nh_{\min}}}\Bigr). 
\eeq
Write for short $X = \sum_{i=1}^n \{ (u'H^{-1/2}z_i)^2 - E[(u'H^{-1/2}z_i)^2]\}$. 
Let $Y_{im}$ be the same as in \eqref{Yim}. Then,
\beq \label{thm-large-1}
u_i'H^{-1/2}z_i = \frac{1}{N}\sum_{m=1}^N Y_{im}, \qquad \mbox{where }|Y_{im}|\leq \frac{2}{\sqrt{h_{\min}}}, \;\mathrm{var}(Y_{im})\leq 1. 
\eeq
Then
\beq \label{thm-large-2}
X = \frac{1}{N^2}\sum_{i=1}^n \sum_{m,s=1}^N (Y_{im}Y_{is}-\mathbb{E}[Y_{im}Y_{is}]). 
\eeq 
Our tool for studying $X$ is the Bernstein inequality for martingales \citep{freedman1975tail}:
\begin{lem}[Bernstein inequality for martingales] \label{lem:martingale}
Let $\{\xi_n\}_{n=1}^\infty$ be a martingale difference sequence with respect to the filtration $\{{\cal F}_n\}_{n=0}^\infty$, where $|\xi_n|\leq b$ for $b>0$. Define the martingale $M_n=\sum_{i=1}^n\xi_i$, and let its variance process be defined as $\langle M\rangle_n= \sum_{i=1}^n E[\xi_i^2|{\cal F}_{i-1}]$. 
Suppose $\tau$ is a finite stopping time with respect to $\{{\cal F}_n\}_{n=0}^\infty$. Then, for any $t>0$ and $\sigma^2>0$,
\[
P\Bigl(\max_{n\leq \tau}M_n>t, \langle M\rangle_n>\sigma^2 \Bigr)\leq 2\exp\Bigl(- \frac{t^2/2}{\sigma^2+bt/3}\Bigr). 
\]
\end{lem}

We construct a martingale as follows:
\[
\theta_{im} = \frac{1}{N^2}\sum_{j=1}^i\sum_{s,k=1}^m (Y_{js}Y_{jk}-\mathbb{E}[Y_{js}Y_{jk}]), \quad 1\leq i\leq n,1\leq m\leq N.
\]
It is seen that $X=\theta_{nN}$, and $\{\theta_{11},\ldots,\theta_{1N},\ldots,\theta_{n1},\ldots,\theta_{nN}\}$ is a martingale with respect to the filtration ${\cal F}_{im}=\sigma\bigl( \{Y_{js}\}_{1\leq j\leq i-1,1\leq s\leq N}\cup \{Y_{is}\}_{s=1}^{m-1} \bigr)$. We study the variance process of this martingale. Let
\[
\Gamma_{im} = \begin{cases}
E[(\theta_{i1}-\theta_{(i-1)N})^2|{\cal F}_{(i-1)N}], & m=1,\\
E[(\theta_{im}-\theta_{i(m-1)})^2|{\cal F}_{i(m-1)}], & m\geq 2. 
\end{cases}
\] 
The variance process is 
\[
\langle\theta\rangle_{im} = \sum_{j=1}^i\sum_{s=1}^m \Gamma_{js}, \qquad 1\leq i\leq n,1\leq m\leq N.  
\]
For $m=1$, $\theta_{i1}-\theta_{(i-1)N}=\frac{1}{N^2}Y^2_{i1}$. Hence, 
\[
\Gamma_{im}\leq \frac{1}{N^4}E(Y^4_{i1})\leq   \frac{4}{N^4h_{\min}}E(Y^2_{i1})\leq \frac{4}{N^4h_{\min}}, 
\]
where we used \eqref{thm-large-1}. For $m\geq 2$, $\theta_{im}-\theta_{i(m-1)}=\frac{1}{N^2}[2(\sum_{s=1}^{m-1}Y_{is})Y_{im} + Y^2_{im} - E(Y^2_{im})]$. It follows that
\begin{align*}
\Gamma_{im}& \leq \frac{C}{N^4}\left[ \Bigl(\sum_{s=1}^{m-1}Y_{is}\Bigr)^2\mathrm{var}(Y_{im}) + \mathrm{var}(Y^2_{im})  \right]\cr
&\leq \frac{C}{N^4}\Bigl(\sum_{s=1}^{m-1}Y_{is}\Bigr)^2 + \frac{C}{N^4h_{\min}}. 
\end{align*}
Combining the above gives
\beq \label{thm-large-3}
\langle\theta\rangle_{nN} \leq \frac{C}{N^4}\sum_{m=1}^N \underbrace{\sum_{i=1}^n\Bigl(\sum_{s=1}^{m-1}Y_{is}\Bigr)^2}_{\equiv S_{m-1}}
+ \frac{Cn}{N^3h_{\min}}.
\eeq
For the variable $S_{m-1}$, note that 
\[
E(S_{m-1})= \sum_{i=1}^n\sum_{s,k=1}^{m-1} E(Y_{is}Y_{ik})= \sum_{i=1}^n\sum_{s=1}^{m-1} E(Y^2_{is})\leq Nn.
\]
To study $S_{m-1}-E(S_{m-1})$, note that $S_N=N^2\cdot u'H^{-1/2}(ZZ'-E[ZZ'])H^{-1/2}u$. Hence, we already gave a bound for $N^{-2}|S_N-E(S_N)|$ in \eqref{lem-ZZ'-1}, which translates to: with probability $1-o(9^{-p}n^{-3})$,
\[
|S_N- E(S_N)|\leq C\Bigl(N+\frac{p}{h_{\min}}\Bigr)\sqrt{np}. 
\]
Note that $S_{m}=\sum_{i=1}^n(\sum_{s=1}^{m}Y_{is})^2$ and $S_N=\sum_{i=1}^n(\sum_{s=1}^{N}Y_{is})^2$ have similar forms: the former involves $nm$ independent multinomial variables (each has a trial number equal to $1$), and the latter involves $nN$ such independent multinomial variables. Therefore, we get a similar bound for $|S_{m}-E(S_{m})|$ by replacing $N$ with $m$ above. It yields that, with probability $1-o(9^{-p}n^{-3}N^{-1})$,
\[
|S_{m-1}-E(S_{m-1})|\leq C \Big(m+\frac{p}{h_{\min}}\Big)\sqrt{np}\leq C\Bigl(N+\frac{p}{h_{\min}}\Bigr)\sqrt{np}. 
\]
If $n\geq (Nh_{\min})^{-2}p^3$, the mean of $S_{m-1}$ dominates its variance. Hence, with probability $1-o(9^{-p}n^{-3})$, $\max_{1\leq m\leq N} S_{m}\leq CNn$. Plugging it into \eqref{thm-large-3}, we conclude that, 
\beq \label{thm-large-4}
\langle\theta\rangle_{nN}\leq \frac{Cn}{N^2} + \frac{Cn}{N^3h_{\min}}\equiv \sigma^2, \qquad \mbox{with probability $1-o(9^{-p}n^{-3})$}. 
\eeq
Moreover, for $m=1$, $|\theta_{i1}-\theta_{(i-1)N}|=\frac{1}{N^2}Y^2_{i1}\leq 2/(N^2h_{\min})$. For $m\geq 2$, 
\begin{align*}
|\theta_{im}-\theta_{i(m-1)}|& \leq \frac{1}{N^2}\bigl( 2|Y_{im}||\sum_{s=1}^{m-1}Y_{is}| + Y^2_{im} \bigr)\leq \frac{C}{Nh_{\min}}\equiv b,
\end{align*}
where we have used the bound for $|Y_{is}|$ in \eqref{thm-large-1}. We now apply Lemma~\ref{lem:martingale} by taking $t = C\sigma \sqrt{p}$, where $\sigma^2$ is as in \eqref{thm-large-4}. If $\sigma^2>b^2p$, then $bt=C\sigma(b\sqrt{p})\leq C\sigma^2$ and the bound in Lemma~\ref{lem:martingale} is determined by $\sigma^2$. For $\sigma^2>b^2p$ to happen, we need $n>p/h^2_{\min}$ and $n>(Np)/h_{\min}$. Under this condition, it follows from Lemma~\ref{lem:martingale} that 
\beq \label{thm-large-5}
P\Bigl( \theta_{nN}> C\sigma\sqrt{p},\; \langle\theta\rangle_{nN}\leq \sigma^2\Bigr)=o(9.1^{-p})=o(9^{-p}n^{-3}). 
\eeq  
Combining \eqref{thm-large-4}-\eqref{thm-large-5}, with probability $1-o(9^{-p}n^{-3})$,
\[
 \theta_{nN}\leq C\sigma\sqrt{p}\leq C\frac{\sqrt{np}}{N}\Bigl(1 +\frac{1}{\sqrt{Nh_{\min}}}\Bigr).
\]
This proves \eqref{thm-large-0}. The proof of Lemma~\ref{lem:new-ZZ'} is now complete. \qed

\section{Entry-wise analysis of singular vectors} \label{sec:Eigen-perturbation} 
We derive row-wise large deviation bounds for singular vectors and prove Theorem~\ref{thm:noise}.

First, we give a lemma that reduces the problem of deriving row-wise bounds for eigenvectors to the problem of studying the perturbation matrix. It has a similar flavor as the sin-theta theorem \cite{sin-theta}, but this result is stronger: It allocates the total error in eigenvectors into individual coordinates, which cannot be obtained from the sin-theta theorem.

\begin{lem}[A row-wise perturbation bound for eigenvectors] \label{lem:EigVecPerturb}
Let $G_0$ and $G$ be $p\times p$ symmetric matrices with $\mathrm{rank}(G_0)=K$. Write $Y=G-G_0=[y_1,y_2,\ldots,y_p]$. For $1\leq k\leq K$, let $\delta^0_k$ and $\delta_k$ be the respective $k$-th largest eigenvalue of $G_0$ and $G$, and let $u_k^0$ and $u_k$ be the respective $k$-th eigenvector of $G_0$ and $G$.  Fix $1\leq s\leq k\leq K$. For some $c\in (0,1)$, suppose (by default, if $s=1$, $\delta^0_{s-1}-\delta^0_{s}=\infty$)
\[
\min\bigl\{ \delta^0_{s-1}-\delta^0_{s},\; \delta^0_{k}-\delta^0_{k+1},\;\min_{1\leq\ell\leq K}|\delta^0_\ell|   \bigr\} \geq c\|G_0\|, \qquad \|Y\|\leq (c/3)\|G_0\|. 
\] 
Write $U_0=[u^0_s,u^0_{s+1},\ldots,u^0_k]$, $U=[u_s,u_{s+1},\ldots,u_k]$ and $U_0^*=[u^0_1, u^0_2,\ldots, u^0_K]$. There exists an orthogonal matrix $O$ such that 
\[
\|e_j'(UO- U_0)\|\leq \frac{5}{c\|G_0\|}\Bigl(\|Y\|\|e_j'U^*_0\| + \sqrt{K}\|y_j\|\Bigr),\quad \mbox{for all }1\leq j\leq p.   
\] 
\end{lem}

{\bf Remark}. In the claim of Lemma~\ref{lem:EigVecPerturb}, if we take the sum of squares for $j=1,2,\ldots,p$ on both hand sides, it yields $\|UO- U_0\|_F\leq C\sqrt{K}\|G_0\|^{-1}(\|Y\| + \|Y\|_F)$. The first term matches with the sin-theta theorem (up to a constant factor) and is tight; but our result is stronger than the sin-theta theorem, as it allocates the error to individual rows. The second term is not tight after taking the sum of squares for $1\leq j\leq p$; however, for bounding each individual row of $UO-U_0$, this term is good enough (at least for our purpose of proving Theorem~\ref{thm:noise}).

Next, we define a particular pair of $(G, G_0)$ that serves to prove Theorem~\ref{thm:noise}. Define
\begin{align} \label{G0G}
& G \equiv M^{-1/2}DD'M^{-1/2}-\frac{n}{N}I_p\cr
& G_0\equiv (1-\frac{1}{N})M_0^{-1/2}D_0D_0'M_0^{-1/2}.   
\end{align}
Recall that $\hat{\xi}_k$ is the $k$-th singular vector of $M^{-1/2}D$ and $\xi_k$ is the $k$-th singular vector of $M_0^{-1/2}D_0$. Equivalently, $\hat{\xi}_k$ and $\xi_k$ are the respective $k$-th eigenvector of $G$ and $G_0$. 

Now, to apply Lemma~\ref{lem:EigVecPerturb}, we need to study $\|G_0\|$, $\|e_j'\Xi\|$, $\|e_j'(G-G_0)\|$ and $\|G-G_0\|$.  The following lemma is about eigenvalues of $G_0$.  
\begin{lem}\label{lem:PopEigVal}
Suppose the conditions of Theorem~\ref{thm:noise} hold. Let $G_0$ be as in \eqref{G0G}. Denote by $\lambda_1\geq \lambda_2\geq \ldots\geq \lambda_K>0$ the nonzero eigenvalues of $G_0$. There exists a constant $C>1$ such that  
\[
C^{-1}n\leq \lambda_k \leq Cn \;\;\mbox{for all }1\leq k\leq K, \quad \mbox{and}\quad \lambda_1 \geq C^{-1}n + \max_{2\leq k\leq K}\lambda_k. 
\] 
\end{lem}

The following lemma is about $\Xi$, which contains the eigenvectors of $G_0$. 

\begin{lem}\label{lem:PopEigVec}
Suppose the conditions of Theorem~\ref{thm:noise} hold. Let $G_0$ be as in \eqref{G0G}. Denote by $\xi_1,\xi_2,\ldots,\xi_K$ be the first $K$ eigenvectors of $G_0$ and write $\Xi=[\xi_1,\ldots,\xi_K]$. There exists a constant $C>0$ such that 
\[
\|\Xi_j\|\leq C\sqrt{h_j}, \qquad \mbox{for all }1\leq j\leq p.  
\] 
\end{lem}

The following lemma is about the column-wise $\ell_2$-norms of  $G-G_0$. 
\begin{lem}\label{lem:Enorm2}
Under conditions of Theorem~\ref{thm:noise}, with probability $1-o(n^{-3})$, for $1\leq j\leq p$,
\[
\frac{\|e_j'(G-G_0)\|}{\sqrt{h_j}}\leq C\sqrt{\frac{np\log(n)}{N}}\times 
\begin{cases}
1, & \mbox{if } N\geq p\log(n),\\
N^{-3/2}p^{3/2}\log(n),& \mbox{if }N< p\log(n). 
\end{cases}
\]
If additionally $n\geq\frac{p}{h^2_{\min}}(1+\frac{p^2}{N^2}+Nh_{\min})$, then 
\[
\frac{\|e_j'(G-G_0)\|}{\sqrt{h_j}} \leq C( 1 + N^{-1}p)\sqrt{\frac{np\log(n)}{N}}.
\]
\end{lem}

The following lemma is about the spectral norm of $G-G_0$.
\begin{lem}\label{lem:Enorm}
Suppose the conditions of Theorem~\ref{thm:noise} hold. With probability $1-o(n^{-3})$, 
\[
\|G- G_0\|\leq  C\bigl(1+N^{-3/2}p^2\bigr) \sqrt{\frac{np \log(n)}{N}}. 
\]
If additionally $n\geq\frac{p}{h^2_{\min}}(1+\frac{p^2}{N^2}+Nh_{\min})$, then 
\[
\|G-G_0\|\leq C(1+N^{-1}p^{1/2})\sqrt{\frac{np\log(n)}{N}}. 
\]
\end{lem}

Below, we first use the above lemmas to show Theorems~\ref{thm:noise}-\ref{thm:hatR} (row-wise large-deviation bounds for singular vectors). We then prove the above lemmas in Sections~\ref{subsec:proof-lem-Taylor}-\ref{subsec:proof-lem-Enorm}. In the proofs, we will need properties of the noise matrix $Z=D-D_0$, which is already carefully analyzed in Section~\ref{sec:Z-analysis}.  

\subsection{Proof of Theorem~\ref{thm:noise}}
Divide the nonzero eigenvalues of $G_0$ into two groups: $\{\lambda_1\}$ and $\{\lambda_2,\lambda_3,\ldots,\lambda_K\}$. Introduce $\Xi^*=[\xi_2,\ldots,\xi_K]$ and $\hat{\Xi}^*=[\hat{\xi}_2,\ldots,\hat{\xi}_K]$, and let $(\Xi^*_j)'$ and $(\hat{\Xi}^*_j)'$ be the respective $j$-th row.  Then, for $\Omega=\mathrm{diag}(\omega,\Omega^*)$, 
\[
\|\Omega\hat{\Xi}_j-\Xi_j\| \leq \|\omega \hat{\xi}_1(j)-\xi_1(j)\|+ \|\Omega^*\hat{\Xi}^*_j - \Xi^*_j\|, \quad 1\leq j\leq p.  
\] 
By Lemma~\ref{lem:PopEigVal}, $\|G_0\|\asymp n$, and the gap between two groups of eigenvalues is $\geq C^{-1}n$. Also, by Lemma~\ref{lem:Enorm},  $\|G-G_0\|=o(n)$ with probability $1-o(n^{-3})$. Combining them, we conclude that the conditions of Lemma~\ref{lem:EigVecPerturb} hold for either group, $\{\lambda_1\}$ or $\{\lambda_2,\lambda_3,\ldots,\lambda_K\}$, with probability $1-o(n^{-3})$. By this lemma, there exists $\omega\in \{\pm 1\}$ such that 
\[
 \|\omega \hat{\xi}_1(j)-\xi_1(j)\| \leq Cn^{-1}\bigl( \|G-G_0\|\|\Xi_j\| + \|e_j'(G-G_0)\|  \bigr), 
\]
and there exists an $(K-1)\times(K-1)$ orthogonal matrix $\Omega^*$ such that 
\[
 \|\Omega^*\hat{\Xi}^*_j - \Xi^*_j\| \leq Cn^{-1}\bigl( \|G-G_0\|\|\Xi_j\| + \|e_j'(G-G_0)\|  \bigr). 
\]
We combine the above inequalities and use $\|\Xi_j\|\leq C\sqrt{h_j}$ to get 
\beq \label{proof-thm-noise-0}
\|\Omega\hat{\Xi}_j-\Xi_j\|\leq Cn^{-1}\Bigl( \sqrt{h_j}\|G-G_0\| + \|e_j'(G-G_0)\|  \Bigr). 
\eeq

First, we apply the first part of results in Lemmas~\ref{lem:Enorm2}-\ref{lem:Enorm}, which do not need additional assumptions on $n$. It yields that 
\[
\|G-G_0\|\leq C\Bigl(1+\frac{p^2}{N\sqrt{N}}\Bigr)\sqrt{\frac{np\log(n)}{N}}, \qquad  \frac{\|e_j'(G-G_0)\|}{\sqrt{h_j}}\leq C\Bigl(1+\frac{p^{3/2}\log(n)}{N\sqrt{N}}  \Bigr)\sqrt{\frac{np\log(n)}{N}}. 
\]
We plug them into \eqref{proof-thm-noise-0} and use the assumption of $\log^2(n)\leq \min\{p, N\}$. It follows that
\beq \label{proof-thm-noise-1}
\|\Omega\hat{\Xi}_j-\Xi_j\|\leq C\sqrt{h_j}\cdot (1+N^{-3/2}p^2)\sqrt{\frac{p\log(n)}{Nn}}. 
\eeq
Next, we impose an additional requirement of $n\geq \max\{Np^2, p^3, N^2p^5\}$. By \eqref{cond-h}, $h_{\min}\asymp p^{-1}$. 
It implies $n\geq C\frac{p}{h^2_{\min}}(1+\frac{p^2}{N^2}+Nh_{\min})$. We apply the second part of Lemmas~\ref{lem:Enorm2}-\ref{lem:Enorm}
to get
\[
\|G-G_0\|\leq C\Bigl(1+\frac{\sqrt{p}}{N}\Bigr)\sqrt{\frac{np\log(n)}{N}}, \qquad \frac{\|e_j'(G-G_0)\|}{\sqrt{h_j}} \leq C\Bigl( 1 +\frac{p}{N}\Bigr)\sqrt{\frac{np\log(n)}{N}}. 
\]
We plug them into \eqref{proof-thm-noise-0} to get 
\beq \label{proof-thm-noise-2}
\|\Omega\hat{\Xi}_j-\Xi_j\|\leq C\sqrt{h_j}\cdot (1+N^{-1}p)\sqrt{\frac{p\log(n)}{Nn}}. 
\eeq
Finally, we combine \eqref{proof-thm-noise-1}-\eqref{proof-thm-noise-2}. When $n\geq \max\{Np^2, p^3, N^2p^5\}$, both upper bounds in \eqref{proof-thm-noise-1}-\eqref{proof-thm-noise-2} are valid, and we take the minimum of them. When  $n< \max\{Np^2, p^3, N^2p^5\}$, we only use the upper bound in \eqref{proof-thm-noise-1}. It follows that
\[
\|\Omega\hat{\Xi}_j-\Xi_j\|\leq \sqrt{h_j}\cdot C\beta_n\sqrt{\frac{p\log(n)}{Nn}}. 
\]
This proves the claim. \qed

\subsection{Proof of Theorem~\ref{thm:hatR}}
Let $\Omega=\mathrm{diag}(\omega,\Omega^*)$ be the same as in Theorem~\ref{thm:noise}. We can always choose the signs of $\xi_1$ and $\hat{\xi}_1$ such that their first coordinates are both positive. Then, $\omega=1$. By definition, 
\[
\begin{pmatrix}
1\\
r_j
\end{pmatrix} = [\xi_1(j)]^{-1}\Xi_j, \qquad \begin{pmatrix}
1\\
\Omega^*\hat{r}_j
\end{pmatrix} = [\hat{\xi}_1(j)]^{-1}\Omega \hat{\Xi}_j. 
\]
It follows that 
\begin{align} \label{thm-hatR-temp}
\|\Omega^*\hat{r}_j - r_j \| &= \|\frac{1}{\hat{\xi}_1(j)}\Omega\hat{\Xi}_j - \frac{1}{\xi_1(j)}\Xi_j \|\cr
&=  \big\|\frac{1}{\hat{\xi}_1(j)}(\Omega \hat{\Xi}_j - \Xi_j) -  \frac{\hat{\xi}_1(j)-\xi_1(j)}{\hat{\xi}_1(j)}r_j\big\|\cr
&\leq |\hat{\xi}_1(j)|^{-1}\big( \|\Omega \hat{\Xi}_j - \Xi_j\| + \|r_j\|\cdot |\hat{\xi}_1(j)-\xi_1(j)|  \big).  
\end{align}
By Theorem~\ref{thm:noise}, with probability $1-o(n^{-3})$, it holds that $|\hat{\xi}_1(j)-\xi_1(j)|\leq \|\Omega \hat{\Xi}_j - \Xi_j\|\leq \sqrt{h_j}\cdot C\beta_n\sqrt{\frac{p\log(n)}{Nn}}$. At the same time, by Lemma~\ref{lem:R}, $\xi_1(j)\geq C\sqrt{h_j}$; since $\beta_n\sqrt{\frac{p\log(n)}{Nn}}\to 0$, it follows that $\hat{\xi}_1(j)\geq \xi_1(j)/2\geq C\sqrt{h_j}$. Also, by Lemma~\ref{lem:R} again, $\|r_j\|\leq C$. Combining these results, we find that
\[
\|\Omega^*\hat{r}_j - r_j\|\leq C\frac{\|\Omega \hat{\Xi}_j - \Xi_j\|}{\sqrt{h_j}} \leq C\beta_n \sqrt{\frac{p\log(n)}{Nn}}. 
\] 
The claim follows. 
\qed

\subsection{Proof of Lemma~\ref{lem:EigVecPerturb}} \label{subsec:proof-lem-Taylor}
We first prove the claim for the special case of $s=1$ and $k=K$. In this case,  
\[
U_0=U_0^*=[u^0_1, u^0_2,\ldots, u^0_K].
\]   
Let $\Delta_0 = \mathrm{diag}(\delta_1^0,\ldots,\delta_K^0)$ and $\Delta=\mathrm{\diag}(\delta_1,\ldots,\delta_K)$. 
By eigen-decomposition, $U\Delta = GU$. Moreover, $G=G_0+Y=U_0\Delta_0U_0'+Y$. It follows that $U\Delta = U_0\Delta_0(U_0'U)+YU$. Rearranging the terms gives
\beq \label{thm-EV-basic}
U\Delta -YU = U_0(\Delta_0 U_0' U). 
\eeq
In particular, for each $1\leq k\leq K$, \eqref{thm-EV-basic} says that $\delta_ku_k-Yu_k=U_0(\Delta_0 U_0'u_k)$, which means $u_k=(\delta_kI_n - Y)^{-1}U_0(\Delta_0 U_0'u_k)$. We now have
\beq \label{lem-sin-1}
u_k = (I_n -\delta_k^{-1} Y)^{-1}\tilde{u}_k, \qquad \mbox{where}\quad \tilde{u}_k= \delta_k^{-1}U_0(\Delta_0 U_0'u_k).  
\eeq
By Weyl's inequality, $|\delta_k|\geq c\|G_0\|-\|Y\|\geq (2c/3)\|G_0\|\geq 2\|Y\|$. Hence, $\|\delta_k^{-1} Y\|\leq 1/2$. It follows from \eqref{lem-sin-1} that 
\[
\|\tilde{u}_k\| = \|(I_n - \delta_k^{-1}Y)u_k\| \leq \|I_n-\delta_k^{-1}Y\|\|u_k\|\leq (3/2)\|u_k\|\leq 3/2.
\] 
Write $\tilde{U}=[\tilde{u}_1,\ldots,\tilde{u}_K]$ and $Q_k=(I_n -\delta_k^{-1} Y)^{-1}-I_n$. Then, $u_k=\tilde{u}_k+Q_k\tilde{u}_k$. It yields
\beq \label{lem-sin-add1}
|e_j'(u_k-\tilde{u}_k)| =|e_j'Q_k\tilde{u}_k|  \leq \|e_j'Q_k\|\|\tilde{u}_k\|\leq (3/2)\|e_j'Q_k\|.  
\eeq
By definition, $(Q_k+I_n)(I_n- \delta_k^{-1}Y)=I_n$. Expanding the left hand side and canceling $I_n$ on both hand sides, we have  $Q_k = \delta_k^{-1}Y + \delta_k^{-1} Q_kY$. As a result, 
\[
\|e_j'Q_k\| = \|\delta_k^{-1}y_j+\delta_k^{-1}e_j'Q_kY \| \leq \delta_k^{-1}\|y_j\| + \delta_k^{-1}\|e_j'Q_k\|\|Y\|.  
\]
Recalling that $\delta_k^{-1}\|Y\|\leq 1/2$ and $|\delta_k|\geq (2c/3)\|G_0\|$, we immediately have 
\beq \label{lem-sin-add2}
\|e_j'Q_k\|\leq \frac{\delta_k^{-1}\|y_j\|}{1-\delta_k^{-1}\|Y\|}\leq 2\delta_k^{-1}\|y_j\|\leq 3c^{-1}\frac{\|y_j\|}{\|G_0\|}. 
\eeq
Combining \eqref{lem-sin-add1}-\eqref{lem-sin-add2}, $|e_j'(u_k-\tilde{u}_k)|\leq (9/2)c^{-1}\frac{\|y_j\|}{\|G_0\|}$, for each $1\leq k\leq K$. It follows that 
\beq  \label{lem-sin-3}
\|e'_j(U-\tilde{U})\|\leq (9/2)c^{-1}\sqrt{K}\frac{\|y_j\|}{\|G_0\|}. 
\eeq
By \eqref{lem-sin-3} and the triangle inequality (below, the minimums are over orthogonal matrices), 
\begin{align} \label{lem-sin-triangle}
\min_{O}\|e_j'(UO-U_0)\|&\leq \min_{O}\bigl\{ \|e_j'(\tilde{U}O-U_0)\| + \|e_j'(U-\tilde{U})O\|\bigr\}\cr
&= \min_{O}\bigl\{ \|e_j'(\tilde{U}O-U_0)\| + \|e_j'(U-\tilde{U})\|\bigr\}\cr
&\leq \min_O \bigl\{\|e_j'(\tilde{U}O-U_0)\|\bigr\} + \frac{9\sqrt{K}}{2c}\frac{\|y_j\|}{\|G_0\|}. 
\end{align}
It remains to bound the first term in \eqref{lem-sin-triangle}. We apply the sin-theta theorem \citep{sin-theta}:
\[
\|UU'-U_0U_0'\|\leq \frac{\|Y\|}{|\delta^0_K-\delta_{K+1}|}. 
\]
Note that $|\delta^0_{K}|\geq c\|G_0\|$, $\delta^0_{K+1}=0$ and $|\delta_{K+1}-\delta^0_{K+1}|\leq \|Y\|\leq (c/3)\|G_0\|$. It follows that $|\delta^0_K-\delta_{K+1}|\geq (2/3)c\|G_0\|$. 
Therefore, $\|UU'-U_0U_0'\|\leq (3/2)c^{-1}\|G_0\|^{-1}\|Y\|$. Moreover, by Lemma 1 of \cite{cai2018rate}, there is an orthogonal matrix $O$ such that $\| UO -U_0\| \leq \sqrt{2}\|UU'-U_0U_0'\|$. Combining the above, there is an orthogonal matrix $O$ such that 
\beq   \label{lem-sin-4}
\|UO-U_0\|\leq (3/\sqrt{2})c^{-1}\|G_0\|^{-1}\|Y\|.
\eeq
Recall the definition of $\tilde{U}=[\tilde{u}_1,\ldots,\tilde{u}_K]$ in \eqref{lem-sin-1}. We can rewrite
\[
\tilde{U}=U_0(\Delta_0 U_0'U)\Delta^{-1}. 
\]
It follows that
\beq \label{lem-sin-5}
\|e_j'(\tilde{U}O-U_0)\| \leq \|e_j'U_0\|\cdot \|\Delta_0 U_0'U\Delta^{-1}O-I_K\|. 
\eeq
In \eqref{thm-EV-basic}, multiplying both sides by $U_0'$ and noticing that $U_0'U_0=I_K$, we have 
\[
U_0'U\Delta-U_0'YU  =  \Delta_0U_0' U. 
\]
It follows that
\begin{align*}
\|\Delta_0 U_0'U\Delta^{-1}O-I_K\| & = \|(U_0'U\Delta-U_0'YU)\Delta^{-1}O-I_K\|\cr
& = \|(U_0'UO - I_K) -  U_0'YU\Delta^{-1}O\|\cr
&\leq \|U_0'UO-U_0'U_0\| + \|U_0'YU\Delta^{-1}O\|\cr
&\leq \|UO - U_0\| + \|Y\|\|\Delta^{-1}\|\cr
&\leq \bigl(3/\sqrt{2}+3/2\bigr)c^{-1}\|G_0\|^{-1}\|Y\|,
\end{align*}
where in the third line, we use the triangle inequality and $U_0'U_0=I_K$, and in the last line, we use \eqref{lem-sin-4} and $\min_k |\delta_k| \geq c\|G_0\|- \|Y\|\geq (2c/3)\|G_0\|$. Plugging it into \eqref{lem-sin-5}, we have
\beq \label{lem-sin-6}
\|e_j'(\tilde{U}O-U_0)\| \leq (3\sqrt{2}+3/2)c^{-1}\frac{\|Y\|\|e_j'U_0\|}{\|G_0\|}. 
\eeq
We combine \eqref{lem-sin-6} with \eqref{lem-sin-triangle} and note that $3/\sqrt{2}+3/2< 5$. It follows that 
\[
\min_{O}\|e_j'(UO-U_0)\| \leq \frac{5}{c\|G_0\|}\bigl(\|Y\|\|e_j'U_0\| + \sqrt{K}\|y_j\|\bigr).
\]
This proves the claim when $(s, k)=(1, K)$ and $U_0=U_0^*$. 

Next, we consider the general $(s,k)$. In our notation, $U^*_0=[u^0_1,\ldots,u^0_K]$, $U^*=[u_1,\ldots,u_K]$, $U_0=[u^0_s,\ldots,u^0_k]$ and $U=[u_s,\ldots,u_k]$. We have proved
\[
\min_{O^*}\|e_j'(U^*O^*-U^*_0)\| \leq \frac{5}{c\|G_0\|}\bigl(\|Y\|\|e_j'U^*_0\| + \sqrt{K}\|y_j\|\bigr),
\]
where the orthogonal matrix $O^*$ is from \eqref{lem-sin-4}. We divide the eigenvalues of $G_0$ into three groups: group 1 contains $\delta_1^0, \ldots,\delta_{s-1}^0$, group 2 contains $\delta_{s}^0, \ldots,\delta_k^0$, and group 3 consists of $\delta_{k+1}^0, \ldots, \delta^0_K$. By our assumption, there is a gap of $\geq c\|G_0\|$ between the eigenvalues in any two distinct groups and between zero and each of these eigenvalues.  
Therefore, by sin-theta theorem, the orthogonal matrix $O^*$ in \eqref{lem-sin-4} can take the form of a blockwise diagonal matrix, with respect to the above group division. Let $O$ be the diagonal block in $O^*$ that corresponds to the index set $\{s,\ldots,k\}$. Then, the $s$th to $k$th columns of $U^*O^*$ are the same as all columns of $UO$. It follows that $e_j'(UO-U_0)$ is a sub-vector of $e_j'(U^*O^*-U_0^*)$. We thus have 
\[
\min_{O}\|e_j'(UO-U_0)\|\leq \min_{O^*}\|e_j'(U^*O^*-U^*_0)\| \leq \frac{5}{c\|G_0\|}\bigl(\|Y\|\|e_j'U^*_0\| + \sqrt{K}\|y_j\|\bigr). 
\]
This proves the claim for general $(s,k)$. \qed


\subsection{Proof of Lemmas~\ref{lem:PopEigVal}}
Consider the first claim. By Lemma~\ref{lem:M0}, $c_2h_j\leq M_0(j,j)\leq h_j$, for all $1\leq j\leq p$. So, 
\beq \label{M0toH}
1\leq \lambda_{\min}(M_0^{-1}H)\leq \lambda_{\max}(M_0^{-1}H)\leq 1/c_2.
\eeq
Let $s_{\min}(\cdot)$ denote the minimum singular value of a matrix. By basic linear algebra, for a matrix $A$ and a positive definite matrix $B$, $s_{\min}(ABA')\geq \lambda_{\min}(B)\cdot s_{\min}(AA')=\lambda_{\min}(B)\cdot s_{\min}(A'A)$. 
It follows that 
\begin{align*}
s_{\min}(G_0) &\gtrsim s_{\min} \big(M_0^{-1/2}AWW'A'M_0^{-1/2}\big)\cr
& \geq s_{\min} \big(H^{-1/2}AWW'A'H^{-1/2}\big)\cdot s_{\min}(H^{1/2}M^{-1}_0H^{1/2})\cr
& \geq  s_{\min}\big(H^{-1/2}AWW'A'H^{-1/2}\big)\cr
&\geq  \lambda_{\min}(WW')\cdot s_{\min}(A'H^{-1}A)\cr
& = n\lambda_{\min}(\Sigma_W)\lambda_{\min}(\Sigma_A)\cr
& \geq c_2^2 n,  
\end{align*}
where the third line is because of \eqref{M0toH} and the last line follows from the condition \eqref{cond-A}. Similarly, since $\|\Sigma_W\|\leq 1$ and $\|\Sigma_A\|\leq C$, we can derive that 
\[
\lambda_{\max}(G_0)\leq (1/c_2)n\lambda_{\max}(\Sigma_W)\lambda_{\max}(\Sigma_A)\leq Cn.
\]
The first claim follows.  

Consider the second claim. Note that, for any matrices $A$ and $B$, the nonzero eigenvalues of $AB$ are the same as the nonzero eigenvalues of $BA$. Then, the nonzero eigenvalues of $G_0=(1-\frac{1}{N})M_0^{-1/2}AWW'A'M_0^{-1/2}$
are the same as the nonzero eigenvalues of 
\[
(1-\frac{1}{N})n\Theta, \qquad \mbox{where }\Theta \equiv \Sigma_W(A'M_0^{-1}A). 
\]
It suffices to show that 
\beq \label{lem-eigenval-0}
\mbox{gap between the first two eigenvalues of  $\Theta$ is $\geq C$.}
\eeq
In the proof of Lemma~\ref{lem:V}, we have studied this matrix $\Theta$; in the paragraph below \eqref{lem-Vprop-7}, we have argued that, given \eqref{cond-A}, 
\[
\mbox{all entries of $\Theta$ are lower bounded by a constant.}
\]
Now, suppose there is a sequence $\Theta=\Theta^{(n)}$ such that the gap between its first two eigenvalues $\to 0$. Then, since $\|\Theta\|\leq C$, we can select a subsequence $\{n_m\}_{m=1}^\infty$ such that as $m\to\infty$, $\Theta^{(n_m)}\to\Theta_0$ for a fixed $K\times K$ matrix $\Theta_0$. Then, $\Theta_0$ must satisfy that (i) all entries of $\Theta_0$ are strictly positive, and (ii) the first two eigenvalues of $\Theta_0$ are equal. However, such a $\Theta_0$ does not exist, due to the Perron's theorem. We then get a contradiction. This proves \eqref{lem-eigenval-0}, and the second claim follows. \qed

\subsection{Proof of Lemma~\ref{lem:PopEigVec}}
Let $\Xi_j'$ denote the $j$-th row of $\Xi=[\xi_1,\ldots,\xi_K]$, $1\leq j\leq p$. We recall that the matrix $V$ in Lemma~\ref{lem:V} is defined by $\Xi=M_0^{-1/2}AV$. As a result, 
\[
\Xi_j=[M_0(j,j)]^{-1/2}(Va_j),
\]
where $a_j'$ is the $j$-th row of $A$. First, by Lemma~\ref{lem:M0}, we have  $c_2h_j\leq M_0(j,j)\leq h_j$. Second, by Lemma~\ref{lem:V}, $(VV')^{-1}=A'M_0^{-1}A$; so, $\|V\|^2=\lambda^{-1}_{\min}(A'M_0^{-1}A)\leq\lambda^{-1}_{\min}(A'H^{-1}A)\leq c^{-1}_2$, where the last inequality comes from the condition \eqref{cond-A}. Last, $\|a_j\|\leq \|a_j\|_1=h_j$. Combing these results, we obtain:
\[
\|\Xi_j\| \leq \frac{\|V\|\|a_j\|}{\sqrt{M_0(j,j)}}\leq \frac{(1/\sqrt{c_2})\cdot h_j}{\sqrt{c_2h_j}}=\frac{\sqrt{h_j}}{c_2}. 
\]
Then, it follows from the Cauchy-Schwarz inequality that 
\[
\sum_{\ell=1}^K|\xi_\ell(j)|=\|\Xi_j\|_1\leq \sqrt{K}\|\Xi_j\|\leq C\sqrt{h_j}.
\]
This proves the claim. \qed

\subsection{Proof of Lemmas~\ref{lem:Enorm2}-\ref{lem:Enorm}} \label{subsec:proof-lem-Enorm}
We prove the two lemmas together, as they share a common proof structure. 
Each lemma has statement for the general case and a statement for the case where $n$ satisfies an extra condition. 
We primarily focus on the general case. The case with an additional requirement of $n$ can be analyzed in a similar way (deferred to the end of the proofs). 
The proofs rely on properties of the random matrix $Z=D-D_0$, which are given by those technical lemmas in Section~\ref{sec:Z-analysis}. 

We first decompose the quantities to bound in Lemmas~\ref{lem:Enorm2}-\ref{lem:Enorm}. 
Write $Z=[z_1,\ldots,z_n]=[Z_1,\ldots,Z_p]'$. 
From basic properties of multinomial distributions, $\mathrm{Cov}(z_i)=N^{-1}\mathrm{diag}(d_i^0)-N^{-1}d_i^0(d_i^0)'$. As a result, 
\[
E[ZZ']=\sum_{i=1}^n\mathrm{Cov}(z_i)=\frac{n}{N}M_0 - \frac{1}{N}D_0D_0'.
\] 
Then, we can write $G-G_0= E_1+E_2+E_3+E_4$, where
\begin{align*}
E_1 &= \frac{n}{N}M^{-1/2}(M_0-M)M^{-1/2},\cr
E_2 &= M^{-1/2}(D_0Z'+ZD_0')M^{-1/2},\cr
E_3 &= M^{-1/2}(ZZ'-E[ZZ'])M^{-1/2},\cr
E_4 &= (1-\frac{1}{N})\bigl(M^{-1/2}D_0D_0'M^{-1/2}-M_0^{-1/2}D_0D_0'M_0^{-1/2}\bigr). 
\end{align*}
Then, $\|e_j'(G-G_0)\|\leq \sum_{m=1}^4\|e_j'E_m\|$ and $\|G-G_0\|\leq \sum_{m=1}^4 \|E_m\|$. Therefore, to show the claims, we only need to study $\|e_j'E_m\|$ and $\|E_m\|$ for each $1\leq m\leq 4$. We first bound these quantities in the general case,  and then tighten the bounds for $n\geq \max\{Np^2, p^3, N^2p^5\}$. 

Consider $E_1$. By Lemma~\ref{lem:M}, $|M(j,j)-M_0(j,j)|\leq C(Nn)^{-1/2}\sqrt{h_j\log(n)}$ simultaneously for all $j$, with probability $1-o(n^{-3})$.
Moreover, by Lemma~\ref{lem:M0}, $c_2h_j\leq M_0(j,j)\leq h_j$. Since $h_j\geq h_{\min}\gg (Nn)^{-1}\log(n)$, the above suggests that $|M(j,j)-M_0(j,j)|\ll M_0(j,j)$; in particular, $M(j,j)\geq M_0(j,j)/2$. 
Therefore, with probability $1-o(n^{-3})$, for all $1\leq j\leq p$, 
\beq \label{lem-Enorm-j-E1}
\|e_j'E_1\|\leq \frac{n}{N}\frac{|M(j,j)-M_0(j,j)|}{M_0(j,j)/2}\leq \frac{C\sqrt{n\log(n)}}{N\sqrt{Nh_j}}. 
\eeq
Also, with probability $1-o(n^{-3})$, 
\beq \label{lem-Enorm-E1}
\|E_1\| \leq \frac{n}{N}\max_{1\leq j\leq p}\Bigl\{\frac{|M(j,j)-M_0(j,j)|}{M_0(j,j)/2}\Bigr\} \leq \frac{C\sqrt{n\log(n)}}{N\sqrt{Nh_{\min}}}. 
\eeq

Consider $E_2$. Denote by $W_k'$ the $k$-th row of $W$, and recall that $A_k$ is the $k$-th column of $A$, $1\leq k\leq K$. Then, $D_0=\sum_{k=1}^K A_kW_k'$. It follows that
\[
E_2 = \sum_{k=1}^K \bigl[ (M^{-1/2}A_k)(M^{-1/2}ZW_k)' + (M^{-1/2}ZW_k)(M^{-1/2}A_k)'\bigr]. 
\]
As a result, with probability $1-o(n^{-3})$, 
\[
\|E_2\|\leq \sum_{k=1}^K 2\|M^{-1/2}A_k\|\cdot\|M^{-1/2}ZW_k\| \leq C\sum_{k=1}^K \|H^{-1/2}A_k\|\cdot \|M_0^{-1/2}ZW_k\|,
\]
where the last inequality is because $M_0(j,j)\geq c_2h_j$ and $M(j,j)\geq M_0(j,j)/2$ with probability $1-o(n^{-3})$. By Lemma~\ref{lem:cross-term}, $\|M_0^{-1/2}ZW_k\|\leq CN^{-1/2}\sqrt{np\log(n)}$. Moreover, $\sum_{k=1}^K \|H^{-1/2}A_k\|^2=\sum_{k=1}^K\sum_{j=1}^ph_j^{-1}A_k^2(j)\leq  \sum_{k=1}^K\sum_{j=1}^p A_k(j)=K$. It then follows from the Cauchy-Schwarz inequality that $\sum_{k=1}^K \|H^{-1/2}A_k\|\leq K$. As a result, with probability $1-o(n^{-3})$, 
\beq  \label{lem-Enorm-E2}
\|E_2\|\leq CN^{-1/2}\sqrt{np\log(n)}. 
\eeq
In addition, with probability $1-o(n^{-3})$, 
\begin{align}  \label{lem-Enorm-j-E2}
\|e_j'E_2\| &\leq \sum_{k=1}^K \frac{A_k(j)}{\sqrt{M(j,j)}}\|M^{-1/2}ZW_k\| +\sum_{k=1}^K \frac{|Z_j'W_k|}{\sqrt{M(j,j)}}\|M^{-1/2}A_k\|\cr
&\leq C\sqrt{h_j}\max_{1\leq k\leq K}\|M_0^{-1/2}ZW_k\| + \frac{C}{\sqrt{h_j}}\max_{1\leq k\leq K}|Z_j'W_k|\cr
&\leq CN^{-1/2}\sqrt{nph_j\log(n)} + CN^{-1/2} \sqrt{n\log(n)}\cr
&\leq C\sqrt{\frac{n\log(n)}{N}}\bigl(1+\sqrt{ph_j}\bigr),
\end{align}
where the second inequality is due to that $M(j,j)\geq M_0(j,j)/2\geq c_2h_j/2$, $\sum_{k=1}^K A_k(j)=h_j$ and $\sum_{k=1}^K \|M^{-1/2}A_k\|\leq \sqrt{2/c_2}\sum_{k=1}^K\|H^{-1/2}A_k\|\leq K\sqrt{2/c_2}$, and the third inequality follows from Lemma~\ref{lem:cross-term}. 

Consider $E_3$. We have seen that $\|M^{-1/2}M_0^{1/2}\|\leq 2$ with probability $1-o(n^{-3})$. Combining it with Lemma~\ref{lem:ZZ'} gives: with probability $1-o(n^{-3})$,
\beq \label{lem-Enorm-E3}
\|E_3\| \leq 2\|M_0^{-1/2}(ZZ'-E[ZZ'])M_0^{-1/2}\|\leq C\Bigl(\frac{1}{N}+\frac{p}{N^2h_{\min}}\Bigr)\sqrt{np}.
\eeq
Furthermore, by Lemma~\ref{lem:ZZ'-2}, with probability $1-o(n^{-3})$, for all $1\leq j,\ell\leq p$,
\begin{align*}
|E_3(j,\ell)| &= \frac{|Z_j'Z_\ell-E[Z_j'Z_\ell]|}{\sqrt{M(j,j)M(\ell,\ell)}}\leq \frac{C}{\sqrt{h_jh_\ell}}\cdot\Bigl(\frac{1}{N}+\frac{\log(n)}{N^2h_{\min}} \Bigr)\sqrt{nh_jh_\ell\log(n)}\cr
&\leq  C\Bigl(\frac{1}{N}+\frac{\log(n)}{N^2h_{\min}} \Bigr)\sqrt{n\log(n)}. 
\end{align*}
It follows that with probability $1-o(n^{-3})$. 
\beq\label{lem-Enorm-j-E3}
\|e_j'E_3\|\leq C\Bigl(\frac{1}{N}+\frac{\log(n)}{N^2h_{\min}} \Bigr)\sqrt{np\log(n)}. 
\eeq

Consider $E_4$. Since $D_0=\sum_{k=1}^K A_kW_k'$, 
\begin{align*}
E_4 &= (1-\frac{1}{N})\sum_{k,\ell=1}^K (W_k'W_\ell)\bigl(M^{-1/2}A_kA_\ell'M^{-1/2} - M_0^{-1/2}A_kA_\ell'M_0^{-1/2}\bigr)\cr
&= (1-\frac{1}{N})\sum_{k,\ell=1}^K  (W_k'W_\ell)\bigl[M^{-1/2}A_kA_\ell'(M^{-1/2}-M_0^{-1/2}) + (M^{-1/2}- M_0^{-1/2})A_kA_\ell'M_0^{-1/2}\bigr].
\end{align*}
In the proof of \eqref{lem-Enorm-E2}-\eqref{lem-Enorm-j-E2}, we have seen that $\sum_{k=1}^K\|M^{-1/2}A_k\|\leq2\sum_{k=1}^K\|M_0^{-1/2}A_k\|\leq C$. It follows that
\begin{align*}
\|E_4\|& \leq n\sum_{k,\ell=1}^K\bigl(\|M^{-1/2}A_k\|\|(M^{-1/2}-M_0^{-1/2})A_\ell\|+\|M_0^{-1/2}A_\ell\|\|(M^{-1/2}-M_0^{-1/2})A_k\|\bigr)\cr
&\leq CnK\cdot \max_{1\leq k\leq K}\| (M^{-1/2}-M_0^{-1/2})A_k \|. 
\end{align*}
By Lemma~\ref{lem:M} and that $M(j,j)\geq M_0(j,j)/2\geq c_2h/2$, with probability $1-o(n^{-3})$,  
$|[M(j,j)]^{-1/2}-[M_0(j,j)]^{-1/2}|\leq h_j^{-1}(Nn)^{-1/2}\sqrt{\log(n)}$. So, with probability $1-o(n^{-3})$, 
\[
\|(M^{-1/2}- M_0^{-1/2})A_k\| \leq \frac{\sqrt{\log(n)}}{\sqrt{Nn}}\sqrt{\sum_{j=1}^p h_j^{-2}A_k^2(j)}\leq  \frac{C\sqrt{p\log(n)}}{\sqrt{Nn}}.
\]
Combining the above, with probability $1-o(n^{-3})$, 
\beq \label{lem-Enorm-E4}
\|E_4\|\leq CN^{-1/2}\sqrt{np\log(n)}. 
\eeq
Moreover, 
\begin{align} \label{lem-Enorm-j-E4}
\|e_j'E_4\|& \leq \frac{n}{\sqrt{M(j,j)}} \cdot \sum_{k,\ell=1}^KA_k(j)\|(M^{-1/2}-M_0^{-1/2})A_\ell\|\cr
&+n\bigl|\frac{1}{\sqrt{M(j,j)}}-\frac{1}{\sqrt{M_0(j,j)}}\bigr|\cdot \sum_{k,\ell=1}^K A_k(j)\|M_0^{-1/2}A_\ell\|\cr
&\leq C\frac{n}{\sqrt{h_j}}\cdot h_j\cdot \frac{\sqrt{p\log(n)}}{\sqrt{Nn}} + Cn\cdot \frac{\sqrt{\log(n)}}{h_j\sqrt{Nn}}\cdot h_j \cr
&\leq C\sqrt{\frac{n\log(n)}{N}}\bigl(1+\sqrt{ph_j}\bigr). 
\end{align}

We combine the results on $E_1$-$E_4$. By \eqref{lem-Enorm-j-E1}, \eqref{lem-Enorm-j-E2}, \eqref{lem-Enorm-j-E3} and \eqref{lem-Enorm-j-E4}, with probability $1-o(n^{-3})$, 
\begin{align*}
\|e_j'(G-G_0)\|&\leq C\sqrt{\frac{n\log(n)}{N}}\Bigl[1+\sqrt{ph_j} + \frac{1}{N\sqrt{h_j}}+\frac{\sqrt{p}}{\sqrt{N}}\Bigl(1+\frac{\log(n)}{Nh_{\min}}\Bigr)\Bigr]\cr
& \leq C\sqrt{\frac{n\log(n)}{N}}\Bigl[\sqrt{ph_j} + \frac{\sqrt{p}}{\sqrt{N}}\Bigl(1+\frac{p\log(n)}{N}\Bigr)\Bigr], 
\end{align*}
where in the last line we have used $h_j\geq c_1h_{\min}\geq c_1\bar{h}=c_1p^{-1}$. Using $h_j\geq c_1p^{-1}$ again, we find that
\beq \label{lem-Enorm-final1}
\frac{\|e_j'(G-G_0)\|}{\sqrt{h_j}}\leq C\sqrt{\frac{np\log(n)}{N}}
\begin{cases}
1, & \mbox{if } N\geq p\log(n),\\
\frac{p^{3/2}\log(n)}{N^{3/2}},& \mbox{if }N< p\log(n). 
\end{cases}
\eeq
This proves Lemma~\ref{lem:Enorm2}. By \eqref{lem-Enorm-E1}, \eqref{lem-Enorm-E2}, \eqref{lem-Enorm-E3} and \eqref{lem-Enorm-E4}, with probability $1-o(n^{-3})$, 
\begin{align*}
\|G-G_0\|& \leq C\sqrt{np}\Bigl[\frac{\sqrt{\log(n)}}{\sqrt{N}}+\frac{\sqrt{\log(n)}}{N\sqrt{Nph_{\min}}} + \Bigl(\frac{1}{N}+\frac{p}{N^2h_{\min}}\Bigr)\Bigr]\cr
&\leq C\sqrt{np} \Bigl(\frac{\sqrt{\log(n)}}{\sqrt{N}}+\frac{p^2}{N^{2}}\Bigr),
\end{align*}
where the last inequality is because $ph_{\min}\geq c_1$ and $N\geq C\log(n)$. It follows that 
\beq \label{lem-Enorm-final2}
\|G-G_0\| \leq C\sqrt{\frac{np\log(n)}{N}}
\begin{cases}
1, & \mbox{if } N\geq p^{4/3},\\
p^2\cdot N^{-3/2},& \mbox{if }N< p^{4/3}. 
\end{cases}
\eeq
This proves Lemma~\ref{lem:Enorm}. 

The above conclusions hold as long as $nN\gg p\log(n)$. If $n\geq \max\{Np^2, p^3, N^2p^5\}$, we can further improve these results. First, we still use \eqref{lem-Enorm-j-E1}, \eqref{lem-Enorm-j-E2} and \eqref{lem-Enorm-j-E4}, but replace \eqref{lem-Enorm-j-E3} with $\sqrt{p}$ times the bound for $(h_jh_\ell)^{-1/2}|Z_j'Z_\ell - E[Z_j'Z_\ell]|$ suggested by Lemma~\ref{lem:new-ZZ'-2}. It follows that with probability $1-o(n^{-3})$, 
\begin{align} \label{thm-large-7}
\|e_j'(G-G_0)\|&\leq C\sqrt{\frac{n\log(n)}{N}}\Bigl[1+\sqrt{ph_j} + \frac{1}{N\sqrt{h_j}}+\frac{\sqrt{p}}{\sqrt{N}}\Bigl(1+\frac{1}{\sqrt{Nh_{\min}}}\Bigr)\Bigr]\cr
&\leq C\sqrt{\frac{n\log(n)}{N}}\Bigl[\sqrt{ph_j} + \frac{\sqrt{p}}{\sqrt{N}}\Bigl(1+\frac{1}{\sqrt{Nh_{\min}}}\Bigr)\Bigr]\cr
&\leq \sqrt{h_j}\cdot C\sqrt{\frac{np\log(n)}{N}}\Bigl( 1 +\frac{p}{N}\Bigr). 
\end{align}
This proves Lemma~\ref{lem:Enorm2} in the case of $n\geq \max\{Np^2, p^3, N^2p^5\}$. Second, we still use \eqref{lem-Enorm-E1}, \eqref{lem-Enorm-E2} and \eqref{lem-Enorm-E4}, but replace \eqref{lem-Enorm-E3} with the result in Lemma~\ref{lem:new-ZZ'}. It follows that  with probability $1-o(n^{-3})$, 
\begin{align} \label{lem-Enorm-final4}
\|G-G_0\|& \leq C\sqrt{np}\Bigl[\frac{\sqrt{\log(n)}}{\sqrt{N}}+\frac{\sqrt{\log(n)}}{N\sqrt{Nph_{\min}}} + \Bigl(\frac{1}{N} +\frac{1}{N\sqrt{Nh_{\min}}}\Bigr)\Bigr]\cr
&\leq C\frac{\sqrt{np}}{\sqrt{N}} \Bigl(\sqrt{\log(n)}+\frac{1}{N\sqrt{h_{\min}}}\Bigr)\cr
&\leq C\Bigl(1+\frac{\sqrt{p}}{N}\Bigr)\sqrt{\frac{np\log(n)}{N}}. 
\end{align}
This proves Lemma~\ref{lem:Enorm} in the case of $n\geq \max\{Np^2, p^3, N^2p^5\}$.\qed

\section{Rates of convergence of Topic-SCORE} \label{sec:UBproof}

\subsection{Proof of Theorem~\ref{thm:UB}} \label{subsec:supp-UB-proof}
Write $Z=D-D_0$, $A=[A_1,A_2,\ldots,A_K]=[a_1,a_2,\ldots,a_p]'$, and $h_j=\|a_j\|_1$, $1\leq j\leq p$. We define two quantities related to $Z$. For $M=\mathrm{diag}(n^{-1}D{\bf 1}_n)$ and $M_0=\mathrm{diag}(n^{-1}D_0{\bf 1}_n)$, let
\beq \label{def-noiselevel1}
\Delta_1(Z, D_0) \equiv \max_{1\leq j\leq p}\bigl\{ h_j^{-1}|M(j,j)-M_0(j,j)|\bigr\},
\eeq
For $1\leq j\leq p$, let $\hat{\Xi}_j'$ and $\Xi_j'$ be the $j$-th row of $\hat{\Xi}=[\hat{\xi}_1,\hat{\xi}_2,\ldots,\hat{\xi}_K]$ and $\Xi=[\xi_1,\xi_2, \ldots,\xi_K]$, respectively. Let $\mathcal{O}_K$ be the set of all matrices of the form $\Omega = \mathrm{diag}(\omega, \Omega^*)\in\mathbb{R}^{K,K}$, where $\omega\in \{\pm 1\}$ and $\Omega^*\in\mathbb{R}^{(K-1)\times (K-1)}$ is an orthogonal matrix. Let
\beq \label{def-noiselevel2}
\Delta_2(Z, D_0)\equiv \min_{\Omega\in\mathcal{O}_K}\max_{1\leq j\leq p} \bigl\{ h_j^{-1/2}\|\Omega\hat{\Xi}_j -\Xi_j\|\bigr\}. 
\eeq
We also introduce a quantity to describe the error of vertex hunting. 
Given any orthogonal matrix $\Omega^*\in\mathbb{R}^{(K-1)\times(K-1)}$, define 
\beq \label{def-VHerr}
Err_{VH}(\Omega^*)\equiv \min_{\substack{\kappa: \text{ a permutation}\\\text{ on }\{1,\ldots,K\}}}\Bigl\{ \max_{1\leq k\leq K}\|\Omega^*\hat{v}_k^*-v^*_{\kappa(k)}\| \Bigr\}. 
\eeq
The key of the proof is hinged on the following lemma, which is proved in Section~\ref{subsec:proof-lem-method}:
\begin{lem}[Non-stochastic error analysis] \label{lem:method}
Consider model \eqref{pLSI}, where $K$ is fixed and \eqref{cond-h}-\eqref{cond-A} are satisfied. Let $\Delta_1(Z,D_0)$, $\Delta_2(Z,D_0)$ and $Err_{VH}(\Omega^*)$ be as defined in \eqref{def-noiselevel1}-\eqref{def-VHerr}. Let $\hat{A}$ be the estimate from Topic-SCORE. Suppose $\Delta_1(Z,D_0)\leq c$, $\Delta_2(Z,D_0)\leq c$ and for the $\Omega=\mathrm{diag}(\omega,\Omega^*)$ that attains the minimum in $\Delta_2(Z,D_0)$, $Err_{VH}(\Omega^*)\leq c$, where $c>0$ is a sufficiently small constant. Then, up to a permutation of columns of $\hat{A}$, 
\[
\max_{1\leq j\leq p}\biggl\{\frac{\|\hat{a}_j-a_j\|_1}{\|a_j\|_1}\biggr\} \leq C\bigl[ \Delta_1(Z,D_0) + \Delta_2(Z,D_0) + Err_{VH}(\Omega^*)\bigr]. 
\]
\end{lem}

We now use Lemma~\ref{lem:method} to prove Theorem~\ref{thm:UB}. By Lemma~\ref{lem:M} and Theorem~\ref{thm:noise}, there exists an event $E$ such that $P(E)=1-o(n^{-3})$ and that on the event $E$, 
\begin{align*}
\max_{1\leq j\leq p}\bigl\{ h_j^{-1/2}|M(j,j)-M_0(j,j)|\bigr\}& \leq C\sqrt{\frac{\log(n)}{Nn}},\cr
\min_{\Omega\in\mathcal{O}_K}\max_{1\leq j\leq p} \bigl\{ h_j^{-1/2}\|\Omega\hat{\Xi}_j -\Xi_j\|\bigr\} & \leq C\beta_n \sqrt{\frac{p \log(n)}{Nn}}.
\end{align*}
The second inequality gives an upper bound for $\Delta_2(Z,D_0)$. Furthermore,  
by the condition \eqref{cond-h}, $h_j\geq h_{\min}\geq  Cp^{-1}$. We thus have
\[
\Delta_1(Z,D_0)\leq  C\sqrt{p}\cdot \max_{1\leq j\leq p}\bigl\{h_j^{-1/2}|M(j,j)-M_0(j,j)|\bigr\}. 
\]
This yields an upper bound for $\Delta_1(Z,D_0)$. Combining the above, on the event $E$, 
\beq \label{proof-UB-temp1}
\Delta_1(Z,D_0)\leq C\sqrt{\frac{p\log(n)}{Nn}},\qquad \Delta_2(Z,D_0) \leq C\beta_n \sqrt{\frac{p \log(n)}{Nn}}. 
\eeq
It remains to bound $Err_{VH}(\Omega^*)$, where $\Omega=\diag(\omega,\Omega^*)$ attains the minimum in $\Delta_2(Z,D_0)$. If we pick the signs of $\xi_1$ and $\hat{\xi}_1$ such that their first coordinates are positive, then it holds that $\omega=1$ (but $\Omega^*$ still depends on noise and is stochastic).  By Assumption~\ref{cond:VH}, the vertex hunting error is controlled by the noise in $\hat{r}_j$'s: up to a permutation of the $K$ vertices, 
\[
\max_{1\leq k\leq K}\|\Omega^*\hat{v}_k^*-v^*_k\|\leq C\max_{1\leq j\leq p}\|\Omega^*\hat{r}_j-r_j\|.  
\] 
In the proof of Theorem~\ref{thm:hatR} (see \eqref{thm-hatR-temp} and the paragraph below), we have shown that 
\[
\|\Omega^*\hat{r}_j-r_j\|\leq Ch_j^{-1/2}\|\Omega \hat{\Xi}_j-\Xi_j\|, \qquad\mbox{for all }1\leq j\leq p. 
\] 
Combining the above, when $\Omega^*$ is from the $\Omega$ that attains the minimum in $\Delta_2(Z,D_0)$, 
\beq \label{proof-UB-temp2}
Err_{VH}(\Omega^*) \leq C\max_{1\leq j\leq p}\bigl\{ h_j^{-1/2}\|\Omega \hat{\Xi}_j-\Xi_j\|\bigr\}=C\Delta_2(Z, D_0). 
\eeq
We plug \eqref{proof-UB-temp1} and \eqref{proof-UB-temp2} into Lemma~\ref{lem:method}. It gives that, with probability $1-o(n^{-3})$,
\[
\max_{1\leq j\leq p}\biggl\{\frac{\|\hat{a}_j-a_j\|_1}{\|a_j\|_1}\biggr\}\leq C\bigl[ \Delta_1(Z,D_0) + \Delta_2(Z,D_0)\bigr]\leq C\beta_n\sqrt{\frac{p\log(n)}{Nn}}. 
\]
This proves the first claim of Theorem~\ref{thm:UB}. Additionally, 
\[
{\cal L}(\hat{A},A) =\sum_{j=1}^p\|\hat{a}_j-a_j\|_1 \leq \biggl( \sum_{j=1}^p\|a_j\|_1\biggr)\max_{1\leq j\leq p}\biggl\{\frac{\|\hat{a}_j-a_j\|_1}{\|a_j\|_1}\biggr\}, 
\]
where on the right hand side, $\sum_{j=1}^p\|a_j\|_1=\sum_{k=1}^K\|A_k\|_1=K$. It follows immediately that, with probability $1-o(n^{-3})$, 
\[
{\cal L}(\hat{A},A)\leq K\max_{1\leq j\leq p}\biggl\{\frac{\|\hat{a}_j-a_j\|_1}{\|a_j\|_1}\biggr\}\leq C\sqrt{\frac{p\log(n)}{Nn}}. 
\]
This proves the second claim of Theorem~\ref{thm:UB}. \qed

\subsection{Proof of Lemma~\ref{lem:method}} \label{subsec:proof-lem-method}

For notation simplicity, in the proof below, we omit the permutation $\kappa(\cdot)$ in the definition of $Err_{VH}$. From the definitions of $\Delta_1(Z,D_0)$, $\Delta_2(Z,D_0)$ and $Err_{VH}$, there exist $\omega\in \{\pm 1\}$ and a $(K-1)\times (K-1)$ orthogonal matrix $\Omega^*$ such that, letting $\Omega=\mathrm{diag}(\omega,\Omega^*)$, for all $1\leq j\leq p, 1\leq k\leq K$, 
\beq \label{lem-construct-0}
\begin{cases}
\|M(j,j)-M_0(j,j)\|\leq \Delta_1(Z,D_0)\cdot h_j,\\
\|\Omega\hat{\Xi}_j - \Xi_j\| \leq \Delta_2(Z,D_0)\cdot \sqrt{h_j}, \\
\|\Omega^*\hat{v}^*_k-v_k^*\|\equiv Err_{VH}(\Omega^*). 
\end{cases}
\eeq
By Lemma~\ref{lem:R}, all entries of $\xi_1$ are positive, and $\xi_1(j)\geq C\sqrt{h_j}$, $1\leq j\leq p$. At the same time, since $|\omega\hat{\xi}_1(j)-\xi_1(j)|\leq \|\Omega\hat{\Xi}_j-\Xi_j\|\leq \Delta_2(Z,D_0)\sqrt{h_j} $, as long as $\Delta_2(Z,D_0)$ is sufficiently small, all entries of $\omega \hat{\xi}_1$ are also positive. Note that in our method we always choose the sign of $\hat{\xi}_1$ such that its sum is positive. Hence, $\omega=1$ here.  

First, we consider the step of recovering $\Pi$. Note that each $\hat{\pi}_j$ is obtained by truncating and renormalizing $\hat{\pi}_j^*$, where $\hat{\pi}_j^*$ solves the linear equation
\[
\begin{pmatrix}
1 & \ldots & 1\\
\hat{v}_1^* & \ldots & \hat{v}_K^*
\end{pmatrix}\hat{\pi}_j^* = \begin{pmatrix}1\\ \hat{r}_j\end{pmatrix} \;\;\Longleftrightarrow\;\; 
\begin{pmatrix}
1 & \ldots & 1\\
\Omega^*\hat{v}_1^* & \ldots & \Omega^*\hat{v}_K^*
\end{pmatrix}\hat{\pi}_j^* = \begin{pmatrix}1\\ \Omega^*\hat{r}_j\end{pmatrix}. 
\]
It follows that
\[
\hat{\pi}_j^* = \hat{Q}^{-1}\begin{pmatrix}1\\ \Omega^*\hat{r}_j\end{pmatrix}, \;\;\mbox{where}\;\; \hat{Q} = \begin{pmatrix}
1 & \ldots & 1\\
\Omega^*\hat{v}_1^* & \ldots & \Omega^*\hat{v}_K^*
\end{pmatrix}. 
\]
Moreover, by Lemma~\ref{lem:IdealSimplex}, $\pi_j$ is a PMF which satisfies that $\sum_{k=1}^K\pi_j(k)v_k^*=r_j$. Similarly, we have 
\[
\pi_j = Q^{-1}\begin{pmatrix}1\\ r_j\end{pmatrix}, \;\;\mbox{where}\;\; Q = \begin{pmatrix}
1 & \ldots & 1\\
v_1^* & \ldots & v_K^*
\end{pmatrix}. 
\]
Consequently, 
\beq 
\|\hat{\pi}_j^*-\pi_j\|\leq \|\hat{Q}^{-1}\|\|\Omega^*\hat{r}_j-r_j\| + \|\hat{Q}^{-1}-Q^{-1}\|\|r_j\|. 
\eeq 
Since $Q'=[\mathrm{diag}(V_1)]^{-1}V$, we have 
\[
\|Q^{-1}\|^2=\|(Q'Q)^{-1}\|^2\leq (\max_{k}|V_1(k)|)^2\cdot \|(VV')^{-1}\|.
\]
By Lemma~\ref{lem:V}, $\max_k |V_1(k)|\leq C$. It remains to bound $\|(VV')^{-1}\|$. 
By Lemma~\ref{lem:V}, $(VV')^{-1}=A'M_0^{-1}A$; by Lemma~\ref{lem:M0}, $\|A'M_0^{-1}A\|\leq c_2^{-1}\|A'H^{-1}A\|$. Recalling that $a_j'$ is the $j$-th row of $A$, we have  $\|A'H^{-1}A\|\leq \|A'H^{-1}A\|_1=\max_k\sum_{\ell=1}^K \sum_{j=1}^p \|a_j\|_1^{-1}a_j(k)a_j(\ell)\leq\max_k\sum_{\ell=1}^K \sum_{j=1}^pa_j(\ell)=K$. Combining the above gives $\|(VV')^{-1}\|\leq C$. We then have  
 \beq \label{lem-construct-temp1}
 \|Q^{-1}\|\leq C.
 \eeq
Additionally, from the way $Q$ and $\hat{Q}$ are defined, $\|\hat{Q}-Q\|\leq \|\hat{Q}-Q\|_1\leq \sqrt{K}\max_{k}\|\Omega^*\hat{v}_k^*- v_k^*\|$. It follows that 
\beq \label{lem-construct-temp3}
\|\hat{Q}^{-1}-Q^{-1}\|\leq \|\hat{Q}^{-1}\|\|Q^{-1}\|\|\hat{Q}-Q\|\leq C\max_{k}\|\Omega^*\hat{v}_k^*- v_k^*\|.
\eeq
Moreover, by Lemma~\ref{lem:R}, $\|r_j\|\leq C$. Combining the above, we find that
\begin{align}\label{lem-construct-1}
\|\hat{\pi}_j^*-\pi_j\| & \leq C\big(\|\Omega^*\hat{r}_j - r_j\| + \max_{1\leq k\leq K}\|\Omega^*\hat{v}_k^*-v_k^*\|\big)\cr
&\leq C\bigl[\|\Omega^*\hat{r}_j - r_j\| + Err_{VH}(\Omega^*)\bigr]. 
\end{align}
We now use \eqref{lem-construct-1} to study $\hat{\pi}_j$. By definition,  
\[
\hat{\pi}_j=\tilde{\pi}_j^*/\|\tilde{\pi}_j^*\|_1, \qquad \mbox{where}\quad \tilde{\pi}_j^*(k)=\max\{\hat{\pi}^*_j(k), 0\}.
\]
It is seen that
\begin{align*}
\|\hat{\pi}_j-\pi_j\|_1 &\leq \|\hat{\pi}_j-\tilde{\pi}_j^*\|_1  +\|\tilde{\pi}_j^*-\pi_j\|_1\cr
&= \|(1-\|\tilde{\pi}_j^*\|_1)\hat{\pi}_j\|_1 +\|\tilde{\pi}_j^*-\pi_j\|_1\cr  
&= |1-\|\tilde{\pi}_j^*\|_1 | + \|\tilde{\pi}_j^*-\pi_j\|_1. 
\end{align*}
Using the triangle inequality, we have $|1-\|\tilde{\pi}_j^*\|_1|=|\|\pi_j\|_1 - \|\tilde{\pi}_j^*\|_1|\leq \|\pi_j-\tilde{\pi}_j^*\|_1$. Plugging this into the above inequality gives $\|\hat{\pi}_j-\pi_j\|_1\leq 2  \|\tilde{\pi}_j^*-\pi_j\|_1$. Furthermore, since 
all entries of $\pi_j$ are nonnegative, $\|\tilde{\pi}_j^*-\pi_j\|_1\leq \|\hat{\pi}^*_j-\pi_j\|_1\leq \sqrt{K}\|\hat{\pi}_j^*-\pi_j\|$. As a result,
\beq \label{lem-construct-2}
\|\hat{\pi}_j-\pi_j\|_1 \leq 2\sqrt{K}\|\hat{\pi}_j^*-\pi_j\|.
\eeq 
We plug \eqref{lem-construct-1} into \eqref{lem-construct-2} to get 
\beq \label{lem-construct-Pi}
\|\hat{\pi}_j- \pi_j\|_1 \leq C\bigl[\|\Omega^*\hat{r}_j - r_j\| + Err_{VH}(\Omega^*)\bigr]. 
\eeq

Next, we consider the step of recovering $A^*\equiv A\cdot\mathrm{diag}(V_1)$ by 
\[
\hat{A}^*=M^{1/2}\cdot \mathrm{diag}(\hat{\xi}_1)\cdot \hat{\Pi}, 
\]
where $M=\mathrm{diag}(n^{-1}D{\bf 1}_n)$ and $\hat{\Pi}=[\hat{\pi}_1,\ldots,\hat{\pi}_p]'$. By Lemma~\ref{lem:Pi-to-A},
\[
A^* =  M_0^{1/2}\cdot \mathrm{diag}(\xi_1)\cdot \Pi. 
\] 
Fix $j$ and let $(\hat{a}_j^*)'$ and $(a_j^*)'$ be the respective $j$-th row of $\hat{A}^*$ and $A^*$. Then,
\begin{align*}
\|\hat{a}_j^*- a_j^*\|_1 &=  \bigl\| [\sqrt{M(j,j)}\hat{\xi}_1(j)] \hat{\pi}_j - [\sqrt{M_0(j,j)}\xi_1(j) ] \pi_j\bigr\|_1\cr
\leq & \sqrt{M(j,j)}\cdot |\hat{\xi}_1(j)|\cdot \|\hat{\pi}_j - \pi_j\|_1 + \sqrt{M(j,j)}\|\pi_j\|_1 \cdot |\hat{\xi}_1(j)-\xi_1(j)|\cr
& + |\xi_1(j)|\|\pi_j\|_1\cdot|\sqrt{M(j,j)}- \sqrt{M_0(j,j)}|. 
\end{align*}
We plug in \eqref{lem-construct-0} and note $\omega=1$.  
First, $|\hat{\xi}_1(j)-\xi_1(j)|\leq \|\Omega\hat{\Xi}_j-\Xi_j\|\leq \sqrt{h_j}\Delta_2(Z,D_0)$. Second, by Lemma~\ref{lem:R}, $|\xi_1(j)|\leq C\sqrt{h_j}$; furthermore, $|\hat{\xi}_1(j)|\leq 2|\xi_1(j)|\leq C\sqrt{h}_j$. 
Third, by \eqref{lem-construct-0} and Lemma~\ref{lem:M0}, $|\sqrt{M(j,j)}-\sqrt{M_0(j,j)}|\leq C\sqrt{h_j}\cdot \Delta_1(Z,D_0)$ and $M(j,j)\leq 2M_0(j,j)\leq Ch_j$.  As a result,
\beq \label{lem-construct-Astar}
\|\hat{a}_j^*- a_j^*\|_1 \leq Ch_j \cdot \|\hat{\pi}_j - \pi_j\|_1+ Ch_j \bigl[\Delta_1(Z,D_0)+\Delta_2(Z,D_0)\bigr]. 
\eeq

Third, we consider the step of estimating $A$ from renormalizing each column of $\hat{A}^*=[\hat{a}^*_1,\hat{a}^*_2,\ldots,\hat{a}^*_p]'$. Write $\hat{A}=[\hat{A}_1,\ldots,\hat{A}_K]$ and $\hat{A}^*=[ \hat{A}^*_1,\ldots,\hat{A}^*_K]$. Then, 
\[
\hat{A}_k = \|\hat{A}_k^*\|_1^{-1 }\hat{A}_k^*, \qquad 1\leq k\leq K. 
\]
By definition, $A^*=A\cdot \mathrm{diag}(V_1)$. It follows that 
\[
\hat{a}_j(k) = \|\hat{A}_k^*\|_1^{-1}\cdot \hat{a}^*_j(k), \qquad a_j(k) = [V_1(k)]^{-1} \cdot a^*_j(k). 
\]
So, 
\beq \label{lem-construct-temp}
|\hat{a}_j(k)- a_j(k)| \leq \frac{1}{\|\hat{A}_k^*\|_1}|\hat{a}^*_j(k)- a^*_j(k)| + \frac{|\|\hat{A}_k^*\|_1 - V_1(k)|}{\|\hat{A}_k^*\|_1}|a_j(k)|.
\eeq
Since $A^*=A\cdot \mathrm{diag}(V_1)$ and $\|A_k\|_1=1$, we immediately have $\|A^*_k\|_1=V_1(k)$. Then, $|\|\hat{A}_k^*\|_1-V_1(k)| = |\|\hat{A}_k^*\|_1-\|A_k^*\|_1|\leq \|\hat{A}^*_k-A^*_k\|_1\leq \sum_{j=1}^p|\hat{a}^*_j(k)-a^*_j(k)|\leq \sum_{j=1}^p\|\hat{a}_j^*-a_j^*\|_1$. We then apply \eqref{lem-construct-Astar} and use the fact that $\sum_{j=1}^p h_j=K$. It yields 
\beq \label{lem-construct-temp2}
|\|\hat{A}_k^*\|_1-V_1(k)| \leq  C\max_{1\leq i\leq p}\|\hat{\pi}_i-\pi_i\| + C\bigl[\Delta_1(Z,D_0)+\Delta_2(Z,D_0)\bigr]. 
\eeq
In particular, since $V_1(k)\geq C^{-1}$ by Lemma~\ref{lem:V}, we have $\|\hat{A}_k^*\|_1\geq V_1(k)/2\geq C$. Plugging these results into \eqref{lem-construct-temp} and taking the sum over $k$, we find that 
\[
\|\hat{a}_j - a_j\|_1 \leq C \|\hat{a}_j^*-a_j^*\|_1 + C|\|\hat{A}_k^*\|_1 - V_1(k)|\cdot \|a_j\|_1. 
\]
By \eqref{lem-construct-temp2} and that $\|a_j\|_1=h_j$, it follows immediately that  
\begin{align} \label{lem-construct-5}
\|\hat{a}_j - a_j\|_1 
&\leq C\|\hat{a}_j^*-a_j^*\|_1 + Ch_j\cdot \max_{1\leq i\leq p}\|\hat{\pi}_i-\pi_i\|\cr
&+ Ch_j\bigl[\Delta_1(Z,D_0)+\Delta_2(Z,D_0)\bigr]. 
\end{align}

Now, we first plug \eqref{lem-construct-Astar} into \eqref{lem-construct-5}, and then plug in \eqref{lem-construct-Pi}. It yields that
\begin{align} \label{lem-construct-6}
\|\hat{a}_j - a_j\|_1 & \leq Ch_j\cdot \max_{1\leq i\leq p}\|\Omega^*\hat{r}_i-r_i\|\cr
& + Ch_j\bigl[ \Delta_1(Z,D_0)+\Delta_2(Z,D_0)+Err_{VH}(\Omega^*)\bigr]. 
\end{align}
It remains to bound $\max_{1\leq i\leq p}\|\Omega^*\hat{r}_i-r_i\|$. This has been studied in the proof of Theorem~\ref{thm:hatR}. By \eqref{thm-hatR-temp} there, 
\[
\|\Omega^*\hat{r}_j - r_j \| \leq  |\hat{\xi}_1(j)|^{-1}\big( \|\Omega \hat{\Xi}_j - \Xi_j\| + \|r_j\|\cdot |\hat{\xi}_1(j)-\xi_1(j)|  \big). 
\]
By \eqref{lem-construct-0}, $|\hat{\xi}_1(j)-\xi_1(j)|\leq \|\Omega \hat{\Xi}_j - \Xi_j\|\leq \Delta_2(Z,D_0)\sqrt{h_j}$. At the same time, by Lemma~\ref{lem:R}, $\xi_1(j)\geq C\sqrt{h_j}$, which further implies $\hat{\xi}_1(j)\geq \xi_1(j)/2\geq C\sqrt{h_j}$. Also, by Lemma~\ref{lem:R} again, $\|r_j\|\leq C$. Combining these results, we find that
\[
\|\Omega^*\hat{r}_j - r_j\|\leq Ch_j^{-1/2}\|\Omega \hat{\Xi}_j - \Xi_j\| \leq C\Delta_2(Z,D_0). 
\] 
We plug it into \eqref{lem-construct-6} to get
\beq \label{lem-construct-final}
\|\hat{a}_j - a_j\|_1 \leq Ch_j\cdot \bigl[ \Delta_1(Z,D_0)+\Delta_2(Z,D_0)+Err_{VH}(\Omega^*)\bigr]. 
\eeq
The claim follows by noting that $h_j=\|a_j\|_1$. 
\qed

\subsection{Proof of Theorem~\ref{thm:W}}
Each $\hat{w}_i$ is obtained by truncating and re-normalizing the $\hat{w}^*_i$ from \eqref{w-hat}. We start from analyzing $\hat{w}_i^*$. The optimization in \eqref{w-hat} can be re-written as to minimize $\|M^{-1/2}d_i-M^{-1/2}\hat{A}b\|^2$ over $b$, which has an explicit solution:
\[
\hat{w}_i^*=(\hat{A}'M^{-1}\hat{A})^{-1}(\hat{A}'M^{-1}d_i), \qquad 1\leq i\leq n.  
\] 
At the same time, write $d_i^0=\mathbb{E}[d_i]$. Since $d_i^0=Aw_i$, we have
\[
w_i = (A'M_0^{-1}A)^{-1}(A'M_0^{-1}d_i^0), \qquad 1\leq i\leq n. 
\]
It follows that
\begin{align} \label{thm-W-1}
\|\hat{w}_i^*-w_i\|_1 & \leq \|(\hat{A}'M^{-1}\hat{A})^{-1}-(A'M_0^{-1}A)^{-1}\|_1\cdot\|A'M_0^{-1}d_i^0 \|_1\cr
&\qquad +  \| (\hat{A}'M^{-1}\hat{A})^{-1}\|_1\cdot \|A'M_0^{-1}d_i - A'M_0^{-1} d_i^0\|_1 \cr
& \qquad + \| (\hat{A}'M^{-1}\hat{A})^{-1}\|_1\cdot \|\hat{A}'M^{-1}d_i - A'M_0^{-1}d_i\|_1\cr
&\equiv I_1+I_2+I_3. 
\end{align}
Below, we bound each term in \eqref{thm-W-1}. 

Consider $I_1$. By Lemma~\ref{lem:M0},  $c_2h_j\leq M_0(j,j)\leq h_j$. It follows that $A'(M_0^{-1}-H^{-1})A$ and $A'(c_2^{-1}H^{-1}-M_0^{-1})A$ are two positive semi-definite matrices. Therefore, $\lambda_{\min}(A'M_0^{-1}A)\geq \lambda_{\min}(A'H^{-1}A)=\lambda_{\min}(\Sigma_A)\geq c_2$, and $\|A'M_0^{-1}A\|\leq c_2^{-1}\|A'H^{-1}A\|=c_2^{-1}\|\Sigma_A\|$. Moreover,  $\|\Sigma_A\|_1=\max_k\{ \sum_{\ell}\Sigma_A(k,\ell)\}=\max_k \{\sum_\ell a_j(k)\sum_{j} a_j(\ell)/h_j\} = \max_k\{\|A_k\|_1\}=1$. It gives $\|\Sigma_A\|\leq \|\Sigma_A\|_1\leq 1$. Combining the above, we have 
\beq \label{thm-W-2}
\|A'M_0^{-1}A\| \leq c_2^{-1}, \qquad \|(A'M_0^{-1}A)^{-1}\|\leq c_2^{-1}.  
\eeq
Since $d_i^0=Aw_i$, we have $A'M_0^{-1}d_i^0=(A'M_0^{-1}A)w_i$. Then, 
\beq \label{thm-W-2(2)}
\|A'M_0^{-1}d_i^0\|_1\leq \|A'M_0^{-1}A\|_1\|w_i\|_1\leq C.
\eeq
Write $G=A'M_0^{-1}A$ and $\hat{G}=\hat{A}'M^{-1}\hat{A}$. We aim to bound $\|\hat{G}^{-1}-G^{-1}\|$. By Lemma~\ref{lem:M} and Theorem~\ref{thm:UB}, with probability $1-o(n^{-3})$, for all $1\leq j\leq p$,
\begin{align} \label{thm-W-quote}
& \|\hat{a}_j-a_j\|_1 \leq  h_j\cdot C\beta_n(Nn)^{-1/2}\sqrt{p\log(n)}, \cr
& |M(j,j)-M_0(j,j)|  \leq \sqrt{h_j}\cdot C(Nn)^{-1/2}\sqrt{\log(n)}. 
\end{align}
In particular, $\|\hat{a}_j-a_j\|=o(h_j)=o(\|a_j\|_1)$. It follows that $\|\hat{a}_j\|_1\leq 2\|a_j\|_1\leq 2h_j$. Similarly, we have $M(j,j)\geq M_0(j,j)/2\geq c_2h_j/2$. 
Now, we use the above results to bound $\|\hat{G}-G\|_1$.  
By direct calculations, 
\begin{align} \label{thm-W-3}
& \|\hat{G}-G\|_1 = \max_{1\leq k\leq K}\biggl\{ \sum_{\ell=1}^K  \biggl| \sum_{j=1}^p\frac{\hat{a}_j(k)\hat{a}_j(\ell)}{M(j,j)} -  \frac{a_j(k) a_j(\ell)}{M_0(j,j)} \biggr|\biggr\}\cr
\leq &\max_{1\leq k\leq K} \biggl\{\sum_{j=1}^p \sum_{\ell=1}^K \frac{\hat{a}_j(k)|\hat{a}_j(\ell)-a_j(\ell)|}{M(j,j)}\biggr\}+ \max_{1\leq k\leq K} \biggl\{\sum_{j=1}^p \sum_{\ell=1}^K \frac{a_j(\ell)|\hat{a}_j(k)-a_j(k)|}{M(j,j)}\biggr\}\cr
&\qquad  +\max_{1\leq k\leq K}\biggl\{ \sum_{j=1}^p \sum_{\ell=1}^K \frac{a_j(k) a_j(\ell) |M(j,j)- M_0(j,j)|}{M(j,j)M_0(j,j)}\biggr\}\cr
&\leq \sum_{j=1}^p \frac{\|\hat{a}_j\|_1\|\hat{a}_j-a_j\|_1}{M(j,j)} + \sum_{j=1}^p \frac{\|a_j\|_1\|\hat{a}_j-a_j\|_1}{M(j,j)} + \sum_{j=1}^p \frac{\|a_j\|_1^2|M(j,j)- M_0(j,j)|}{M(j,j)M_0(j,j)}\cr
&\leq C\sum_{j=1}^p\|\hat{a}_j-a_j\|_1 + C\sum_{j=1}^p |M(j,j)-M_0(j,j)|\cr
&\leq C\beta_n\sqrt{\frac{p\log(n)}{Nn}} \sum_{j=1}^p h_j + C\sqrt{\frac{\log(n)}{Nn}}\sum_{j=1}^p \sqrt{h_j}\cr
&\leq C\beta_n\sqrt{\frac{p\log(n)}{Nn}}  + C\sqrt{\frac{p\log(n)}{Nn}}, 
\end{align}
where in the last line we use $\sum_jh_j=\sum_j\|a_j\|_1=K$, and by the Cauchy-Schwarz inequality, $\sum_j\sqrt{h_j}\leq \sqrt{p\sum_jh_j}\leq \sqrt{pK}$. It suggests that $\|\hat{G}-G\|\leq \|\hat{G}-G\|_1=o(1)$. We combine it with \eqref{thm-W-2} to get $\|\hat{G}^{-1}\|\leq C$. It follows that, with probability $1-o(n^{-3})$, 
\begin{align} \label{thm-W-4}
\|(\hat{A}'M^{-1}\hat{A})^{-1} &- (A'M_0^{-1}A)^{-1}\|_1 = \|\hat{G}^{-1}-G^{-1}\|_1 \cr
&\leq \sqrt{K}\|\hat{G}^{-1}\|\cdot \|\hat{G}-G\|\cdot \|G^{-1}\| \cr
&\leq C\beta_n\sqrt{\frac{p\log(n)}{Nn}} + C\sqrt{\frac{p\log(n)}{Nn}}. 
\end{align}
By \eqref{thm-W-2(2)}, \eqref{thm-W-4} and $\beta_n\geq 1$, we have that, with probability $1-o(n^{-3})$, 
\beq \label{thm-W-5}
I_1\leq C\beta_n\sqrt{\frac{p\log(n)}{Nn}}. 
\eeq

Consider $I_2$. By our model, $Nd_i\sim \mathrm{Multinomial}(N, d_i^0)$. Introduce $T_{im}\overset{iid}{\sim} \mathrm{Multinomial}(1, d_i^0)$, for $1\leq m\leq N$. Then,  $d_i\overset{(d)}{=}N^{-1}\sum_{m=1}^N T_{im}$. As a result, for for each $1\leq k\leq K$, 
\[
A_k'M_0^{-1}(d_i-d_i^0)\;\; \overset{(d)}{=}\;\; N^{-1}\sum_{m=1}^N X_{im}, \qquad \mbox{where}\quad X_{im}\equiv (M_0^{-1}A_k)'\bigl(T_{im}-\mathbb{E}[T_{im}]\bigr). 
\]
Note that $X_{i1},\ldots,X_{iN}$ are iid random variables, with $|X_{im}|\leq \|M_0^{-1}A_k\|_\infty \|T_{im}-\mathbb{E}[T_{im}]\|_1\leq \|M_0^{-1}A_k\|_\infty$. Since $M_0(j,j)\geq c_2h_j$ (by Lemma~\ref{lem:M0}) and $A_k(j)\leq\|a_j\|_1\leq h_j$, we have
\[
|X_{im}|\leq c_2^{-1}, \qquad\mbox{for all }1\leq m\leq N. 
\]
By Hoeffding's inequality, for any $\delta\in (0,1)$, with probability $1-\delta/K$, 
\[
|A_k'M_0^{-1}(d_i-d_i^0)|\leq CN^{-1/2}\sqrt{\log(K/\delta)}. 
\]
Combining it with the probability union bound and the fact that $K$ is fixed, we have: with probability $1-\delta$,
\beq \label{thm-W-6}
\|A'M_0^{-1}(d_i-d_i^0)\|_1\leq CN^{-1/2}\sqrt{\log(1/\delta)}. 
\eeq
Furthermore, in the paragraph below \eqref{thm-W-3}, we have shown that $\|(\hat{A}'M^{-1}\hat{A})^{-1}\|\leq C$. We plug this inequality and \eqref{thm-W-6} into $I_2$ to get
\beq \label{thm-W-7}
I_2\leq C\sqrt{\frac{\log(1/\delta)}{N}},\qquad \mbox{with probability }1-\delta. 
\eeq

Consider $I_3$. We have seen that $M(j,j)\geq M_0(j,j)\geq c_2h_j/2$ and $\|\hat{a}_j\|_1\leq 2\|a_j\|_1\leq 2h_j$. Moreover, from how the corpus matrix $D$ is defined, each of its columns is self-normalized, i.e., $\sum_{j=1}^p D(j,i)=1$. By direct calculations,  
\begin{align} 
 \|\hat{A}'M^{-1}d_i-A'M_0^{-1}d_i\|_1 &= \sum_{k=1}^K \left|\sum_{j=1}^p\biggl[\frac{\hat{a}_j(k)}{M(j,j)}-\frac{a_j(k)}{M_0(j,j)}\biggr]D(j,i) \right|\cr
 &\leq \Bigl[\sum_{j=1}^pD(j,i)\Bigr] \max_{1\leq j\leq p}\biggl\{ \sum_{k=1}^K \Bigl| \frac{\hat{a}_j(k)}{M(j,j)}-\frac{a_j(k)}{M_0(j,j)}\Bigr| \biggr\}\cr
 &\leq \max_{1\leq j\leq p}\biggl\{ \sum_{k=1}^K \Bigl| \frac{\hat{a}_j(k)}{M(j,j)}-\frac{a_j(k)}{M_0(j,j)}\Bigr| \biggr\}\cr
&\leq \max_{1\leq j\leq p} \Bigl\{h_j^{-1}\|\hat{a}_j-a_j\|_1 +h_j^{-1}|M(j,j)-M_0(j,j)|\Bigr\} \cr
&\leq C\beta_n\sqrt{\frac{p\log(n)}{Nn}} + C\sqrt{\frac{\log(n)}{Nn}}\max_{1\leq j\leq p}\{h_j^{-1/2}\}\cr
&\leq C\beta_n\sqrt{\frac{p\log(n)}{Nn}}  + C\sqrt{\frac{p\log(n)}{Nn}}, 
\end{align}
where the fifth line is from \eqref{thm-W-quote}. We combine it with the fact of $\|(\hat{A}'M^{-1}\hat{A})^{-1}\|\leq C$ and $\beta_n\geq 1$ to get, with probability $1-o(n^{-3})$,
\beq \label{thm-W-8}
I_3\leq C\beta_n\sqrt{\frac{p\log(n)}{Nn}}. 
\eeq

We now plug \eqref{thm-W-5}, \eqref{thm-W-7} and \eqref{thm-W-8} into \eqref{thm-W-1}. It follows that, 
\beq \label{thm-W-9}
\|\hat{w}_i^*-w_i\|_1 \leq C \left(\beta_n\sqrt{\frac{p\log(n)}{Nn}} + \sqrt{\frac{\log(1/\delta)}{N}}\right), \quad\mbox{with probability }1-\delta. 
\eeq

It remains to bound $\|\hat{w}_i-w_i\|_1$ in terms of $\|\hat{w}_i^*-w_i\|_1$. Let $\tilde{w}_i$ be the vector obtained by setting the negative entries in $\hat{w}_i^*$ to zero. Since $w_i$ is a nonnegative vector, we have
\beq \label{thm-W-10}
\|\tilde{w}_i-w_i\|_1\leq \|\hat{w}^*_i-w_i\|_1. 
\eeq
Note that $\hat{w}_i=\tilde{w}_i/\|\tilde{w}_i\|_1$. For each $1\leq k\leq K$, we have 
\begin{align*}
|\hat{w}_i(k)-w_i(k)| & \leq |\tilde{w}_i(k)-w_i(k)| + \tilde{w}_i(k)\Bigl| \frac{1}{\|\tilde{w}_i\|_1}-1 \Bigr| \cr
&=   |\tilde{w}_i(k)-w_i(k)|  + \hat{w}_i(k) \bigl|1-\|\tilde{w}_i\|_1\bigr|\cr
&=   |\tilde{w}_i(k)-w_i(k)|  + \hat{w}_i(k) \bigl|\|w_i\|_1-\|\tilde{w}_i\|_1\bigr|\cr
&\leq  |\tilde{w}_i(k)-w_i(k)|  + \hat{w}_i(k) \|w_i-\tilde{w}_i\|_1
\end{align*}
Summing over $k$ on both sides and using the self-normalization of $\|\hat{w}_i\|_1=1$, we have
\[
\|\hat{w}_i-w_i\|_1\leq \|\tilde{w}_i-w_i\|_1 + \|\hat{w}_i\|_1 \|\tilde{w}_i-w_i\|_1\leq 2\|\tilde{w}_i-w_i\|_1. 
\]
We combine it with \eqref{thm-W-9} to get
\beq \label{thm-W-11}
\|\hat{w}_i-w_i\|_1 \leq 2\|\hat{w}_i^*-w_i\|_1.  
\eeq
The claim follows by plugging \eqref{thm-W-9} into \eqref{thm-W-11}. \qed

\subsection{Proof of Theorem~\ref{thm:K}}
Define $G$ and $G_0$ in the same way as in \eqref{G0G}: 
\[
G \equiv M^{-1/2}DD'M^{-1/2}-\frac{n}{N}I_p, \qquad G_0 = (1-\frac{1}{N})M_0^{-1/2}D_0D_0'M_0^{-1/2}.
\]
Let $\hat{\lambda}_k$ and $\lambda_k$ be the $k$th largest eigenvalue of $G$ and $G_0$, respectively. By definition,
\beq \label{proof-estK-1}
\hat{\sigma}_k^2 = \hat{\lambda}_k + \frac{n}{N}, \qquad \mbox{for } 1\leq k\leq (p\wedge n). 
\eeq
First, by Weyl's inequality and Lemma~\ref{lem:Enorm}, $\max_{k}|\hat{\lambda}_k-\lambda_k|\leq \|G-G_0\|\leq C\beta_n\sqrt{\frac{np\log(n)}{N}}$, with probability $1-o(n^{-3})$. Next, since $G_0$ has a rank $K$, it holds that $\lambda_{k}=0$ for $k\geq K+1$. Last, by Lemma~\ref{lem:PopEigVal}, $\lambda_k\geq Cn$, for $1\leq k\leq K$. Therefore, with probability $1-o(n^{-3})$,  
\beq \label{proof-estK-2}
\hat{\lambda}_k
\begin{cases}
\geq Cn, &\mbox{for }1\leq k\leq K, \cr
\leq C\beta_n\sqrt{\frac{np\log(n)}{N}}, &\mbox{for }k\geq K+1.  
\end{cases}
\eeq
Write $T_n=\beta_n\sqrt{\frac{np\log(n)}{N}}\cdot g_n$. It follows from the conditions on $g_n$ that $\beta_n\sqrt{\frac{np\log(n)}{N}}\ll T_n\ll n$. Combining it with \eqref{proof-estK-1}-\eqref{proof-estK-2}, we have
\[
\hat{\sigma}^2_k - \frac{n}{N}
\begin{cases}
\gg T_n, &\mbox{for }1\leq k\leq K, \cr
=o(T_n), &\mbox{for }k\geq K+1,  
\end{cases}
\]
The claim follows immediately. 
\qed

\section{Proof of Theorem~\ref{thm:LB} (lower bound)}

At the heart of the proof of Theorem~\ref{thm:LB} is the {\it least favorable configurations}, which live in a smaller parameter space: Fixing constants $\gamma_1,\gamma_2\in (0,1/K)$ and a weight vector $\eta^*\in\mathbb{R}^K$ that is in the interior of the standard simplex, define ($w_i$ is called a pure column of $W$ for topic $k$ if $w_i(k)=1$)
\begin{align*}
& \;\; \Phi^*_{n,N,p}(K,c_1,c_2,\gamma_1,\gamma_2, \eta^*)\cr
 = & \left\{
\begin{array}{lc}
\mbox{$(A,W)$}:&  \mbox{\eqref{cond-h}-\eqref{cond-A} are satisfied; $A$ has $\geq \gamma_1p$ anchor rows for each}\\
& \mbox{topic; $W$ has $\geq \gamma_2n$ pure columns for each topic; for }\cr
&\mbox{any non-anchor row of $A$, $\|\frac{a_j}{\|a_j\|_1} - \eta^*\|\leq C\sqrt{p/(Nn)}$ }
\end{array} \right\}. 
\end{align*}

\begin{lem}[Minimax lower bound for a smaller class] \label{lem:LB2}
Suppose the conditions of Theorem~\ref{thm:LB} hold, except that $(A, W)$ live in $\Phi^*_{n, N, p}(K, c_1, c_2,\gamma_1,\gamma_2,\eta^*)$ for given constants  $0 < c_1, c_2 < 1$ and $0<\gamma_1,\gamma_2<1/K$ and a given positive vector $\eta^*\in\mathbb{R}^K$ where $\|\eta^*\|_1=1$, $\eta_1^*,\ldots,\eta_K^*$ are distinct, and $1/(2K)\leq \eta_K^*\leq 3/(2K)$ for $1\leq k\leq K$. There exist constants $C_0>0$ and $\delta_0\in (0,1)$ such that, for all large enough $n$, 
\[
\inf_{\hat{A}}\sup_{(A,W)\in \Phi^*_{n,N,p}(K,c_1,c_2,\gamma_1,\gamma_2, \eta^*)}\mathbb{P}\biggl( \mathcal{L}(\hat{A},A)\geq C_0\sqrt{\frac{p}{Nn}}\biggr) \geq \delta_0. 
\]
\end{lem}

\noindent
Since the lower bound can only increase when the parameter space is enlarged, Theorem~\ref{thm:LB} follows immediately from Lemma~\ref{lem:LB2}.

\subsection{Proof of Lemma~\ref{lem:LB2}}
We need a useful lemma, which is proved in Section~\ref{subsec:lem-proof-KL}. 
\begin{lem}[Kullback-Leibler divergence] \label{lem:KL}
Let $D_0,\tilde{D}_0$ be two $p\times n$ matrices such that each column of them is a weight vector. Under Model \eqref{pLSI}, let $\mathbb{P}$ and $\tilde{\mathbb{P}}$ be the probability measures associated with $D_0$ and $\tilde{D}_0$, respectively, and let $KL(\tilde{\mathbb{P}}, \mathbb{P})$ be the Kullback-Leibler divergence between them. Suppose $D_0$ is a positive matrix. Let $\delta=\max_{1\leq j\leq p,1\leq i\leq n}\frac{|\tilde{D}_0(j,i)-D_0(j,i)|}{D_0(j,i)}$ and assume $\delta<1$. There exists a universal constant  $C>0$ such that 
\[
KL(\tilde{\mathbb{P}},\mathbb{P})\leq (1+C\delta)N\sum_{i=1}^n\sum_{j=1}^p \frac{|\tilde{D}_0(j,i)-D_0(j,i)|^2}{D_0(j,i)}. 
\]
\end{lem}

We now show the claim. Write $\Phi_n^*=\Phi^*_{n, N, p}(K, c_1, c_2,\gamma_1,\gamma_2,\eta^*)$ for short. 
Our proof is based a standard argument in minimax analysis. By Theorem 2.5 of \cite{tsybakov2009introduction}: If there exist $(A^{(0)}, W^{(0)})$, $ (A^{(1)}, W^{(1)})$, $\ldots$, $(A^{(J)}, W^{(J)})\in\Phi_{n}^*$ such that:
\begin{itemize}
\item[(i)] $\mathcal{L}(A^{(j)}, A^{(k)})\geq 2C_0 \sqrt{\frac{p}{Nn}}$ for all $0\leq j\neq k\leq J$, 
\item[(ii)] $KL(\mathcal{P}_j,\mathcal{P}_0)\leq \beta \log(J)$ for all $1\leq j\leq J$,  
\end{itemize} 
where $C_0>0$, $\beta\in (0, 1/8)$, and $\mathcal{P}_j$ denotes the probability measure associated with $(A^{(j)}, W^{(j)})$, then
\[
\inf_{\hat{A}}\sup_{(A,W)\in\Phi_{n,N,p}(K,c)}\mathbb{P}\Bigl( \mathcal{L}(\hat{A},A)\geq C_0\sqrt{\tfrac{p}{Nn}}\Bigr) \geq \tfrac{\sqrt{J}}{1+\sqrt{J}}\Big(1-2\beta-\sqrt{\tfrac{2\beta}{\log(J)}}\Big). 
\]
As long as $J\to\infty$ as $(n,N,p)\to\infty$, the right hand side is lower bounded by a constant, and the claim follows. 

What remains is to construct $(A^{(0)}, W^{(0)}), (A^{(1)}, W^{(1)}), \ldots, (A^{(J)}, W^{(J)})$ that  are in $\Phi_n^*$ and satisfy (i) and (ii). First, we construct $(A^{(0)},W^{(0)})$. Write $A^{(0)}=A$ and $W^{(0)}=W$ for short. In all steps below, for an index $j$ and real values $a$ and $b$, the inequality $a< j\leq b$ means that we first round $a$ and $b$ to the closest integers $a^*$ and $b^*$ and then let $a^*< j\leq b^*$. 
Recall that $e_1,\ldots,e_K$ are the standard basis vectors of $\mathbb{R}^K$. We construct $W=[w_1,\ldots,w_n]$ by
\beq \label{thm-LB-construct1}
w_i = e_k, \qquad \mbox{for all $1\leq k\leq K$ and } (k-1)\frac{n}{K}< i\leq k\frac{n}{K}. 
\eeq  
To construct $A$, we recall that $\Phi_n^*$ is defined using a vector $\eta^*$. We first consider
\[
\eta = K\cdot \eta^*.  
\]
Write $\eta=(\eta_1,\eta_2,\ldots,\eta_K)'$. It can be shown that
\begin{itemize}
\item $\eta_1,\eta_2,\ldots,\eta_K\in [1/2,3/2]$, and they are distinct from each other; 
\item $\bar{\eta}\equiv (1/K)\sum_{k=1}^K\eta_k=1$;
\end{itemize}
For two constants $b_1>0$ and $b_2\in (0,1)$ to be determined, we construct $A=[A_1,\ldots,A_K]=[a_1,\ldots,a_p]'$ as follows. Introduce
\[
\theta_k = \frac{1}{Kb_1b_2}[1-(1-b_1b_2)\eta_k], \qquad 1\leq k\leq K.  
\]
Note that $\eta_k\leq 3/2$ and $\bar{\eta}=1$. Hence, when $3(1-b_1b_2)/2<1$, it holds that $\theta_1,\ldots,\theta_K$ are positive, they are distinct from each other, and 
$\sum_{k=1}^K \theta_k=1$. We construct the first $b_2p$ rows of $A$ as follows: For $1\leq k\leq K$, 
\beq  \label{thm-LB-construct2}
a_j = \frac{b_1K}{p} e_k,\qquad  (\theta_1+\ldots+\theta_{k-1}) b_2p< j\leq (\theta_1+\ldots+\theta_{k}) b_2p. 
\eeq
We then construct the remaining $(1-b_2)p$ rows of $A$ as follows:
\beq \label{thm-LB-construct3}
a_j = \frac{1-b_1b_2}{(1-b_2)p}\cdot (\eta_1,\eta_2,\ldots,\eta_K)', \qquad b_2p<j\leq p.   
\eeq
It can be verified that each column of $A$ has a sum of $1$. The next lemma confirms that we can find $(b_1, b_2)$ to make the $(A^{(0)},W^{(0)})$ constructed above to belong to $\Phi_n^*$. It is proved in Section~\ref{subsec:lem-proof-LBconstruct1}. 
\begin{lem} \label{lem:LBconstruct1}
Given any $c_1,c_2,\gamma_1,\gamma_2\in (0,1)$ and $\eta^*\in \mathbb{R}^K$ as in Lemma~\ref{lem:LB2}, there always exist $b_1>0$ and $b_2\in (0,1)$ such that $(A,W)$ constructed from \eqref{thm-LB-construct1}-\eqref{thm-LB-construct3} is contained in $\Phi^*_{n,N,p}(K,c_1,c_2,\gamma_1,\gamma_2,\eta^*)$. 
\end{lem}

Next, we construct $(A^{(1)}, W^{(1)}), \ldots, (A^{(J)}, W^{(J)})$. Recall that $(b_1,b_2)$ are the same as above. Let $p_1$ be the largest integer such that $p_1\leq (1-b_2)p$. 
Let $m=p_1/2$ if $p_1$ is even and $m=(p_1-1)/2$ if $p_1$ is odd. The Varshamov-Gilbert bound for the packing numbers \citep[Lemma 2.9]{tsybakov2009introduction} guarantees that there exist $J\geq 2^{m/8}$ and $\omega^{(0)},\omega^{(1)},\ldots,\omega^{(J)}\in \{0,1\}^{m}$ such that $\omega^{(0)}=(0,\ldots,0)$ and  
\[
\sum_{j=1}^m 1\{\omega^{(s)}_j\neq \omega^{(\ell)}_j\} \geq \frac{m}{8}, \qquad \mbox{for any }0\leq s\neq \ell\leq J.  
\]
Let $\alpha_n=\frac{C_1}{K}\frac{1}{\sqrt{Nnp_1}}$ for a positive constant $C_1$ to be determined. We construct $A^{(1)},\ldots,A^{(J)}$ as follows: 
\[
A_k^{(s)}  = A_k^{(0)} +  \alpha_n \begin{cases}
({\bf 0}_{p-p_1},\; \omega^{(s)},\; -\omega^{(s)})', & \mbox{if $p_1$ is even},\\
({\bf 0}_{p-p_1},\; \omega^{(s)},\; -\omega^{(s)}, 0)',  & \mbox{if $p_1$ is odd}, 
\end{cases}
\quad 1\leq k\leq K,1\leq s\leq J, 
\]
where ${\bf 0}_{p-p_1}$ is a zero vector of length $(p-p_1)$. 
It is easy to see that $A^{(s)}$ is still a valid topic matrix. 
We then let $W^{(s)}=W^{(0)}$ for all $1\leq s\leq J$. The following lemma is proved in Section~\ref{subsec:lem-proof-LBconstruct2}. 
\begin{lem} \label{lem:LBconstruct2}
Given any $c_1,c_2,\gamma_1,\gamma_2\in (0,1)$ and $\eta^*\in \mathbb{R}^K$ as in Lemma~\ref{lem:LB2}, for the $b_1>0$ and $b_2\in (0,1)$in Lemma~\ref{lem:LBconstruct1},  the $\{(A^{(s)},W^{(s)})\}_{1\leq s\leq J}$ constructed above are all contained in $\Phi^*_{n,N,p}(K,c_1,c_2,\gamma_1,\gamma_2,\eta^*)$. 
\end{lem}

Last, we check that (i)-(ii) are satisfied. For any $0\leq s\neq \ell\leq J$, we have $\mathcal{L}(A^{(s)}, A^{(\ell)})=\sum_{k=1}^K \|A^{(s)}_k-A^{(\ell)}_k\|_1$, without minimizing over permutation of columns. This is because the first $b_2p$ rows are anchor rows and they are the same for both matrices. It follows that 
\beq
\mathcal{L}(A^{(s)}, A^{(\ell)})=\alpha_n \cdot 2K\|\omega^{(s)}-\omega^{(\ell)}\|_1\geq \frac{1}{4}K\alpha_nm\gtrsim \tfrac{C_1\sqrt{1-b_2}}{8}\sqrt{\tfrac{p}{Nn}},
\eeq
where we have used that $\|\omega^{(s)}-\omega^{(\ell)}\|_1\geq m/8$ and $m\gtrsim p_1/2\gtrsim (1-b_2)p/2$. So (i) is satisfied for $C_0=\frac{C_1}{16}\sqrt{1-b_2}$. 

We then verify (ii). Fix $s$ and write $W^{(0)}=W_*$ for short. By construction, $W^{(s)}=W_*$. 
The key of characterizing the KL distance is to study the matrix $D_0^{(s)}-D_0^{(0)}=(A^{(s)}-A^{(0)})W_*$. Let $F\subset\{1,2,\ldots,m\}$ be the support of $\omega^{(s)}$. Denote by $(a_j^{(s)})'$ and $(a_j^{(0)})'$ the $j$-th row of $A^{(0)}$ and $A^{(s)}$, respectively. It is seen that 
\[
a_j^{(s)} - a_j^{(0)} =\begin{cases}
(\alpha_n,\alpha_n,\ldots,\alpha_n), & j=p-p_1+i \mbox{ for some $i\in F$},\\
-(\alpha_n,\alpha_n,\ldots,\alpha_n), & j=p-p_1+m+i, \mbox{ for some $i\in F$},\\
(0,0,\ldots,0), & \mbox{otherwise}. 
 \end{cases}
\] 
Therefore, the $j$-th row of $D_0^{(s)}-D_0^{(0)}$ is either a zero vector or $\pm \alpha_n$ times the sum of the rows in $W_*$. By direct calculations, 
\[
\sum_{i=1}^n\sum_{j=1}^p |D_0^{(s)}(j,i)-D_0^{(0)}(j,i)|^2 = n\alpha_n^2 \cdot 2\|\omega^{(s)}-\omega^{(0)}\|_1\leq np_1\alpha_n^2.
\]
Additionally, each entry of $D_0^{(0)}$ is lower bounded by $C^{-1}p^{-1}$ from the construction above, and $\max_{i,j}\tfrac{|D_0^{(s)}(j,i)-D_0^{(0)}(j,i)|}{D^{(0)}_0(j,i)}=O(p\alpha_n)=O(\sqrt{\tfrac{p}{Nn}})=o(1)$. We plug the above results into Lemma~\ref{lem:KL} and obtain that   
\beq
KL(\mathcal{P}_j,\mathcal{P}_0)\leq [1+o(1)]Np\sum_{i=1}^n\sum_{j=1}^p |D_0^{(s)}(j,i)-D_0^{(0)}(j,i)|^2 \lesssim  \frac{C_1^2}{K}p. 
\eeq
At the same time, $\beta\log(J)\geq \beta\tfrac{m}{8}\log(2)\gtrsim \tfrac{\beta(1-b_2)\log(2)}{16}p$. So (ii) is satisfied if we choose $C_1$ appropriately small. The proof is now complete. \qed

\subsection{Proof of Lemma~\ref{lem:KL}} \label{subsec:lem-proof-KL}
Write for short $a_{ji}=D_0(j,i)$, $\tilde{a}_{ji}=\tilde{D}_0(j,i)$, and $\delta_{ji}=\frac{\tilde{a}_{ji}-a_{ji}}{a_{ji}}$. Then, $\delta=\max_{i,j}|\delta_{ji}|$. Note that the KL-divergence between $\mathrm{Multinomial}(N, \eta_1)$ and $\mathrm{Multinomial}(N,\eta_2)$ is equal to $N\sum_{j=1}^p\eta_{1j}\log(\eta_{1j}/\eta_{2j})$. It follows that
\[
KL(\tilde{\mathbb{P}},\mathbb{P}) = N  \sum_{i=1}^n\sum_{j=1}^p \tilde{a}_{ji}\log(1+\delta_{ji}). 
\]
By Taylor expansion, $\log(1+\delta_{ji})\leq\delta_{ji}-\frac{1}{2}\delta^2_{ji}+C\delta^3_{ji}$ for a constant $C>0$. Moreover, since each column of $D_0$ and $\tilde{D}_0$ has a sum of $1$, we have $\sum_{i,j}a_{ji}=\sum_{i,j}\tilde{a}_{ji}$, which implies that $\sum_{i,j}a_{ji}\delta_{ji}=0$. As a result,
\begin{align*}
KL(\tilde{\mathbb{P}},\mathbb{P}) &\leq N\sum_{i,j} (a_{ji}+a_{ji}\delta_{ji})(\delta_{ji}-\frac{1}{2}\delta^2_{ji}+C\delta^3_{ji})\cr
&=N \sum_{i,j}a_{ji}\delta_{ji} + N\sum_{i,j}a_{ji}\delta^2_{ji} - \frac{N}{2}\sum_{i,j}a_{ji}\delta^2_{ji} + O\Big(N\sum_{i,j}a_{ij}\delta^3_{ji}\Big) \cr
& = \frac{N}{2}\sum_{i,j}a_{ji}\delta^2_{ji}  + O\Big(\delta \cdot N\sum_{i,j}a_{ij}\delta^2_{ji}\Big). 
\end{align*}
Then, Lemma~\ref{lem:KL} follows.  \qed

\subsection{Proof of Lemma~\ref{lem:LBconstruct1}} \label{subsec:lem-proof-LBconstruct1}
Without loss of generality, we assume $n/K$, $b_2p\theta_k$, and $(1-b_2)p$ are all integers. If some of them are not integers, the expressions of $\Sigma_W$ and $\Sigma_A$ only change by $O(1/p)$ in individual entries, and the claims continue to hold. 

We first calculate the matrices $\Sigma_W$ and $\Sigma_A$. We claim that 
\beq \label{lem-LBconstruct-1}
\Sigma_W = K^{-1} I_K,\qquad \Sigma_A =  I_K - (1-b_1b_2)\cdot [\mathrm{diag}(\eta) - K^{-1}\eta\eta'].
\eeq
The first equality follows directly from the way $W$ is constructed. 
To show the second equality, we note that 
\[
a_j = \frac{1}{p}\begin{cases}
Kb_1 \cdot e_k, & (\theta_1+\ldots+\theta_{k-1})b_2p<j\leq (\theta_1+\ldots+\theta_{k})b_2p,\\
\frac{1-b_1b_2}{1-b_2}(\eta_1,\eta_2,\ldots,\eta_K)', & b_2p<j\leq p. 
\end{cases}
\]
Write $G=H^{-1/2}A$, where $H(j,j)=\|a_j\|_1$. Denote by $g_j'$ the $j$-th row of $G$. By direct calculations and the fact that $\bar{\eta}=1$, we have  
\[
g_j = \frac{1}{\sqrt{p}}\left\{\begin{array}{ll}
\sqrt{Kb_1}\cdot e_k, &  (\theta_1+\ldots+\theta_{k-1})b_2p< j\leq (\theta_1+\ldots+\theta_{k}) b_2p, \\
\sqrt{\frac{1-b_1b_2}{(1-b_2)K}}\cdot (\eta_1, \ldots, \eta_K)', & b_2p<j\leq p.
\end{array}\right.
\]
Since $\Sigma_A =A'H^{-1}A= \sum_{j=1}^p g_jg_j'$, by direct calculations, we have  
\beq \label{lem-LBconstruct-4}
\Sigma_A = Kb_1b_2\cdot \mathrm{diag}(\theta_1,\ldots,\theta_K) + K^{-1}(1-b_1b_2)\eta\eta'. 
\eeq
By definition of $\theta_k$, it holds that $Kb_1b_2\theta_k=1- (1-b_1b_2)\eta_k$. Plugging it into \eqref{lem-LBconstruct-4} gives the second equality in \eqref{lem-LBconstruct-1}.

We now prove the claim. We need to verify \eqref{cond-h}-\eqref{cond-A}, and show $\|\frac{a_j}{\|a_j\|_1} - \eta^*\|\leq C\sqrt{p/(Nn)}$ for non-anchor rows. First, since $a_j\propto \eta^*$ for non-anchor rows and $\|\eta^*\|_1=1$, we immediately have $\|\frac{a_j}{\|a_j\|_1} - \eta^*\|=0$. Next, consider \eqref{cond-h}. It is easy to see that 
\[
h_{\min}=p^{-1}\min\big\{Kb_1, \tfrac{1-b_1b_2}{1-b_2}\eta_{\min}\big\},  \qquad\mbox{where }\eta_{\min}\geq 1/2. 
\]
This gives \eqref{cond-h}. Last, consider \eqref{cond-A}. From \eqref{lem-LBconstruct-1}, $\lambda_{\min}(\Sigma_W)\geq K^{-1}$. Also, by \eqref{lem-LBconstruct-4}, 
\[
\lambda_{\min}(\Sigma_A)\geq K b_1b_2\theta_{\min}, \qquad \min_{1\leq k,\ell\leq K}\Sigma_A(k,\ell)\geq K^{-1}(1-b_1b_2)\eta^2_{\min}, 
\] 
where $\eta_{\min}\geq 1/2$ and $Kb_1b_2\theta_{\min}=1-(1-b_1b_2)\eta_{\max}\geq 1-3(1-b_1b_2)2>0$. Then, \eqref{cond-A} follows immediately. \qed


%
%
%

\subsection{Proof of Lemma~\ref{lem:LBconstruct2}} \label{subsec:lem-proof-LBconstruct2}

For each $(A^{(s)}, W^{(s)})$, we need to verify the conditions \eqref{cond-h}-\eqref{cond-A} and show that $\|\frac{a^{(s)}_j}{\|a^{(s)}_j\|_1} - \eta^*\|\leq C\sqrt{p/(Nn)}$ for non-anchor rows. 
Each $A^{(s)}$ is obtained by perturbing some non-anchor rows of $A^{(0)}$ with $\pm (\alpha_n ,\alpha_n ,\ldots,\alpha_n )$. Since $h_{\min}\geq C^{-1}p$ for $A^{(0)}$ and $\alpha_n=O(\tfrac{1}{\sqrt{Nnp}})\ll \tfrac{1}{p}$, we still have $h_{\min}\geq C^{-1}p^{-1}$ for $A^{(s)}$. This gives \eqref{cond-h}.  

To verify \eqref{cond-A}, we first notice that $\Sigma_W$ remains unchanged. As a result, it suffices to prove that   
\beq \label{lem-LBconstruct-2}
\|\Sigma_A^{(s)}-\Sigma_A^{(0)}\|_{\max} = O\Bigl(\sqrt{\frac{p}{Nn}}\Bigr). 
\eeq 
Once \eqref{lem-LBconstruct-2} is true, since $K$ is finite and $p/(Nn)=o(1)$, the quantities about $\Sigma_A$ in \eqref{cond-A} change by $o(1)$ when we perturb $A^{(0)}$ to $A^{(s)}$. Hence, \eqref{cond-A} continues to hold. Below, we show \eqref{lem-LBconstruct-2}. 
Fix $s$. By definition, for each $j$ with $\omega_j^{(s)}\neq 0$,    
\beq \label{lem-LBconstruct-3}
\left\{ \begin{array}{l}
a_{p-p_1+j}^{(s)} = \tfrac{1-b_1b_2}{p(1-b_2)}\cdot 
(\eta_1+\epsilon_n, \eta_2+\epsilon_n,\ldots, \eta_K+\epsilon_n),\\
 a_{p-p_1+j+m}^{(s)} = \tfrac{1-b_1b_2}{p(1-b_2)}\cdot 
(\eta_1-\epsilon_n, \eta_2+\epsilon_n,\ldots, \eta_K-\epsilon_n),  
\end{array}\right. \mbox{where $\epsilon_n\equiv \frac{p(1-b_2)\alpha_n}{1-b_1b_2}$}. 
\eeq
Hence, the $(p-p_1+j)$-th row of the matrix $H^{-1/2}A$ is equal to $\sqrt{\tfrac{1-b_1b_2}{p(1-b_2)(K+K\epsilon_n)}}\cdot (\eta_1+\epsilon_n, \eta_2+\epsilon_n, \ldots, \eta_K+\epsilon_n)$. The contribution of this row to the change of the $(k,\ell)$-th entry of $\Sigma_A$ is 
\[
\frac{1-b_1b_2}{pK(1-b_2)}\cdot \Big[ \frac{(\eta_k+\epsilon_n)(\eta_\ell+\epsilon_n)}{(1+\epsilon_n)} - \eta_k\eta_\ell \Big] = O(p^{-1}\epsilon_n). 
\]
Similarly, the $(p-p_1+j+m)$-th row contributes a change of $O(p^{-1}\epsilon_n)$ to each entry of $\Sigma_A$. Since at most $(1-b_2)p$ rows are perturbed when we construct $A^{(s)}$ from $A^{(0)}$, the total change on $\Sigma_A(k,\ell)$ is $O(\epsilon_n)=O(p\alpha_n)=o(1)$. This proves \eqref{lem-LBconstruct-2}.  

To show $\|\frac{a^{(s)}_j}{\|a^{(s)}_j\|_1} - \eta^*\|\leq C\sqrt{p/(Nn)}$ for non-anchor rows, we note by \eqref{lem-LBconstruct-3}, $\tilde{a}_{j}^{(s)}=\frac{1}{K(1\pm \epsilon_n)}(\eta_1\pm\epsilon_n,\eta_2\pm\epsilon_n,\ldots,\eta_K\pm\epsilon_n)$ for those perturbed rows. It follows that $\|\tilde{a}_{j}^{(s)}-\tilde{a}_{j}^{(0)}\|=O(\epsilon_n)$, where $\epsilon_n=O([p/(Nn)]^{1/2})$.  
\qed

\section{Proof of Proposition~\ref{prop:under-fitting} (misspecified $K$)}
By Lemma~\ref{lem:IdealSimplex}, $r_j = \sum_{k=1}^K \pi_j(k)v_k^*$, where $\pi_j$ is a nonnegative vector with a unit sum. We restrict this vector equation to the first $(m-1)$ coordinates. It gives
\[
r_j^{(m)} = \sum_{k=1}^K \pi_j(k) v_k^{(m)}. 
\]
Therefore, each $r_j^{(m)}$ is in the convex hull of $v_1^{(m)},\ldots,v_K^{(m)}$. Furthermore, by the anchor-word condition and Lemma~\ref{lem:IdealSimplex}, each $v_k^{(m)}$ is equal to $r_j^{(m)}$ for some anchor word $j$. 
This proves that the convex hull of $r_1^{(m)},\ldots,r_p^{(m)}$ is exactly the convex hull of $v_1^{(m)},\ldots,v_K^{(m)}$, which is a simplex with $K$ vertices. 
However, this simplex may be degenerate. There always exists a unique $K_m\leq K$ such that this $K$-vertex simplex is a non-degenerate $K_m$-vertex simplex.  We now show that each vertex of this non-degenerate simplex must be one of $v_1^{(m)},\ldots, v_K^{(m)}$. 
If this is not true, there exists a point $x\in\mathbb{R}^{m-1}$ in this non-degenerate simplex, such that it cannot be expressed as a convex combination of  $v_1^{(m)},\ldots, v_K^{(m)}$. However, since restricting a vector to the first $(m-1)$ coordinates is a linear projection, there must exist a point $y\in\mathbb{R}^{K-1}$ in the original Ideal Simplex such that $x$ is obtained from restricting $y$ to its first $(m-1)$ coordinates. Note that $y$ is a convex combination of $v_1^*,\ldots, v_K^*$. It follows that $x$ must be a convex combination of $v_1^{(m)},\ldots,v_K^{(m)}$. This yields a contradiction. 

We then study the output of Topic-SCORE when $K_m=m$ and $v_1^{(m)},\ldots,v_m^{(m)}$ are the vertices of the non-degenerate simplex. Define
\[
Q =\begin{pmatrix}
1 & \ldots & 1\\
v_1^* & \ldots & v_K^*
\end{pmatrix}, \quad Q_1 = \begin{pmatrix}
1 & \ldots & 1\\
v_1^{(m)} & \ldots & v_K^{(m)}
\end{pmatrix}, \quad Q_2 = \begin{pmatrix}
1 & \ldots & 1\\
v_1^{(m)} & \ldots & v_m^{(m)}
\end{pmatrix}. 
\]
Here, $Q_1$ is a sub-matrix of $Q$ by restricting to the first $m$ rows, and $Q_2$ is a sub-matrix of $Q_1$ by restricting to the first $m$ columns. For each $k>m$, using the notation $\beta_k$, we have $v_k^{(m)}=\sum_{\ell=1}^m \beta_k(\ell)v_\ell^{(m)}$. It follows that
\beq \label{proof-misK-1}
Q_1  = Q_2\cdot[ I_m,\; B].  
\eeq
In Topic-SCORE, we conduct vertex hunting on rows of $R^{(m)}$ and express each $r_j^{(m)}$ as a  convex combination of $m$ vertices of the non-degenerate simplex, where the convex combination coefficient vector is denoted by $\pi_j^{(m)}$. 
In matrix form, these operations are equivalent to letting  $\pi_j^{(m)}=Q_2^{-1}[1, (r_j^{(m)})']'$. 
Letting $\Pi^{(m)}$ be the $p\times m$ matrix by stacking $\pi_1^{(m)},\ldots,\pi_p^{(m)}$ together, we have
\beq \label{proof-misK-2}
\Pi^{(m)} = \bigl[ {\bf 1}_p,\; R^{(m)}\bigr]\cdot (Q_2^{-1})'. 
\eeq
In Section~\ref{subsec:proof-Section2}, we have shown the following equation (see \eqref{oracle-key-4}, where the $\diag(V_1)$ there is indeed the $\diag(q)$ in Lemma~\ref{lem:Pi-to-A}):
\[
[ {\bf 1}_p,\; R] = [\mathrm{diag}(\xi_1)]^{-1}M_0^{-1/2}A \cdot \mathrm{diag}(q)\cdot Q'. 
\]
Restricting to the first $m$ columns, we get
\beq \label{proof-misK-3}
\bigl[ {\bf 1}_p,\; R^{(m)}\bigr]  = [\mathrm{diag}(\xi_1)]^{-1}M_0^{-1/2}A \cdot \mathrm{diag}(q)\cdot Q_1'.
\eeq
We first plug \eqref{proof-misK-3} into \eqref{proof-misK-2} and then use the expression of $Q_1$ in \eqref{proof-misK-1}. It gives
\begin{align*}
\Pi^{(m)} & =  [\mathrm{diag}(\xi_1)]^{-1}M_0^{-1/2}A \cdot \mathrm{diag}(q)\cdot Q_1'\cdot (Q_2^{-1})'\cr
& = [\mathrm{diag}(\xi_1)]^{-1}M_0^{-1/2}A \cdot \mathrm{diag}(q)\cdot \begin{bmatrix} I_m\\ B'\end{bmatrix} Q_2'\cdot (Q_2^{-1})'\cr
&= [\mathrm{diag}(\xi_1)]^{-1}M_0^{-1/2}A \cdot \mathrm{diag}(q)\begin{bmatrix} I_m\\ B'\end{bmatrix}. 
\end{align*}
Equivalently, 
\[
M_0^{1/2}\cdot \diag(\xi_1)\cdot \Pi^{(m)} = A \cdot \mathrm{diag}(q)\begin{bmatrix} I_m\\ B'\end{bmatrix}. 
\]
In Topic-SCORE, the $j$th column of $A^{(m)}$ is obtained by re-normalizing the $j$th column of $M_0^{1/2}[\diag(\xi_1)] \Pi^{(m)}$. The claim follows immediately. \qed

\end{document}